\documentstyle[emulateapj,epsf,psfig]{article}

\begin{document}
\newcommand{\COBE}{{\sl COBE\/}}
\newcommand{\MAP}{{\sl MAP\/}}
\newcommand{\Planck}{{\sl Planck\/}}
\slugcomment{Accepted by The Astrophysical Journal}
\lefthead{Seager, Sasselov \& Scott}
\righthead{How exactly did the Universe become neutral?}

\hyphenation{an-is-o-tro-pies}
\hyphenation{an-is-o-tro-py}



\title{How exactly did the Universe become neutral?}

\author{Sara Seager\altaffilmark{1}, Dimitar D. Sasselov,}
\affil{Astronomy Department, Harvard University,\\
60 Garden Street, Cambridge, MA, 02138}

\and

\author{Douglas Scott}
\affil{Department of Physics \& Astronomy,\\
University of British Columbia, Vancouver, BC, V6T 1Z1}

\altaffiltext{1}{Present address: Institute for Advanced Study,
Olden Lane, Princeton, NJ, 08540}
 


\begin{abstract}
We present a refined treatment of H, \ion{He}{1}, and \ion{He}{2}
recombination in the early 
Universe. The difference from previous calculations is that we use
multi-level atoms and evolve the population of each level with
redshift by including all bound-bound and bound-free
transitions.  In this framework we follow several hundred atomic
energy levels for H, \ion{He}{1}, and \ion{He}{2} combined.

The main improvements of this method over
previous recombination calculations are: (1) allowing excited
atomic level populations to
depart from an equilibrium distribution;
(2) replacing
the total recombination coefficient with recombination to and
photoionization from
each level calculated directly at each redshift step;
and (3) correct treatment of the \ion{He}{1} atom, including the
triplet and singlet states.

We find that the ionization fraction $x_{\rm e} \equiv n_{\rm e}/n_{\rm H}$
is approximately 10\%
smaller at redshifts $\lesssim800$ than in previous calculations,
due to the non-equilibrium of the excited
states of H, which is caused by the strong but cool radiation field
at those redshifts.  In addition we find that \ion{He}{1} recombination is
delayed compared with previous calculations, and occurs only just before
H recombination.
These changes in turn can affect the predicted power spectrum of microwave
anisotropies at the few percent level.

Other improvements such as including 
molecular and ionic species of H, including complete heating and cooling terms
for the evolution of the matter temperature, including 
collisional rates, and including feedback of the secondary spectral
distortions on the radiation field, produce negligible change to
$x_{\rm e}$.  The lower $x_{\rm e}$ at low $z$
found in this work affects the abundances of H molecular and ionic
species by 10--25\%.  However this difference is probably not larger
than other uncertainties in the reaction rates.
\end{abstract}

\section{Introduction}
The photons that Penzias \& Wilson (1965) detected coming from all
directions with a temperature of about $3\,$K have traveled freely since their
last Thomson scattering, when the Universe became cool enough for the
ions and electrons to form neutral atoms.
During this recombination epoch, the
opacity dropped precipitously, matter and radiation were decoupled,
and the anisotropies of the Cosmic Microwave Background radiation (CMB) were
essentially
frozen in. These anisotropies of the CMB have now been detected 
on a range of scales (e.g.~White, Scott, \& Silk 1994;
Smoot \& Scott 1998) and developments in the field have been very rapid.
Recently, two post-\COBE\ missions, the Microwave Anisotropy Probe
(\MAP) and the \Planck\ satellite, were approved with the major science goal of
determining
the shape of the power spectrum of anisotropies with experimental
precision at a level similar to current theoretical predictions.

Detailed understanding of the recombination process is crucial for
modeling the power spectrum of CMB anisotropies. Since the seminal work
of the late 1960s (Peebles 1968; Zel'dovich et al.~1968), several
refinements have been introduced by, for example,
Matsuda, Sato, \& Takeda (1971), Zabotin \& Nasel'skii (1982),
Lyubarsky \& Sunyaev (1983), Jones \& Wyse (1985), Krolik (1990),
and others, but in fact little has changed (a fairly comprehensive overview
of earlier work on recombination is to be found in Section~IIIC of
Hu et al.~1995, hereafter HSSW).
More recently, refinements have been made independently in
the radiative transfer to calculate secondary spectral distortions
(Dell'Antonio \& Rybicki 1993; Rybicki \& Dell'Antonio 1994),
and in the chemistry (Stancil, Lepp, \& Dalgarno 1996b). These
improvements may have noticeable effects (at the 1\% level) on the
calculated shapes of the power spectrum of anisotropies. Given the
potential to measure important cosmological parameters
with \MAP\ and \Planck\
(e.g.~Jungman et al.~1996; Bond, Efstathiou, \& Tegmark~1997;
Zaldarriaga, Spergel, \& Seljak~1997; Eisenstein, Hu, \& Tegmark~1998;
Bond \& Efstathiou~1998) it is of
great interest to make a complete and detailed calculation of the
process of recombination.  Our view is that this is in principle such a
simple process that it should be so well understood that it could never
affect the parameter estimation endeavor.

Our motivation is to carry out a `modern' calculation of the cosmic
recombination process.  All of the physics is well understood, and so it
is surprising that cosmologists have not moved much beyond the solution of a
single ODE, as introduced in the late 1960s.  With today's computing power,
there is no need to make the sweeping approximations that were implemented
30 years ago. We therefore attempt to calculate to as great an extent
as possible the {\it full\/} recombination problem.  The other difference compared with
three decades ago is that we are now concerned with high precision
calculations, because of the imminent prospect of high fidelity data.

It is our intention to present here a coupled treatment of the
non-equilibrium radiative transfer and the detailed chemistry.
The present investigation was motivated by indications (HSSW) that multi-level
non-equilibrium effects in H and He, as well as in some molecular
species, may have measurable effects on the power spectrum of CMB
anisotropies by affecting the low-$z$ and high-$z$ tails
of the visibility function 
${\rm e}^{-\tau}\slantfrac{{d\tau}}{{dz}}$ (where $\tau$ is the optical depth).

To that effect this paper presents a study of the recombination era by
evolving neutral and ionized species of H and He, and molecular species of H
simultaneously with the matter temperature.  We believe our work represents
the most accurate picture to date of how exactly the Universe as a whole
became neutral.

\section{Basic Theory}
\subsection{The Cosmological Picture}

We will assume that we live in a homogeneous, expanding Universe within
the context of the canonical Hot Big Bang paradigm.  The general picture
is that at some sufficiently early time the Universe can be regarded as
an expanding plasma of hydrogen plus some helium, with
around $10^9$ photons per
baryon and perhaps some non-baryonic matter.  As it expanded
and cooled there came a time when the protons were able to keep hold of the
electrons and the Universe became neutral.  This is the period of cosmic
recombination.

In cosmology it is standard to use redshift $z$ as a time coordinate, so
that high redshift represents earlier times.  Explicitly, $a(t)=1/(1+z)$
is the scale factor of the Universe, normalized to be unity today, and
with the relationship between scale factor and time depending on the
particular cosmological model.  It is sufficient to consider only hydrogen
and helium recombination, since the other elements exist in minute amounts.
The relevant range of redshift is then $\lesssim10{,}000$, during which the
densities are typical of those astrophysicists deal with every day, and
the temperatures are low enough that there are no relativistic effects.

\subsection{The Radiation Field}
\label{sec-radfield}
In describing the radiation field and its interaction with matter,
we must use a specific form of the radiative transfer equation. It
should describe how radiation is absorbed, emitted and scattered as it
passes through matter in a medium which is homogeneous, isotropic,
infinite, and expanding. The basic time-dependent form
of the equation of transfer is
\begin{eqnarray} 
\lefteqn{\frac{1}{c}\frac{\partial{I(\vec{r},\hat{n},{\nu},t)}}{\partial{t}}
 + \frac{\partial{I(\vec{r},\hat{n},{\nu},t)}}{\partial{l}}=}\nonumber \\
 & & j(\vec{r},\hat{n},{\nu},t) - \kappa(\vec{r},\hat{n},{\nu},t)
 I(\vec{r},\hat{n},{\nu},t).
\label{eq:radtrans}
\end{eqnarray}  
Here the symbols are:
$I(\vec{r},\hat{n},{\nu},t)$ is the specific intensity of
radiation at
position $\vec{r}$, traveling in direction $\hat{n}$ (the unit
direction vector), with
frequency $\nu$
at a time $t$ (in units of ergs s$^{-1}$cm$^{-2}$Hz$^{-1}$sr$^{-1}$);
$l$ is the path length along the ray (and is a coordinate-independent
path length); $j$ is the emissivity,
which is calculated by summing products of upper excitation state
populations and transition probabilities over all relevant processes
that can release a photon at frequency $\nu$, including electron
scattering; and $\kappa$ is the absorption coefficient, which is the
product of an atomic absorption cross section and the number density
of absorbers summed over all states that can interact with photons of
frequency $\nu$.
 
In the homogeneous and isotropic medium of the early Universe, we
can integrate equation~(\ref{eq:radtrans}) over all solid angles
$\omega$ (i.e.~integrating over the unit direction vector $\hat{n}$):
\begin{eqnarray} 
\lefteqn{\frac{4\pi}{c}\frac{\partial{J(\vec{r},\nu,t)}}{\partial{t}} +
\vec{\nabla}{\cdot}\vec{F}(\vec{r},\nu,t) =} \nonumber\\
& & \mbox{} - \oint\Big[j(\vec{r},\hat{n},{\nu},t) -
\kappa(\vec{r},\hat{n},{\nu},t)I(\vec{r},\hat{n},{\nu},t)\Big]\, d\omega.
\label{eq:angleav}
\end{eqnarray}
Here $J(\nu,t)$ is the mean intensity, the zeroth order moment of
the specific intensity over all angles (in units of
ergs ${\rm s}^{-1}{\rm cm}^{-2}{\rm Hz}^{-1}$), $\vec{F}$ is the flux of
radiation, which is the net rate of radiant energy flow across an
arbitrarily oriented surface per unit time and frequency interval, and
$c$ is the speed of light. If
the radiation field is isotropic there is a ray-by-ray cancellation
in the net energy transport across a surface and the net flux is
zero.
Also, because of the isotropy of the radiation field and the medium
being static, we can drop the dependence upon angle of $j$ and
$\kappa$ in equation~(\ref{eq:angleav}).
With the definition of $J$,
$\oint Idw = 4{\pi}J$, this simplifies equation~(\ref{eq:angleav}) to:
\begin{equation}
\label{eq:rt-2}
\frac{1}{c}\frac{\partial{J(\nu,t)}}{\partial{t}} =
j(\nu,t) - \kappa(\nu,t)J(\nu,t).
\end{equation}

The above equation is for a static medium. An isotropically expanding
medium would reduce the number
density of photons due to the expanding volume, and reduce their
frequencies due to redshifting. The term due to the density change
will be simply a $3\frac{\dot{a}(t)}{a(t)}$ factor, while the redshifting
term will involve the frequency derivative of $J(\nu,t)$ and hence a 
$\nu\frac{\dot{a}(t)}{a(t)}$ factor.

Then the equation for the evolution
of the radiation field as affected by the expansion and the sources and
sinks of radiation becomes
\begin{eqnarray}
\lefteqn{\frac{dJ(\nu,t)}{dt} =
 \frac{\partial{J(\nu,t)}}{\partial{t}} -
 \nu H(t)\frac{\partial{J(\nu,t)}}{\partial{\nu}}}\nonumber \\
 & & \quad = -3H(t)J(\nu,t) + c\left[j(\nu,t) - \kappa(\nu,t)J(\nu,t)\right],
\label{eq:radfield}
\end{eqnarray}
where $H(t)\equiv {\dot{a}}/a$.

This equation is in its most general form and difficult to solve; 
fortunately we can make two significant
simplifications, because the primary spectral
distortions{\footnote{Not to be confused here with the power spectrum of
spatial anisotropies.}} are of
negligible intensity (Dell'Antonio \& Rybicki 1993) and the quasi-static
solution for spectral line profiles is valid (Rybicki \& Dell'Antonio 1994).
The first simplification is that 
for the purposes of this paper (in which we do not study secondary spectral
distortions),
we set $J(\nu,t)= B(\nu,t)$, the Planck function, which is observed to
approximate $J(\nu)$ to at least 1 part in 10$^4$ (Fixsen et
al.~1996). Thus
we eliminate explicit frequency integration from the simultaneous integration
of all equations (\S2.6). The validity of this assumption is shown in
\S\ref{sec-Distortions} where we follow the dominant secondary
distortions of H Ly$\,\alpha$ and H
2-photon by including their feedback on the recombination process, and find
that the secondary spectral distortions from the other Lyman lines and He are
not strong enough to feed back on the recombination process. 

The second simplification is in the treatment of the evolution of the
resonance lines
(Ly$\,\alpha$, etc.) which must still be treated explicitly -- because
of cosmological redshifting they cause $J(\nu,t) \neq B(\nu,t)$ in the
lines. These we call the primary distortions. We use escape 
probability methods for moving (expanding) media (\S\ref{sec-Sobolev} and
\ref{sec-3level}).
This simplification is not an approximation, but is an exact
treatment -- a simple solution to the multi-level radiative transfer
problem afforded by the physics of the expanding early Universe.

Not only does using $B(\nu,t)$ with the escape probability method
instead of $J(\nu,t)$ simplify the 
calculation and reduce computing time
enormously, but the effects from following the actual radiation
field will be small compared to the main improvements of our
recombination calculation which are the level-by-level treatment of H,
\ion{He}{1} and \ion{He}{2},
calculating recombination directly, and the correct treatment of
\ion{He}{1} triplet and singlet states.

\subsection{The Rate Equations}
\label{sec-species}
The species we evolve in the expanding Universe are \ion{H}{1}, \ion{H}{2}, 
 \ion{He}{1} , \ion{He}{2}, \ion{He}{3},  e${^-}$, H${^-}$, H$_{2}$,
and H$_{2}{^+}$. 
The chemistry of the early
Universe involves the reactions of association and dissociation among
these species, facilitated by interactions with the radiation field,
$J(\nu,t)$.
The rate equations
for an atomic system with $N$ energy levels can be described as
\begin{eqnarray}
a(t)^{-3}\frac{d\left(n_{i}(t)a(t)^{3}\right)}{dt}
 \,{=}&\Big[n_{\rm e}(t) n_{\rm c}(t) P_{{\rm c}i}
   - n_i(t) P_{i{\rm c}}\Big] \nonumber \\
 {+}&\!\! \sum_{j=1}^N \Big[n_j(t) P_{ji} - n_i(t) P_{ij}\Big],
\end{eqnarray}
where the $P_{ij}$ are the rate coefficients between bound levels $i$ and
$j$, and the $P_{i{\rm c}}$ are the rate coefficients between bound levels and
the continuum ${\rm c}$:
$P_{ij} = R_{ij} + n_{\rm e} C_{ij}$ and $P_{i{\rm c}}
 = R_{i{\rm c}} + n_{\rm e} C_{i{\rm c}}$,
where $R$ refers to radiative rates and $C$ to collisional rates.
Here the $n$s are physical (as opposed to comoving) number
densities: $n_i$ refers to number density of the $i$th excited atomic
state, $n_{\rm e}$ to the number density of electrons, and
$n_{\rm c}$ to the number density of a continuum particle such as a proton,
\ion{He}{2} or \ion{He}{3}.
$a(t)$ is the cosmological scale factor.  
The rate equations for molecules take
a slightly different form because their formation and destruction
depends on the rate coefficients for the reactions discussed in
\S\ref{sec-Chem}, 
and molecular bound states are not included.

\subsubsection{Photoionization and Photorecombination}
\label{sec-Recomb}
By calculating photorecombination rates $R_{{\rm c}i}$ directly to each level
for multi-level H,
\ion{He}{1}, and \ion{He}{2} atoms, we avoid the problem of finding an accurate
recombination coefficient, the choice of which has a large effect on
the power spectra (HSSW).

Photoionization rates are calculated by
integrals of the incident radiation field $J(\nu,t)$ and the bound-free cross
section $\sigma_{i{\rm c}}(\nu)$. The photoionization rate 
in s$^{-1}$ is
\begin{equation}
\label{eq:defRic}
 R_{i{\rm c}} = 4\pi\int_{\nu_{0}}^{\infty}
  \frac{\sigma_{i{\rm c}}(\nu)}{h_{\rm P}\nu}J(\nu,t)d\nu.
\end{equation}
Here $i$ refers to the $i$th excited state and
$c$ refers to the continuum. $\nu_0$ is the threshold frequency for
ionization from the $i$th excited state. The radiation field
$J(\nu,t)$ depends on frequency $\nu$ and time $t$. With $n_i$ as the
number density of the $i$th excited state, the number of
photoionizations per unit volume per unit time (hereafter
photoionization rate) is $n_i R_{i{\rm c}}$. 

By using the principle of detailed balance
in the case of local thermodynamic equilibrium (LTE), the radiative
recombination rate
can be calculated from the photoionization rate.
Then, as described below, the photorecombination
rate can be generalized to the non-LTE case by scaling the LTE
populations with the actual populations and substituting the actual
radiation field for the LTE radiation field.
In LTE the radiation field $J(\nu,t)$ is the Planck
function $B(\nu,t)$.
$B(\nu,t)$ is a function of time $t$ during recombination because
$T_{\rm R} = 2.728 (1 + z(t))\,$K. We
will call the LTE temperature $T$ ($T = T_{\rm R} = T_{\rm M}$ at
early times), where $T_{\rm R}$ is the radiation temperature, and
$T_{\rm M}$ the matter temperature.
To emphasize the Planck function's dependence on temperature,
we will use $B(\nu,T)$ where $T$ is a function of time.

By detailed balance in LTE we have 
\begin{displaymath}
\label{eq:db1}
 \left[n_{\rm e} n_{\rm c}R_{{\rm c}i}\right]^{\rm LTE} =
 \left[n_i R_{i{\rm c}}\right]^{\rm LTE}\!.
\end{displaymath}

Radiative recombination includes spontaneous and stimulated
recombination, so we must rewrite the above equation as
\begin{eqnarray}
\label{eq:db2}
\lefteqn{\left[n_{\rm e} n_{\rm c}R_{{\rm c}i}\right]^{\rm LTE} =
 \left[n_{\rm e} n_{\rm c}R_{{\rm c}i}^{\rm spon}\right]^{\rm LTE}
 + \left[n_{\rm e} n_{\rm c}R_{{\rm c}i}^{\rm stim}\right]^{\rm LTE}} \\
  & & = \left( \left[n_i R_{i{\rm c}}\right]^{\rm LTE}
       -  \left[n_i R_{i{\rm c}}^{\rm stim}\right]^{\rm LTE}\right)
 +  \left[n_i R_{i{\rm c}}^{\rm stim}\right]^{\rm LTE}. \nonumber
\end{eqnarray}
Using the definition of $R_{i{\rm c}}$ in equation~(\ref{eq:defRic}), 
\begin{displaymath}
 \left[n_{\rm e} n_{c}R_{{\rm c}i}\right]^{\rm LTE}\! =
 4\pi n_{i}^{\rm LTE}\!\!\int_{\nu_{0}}^{\infty}\!
  \frac{\sigma_{i{\rm c}}(\nu)}{h_{\rm P}\nu}B(\nu,T)\!
 \left(1 - {\rm e}^{-h_{\rm P}\nu/k_{\rm B}T}\right)\!d\nu \nonumber
\end{displaymath}
\vspace{-3mm}
\begin{equation}
 \qquad\qquad {+}\ 4\pi n_{i}^{\rm LTE}\!\int_{\nu_{0}}^{\infty}
  \frac{\sigma_{i{\rm c}}(\nu)}{h_{\rm P}\nu}B(\nu,T)
   {\rm e}^{-h_{\rm P}\nu/k_{\rm B}T}d\nu.
\label{eq:recombLTE}
\end{equation}
The first term on the right hand side is the spontaneous recombination rate and the second term on the right hand side is the stimulated recombination rate.
Here $h_{\rm P}$ is Planck's constant and $k_{\rm B}$ is Boltzmann's constant.
The factor $(1 - {\rm e}^{-h_{\rm P}\nu/k_{\rm B}T})$ is the correction for
stimulated recombination (see Mihalas 1978, $\S$4--3 for a derivation of this
factor). 
Stimulated recombination can be treated as either negative ionization or as
positive recombination; the physics is the same (see Seager \& Sasselov,
in preparation, for some subtleties). 
With the LTE expression for recombination (equation~(\ref{eq:recombLTE})),
it is easy
to generalize to the non-LTE case, considering spontaneous
and stimulated recombination separately. Because the matter
temperature $T_{\rm M}$ and the radiation temperature $T_{\rm R}$
differ at low $z$, it is important to understand how recombination
depends on each of these separately.

Spontaneous recombination involves a free electron but its calculation
requires no knowledge of the local radiation field,
because the photon energy is derived from the electron's kinetic energy. In
other words, whether or not LTE is valid, the LTE spontaneous recombination
rate holds per ion, as long as the velocity distribution is Maxwellian. The
local Planck function (as representing the Maxwell distribution) depends on
$T_{\rm M}$, because the Maxwell distribution describes a collisional process.
Furthermore, since the 
Maxwellian distribution depends on $T_{\rm M}$, so does the
spontaneous rate. To get the non-LTE rate, we only have
to rescale the LTE ion density to the actual ion density:
\begin{eqnarray}
\lefteqn{n_{\rm e} n_{\rm c}R_{{\rm c}i}^{\rm spon} =
 4\pi \frac{n_{\rm e} n_{\rm c}}{(n_{\rm e} n_{\rm c})^{\rm LTE}}n_{i}^{\rm LTE}
\ \times}\nonumber \\
 & & \qquad\int_{\nu_{0}}^{\infty}\frac{\sigma_{i{\rm c}}(\nu)}
 {h_{\rm P}\nu}B(\nu,T_{\rm M})
 \left(1 - {\rm e}^{-h_{\rm P}\nu/k_{\rm B}T_{\rm M}}\right)d\nu
\end{eqnarray}
\vspace{-3mm}
\begin{equation}
{\!{=}\, 4\pi n_{\rm e} n_{\rm c}
 \left(\frac{n_{i}}{n_{\rm e} n_{\rm c}}\right)^{\rm LTE}\!\!
 \int_{\nu_{0}}^{\infty}\frac{\sigma_{i{\rm c}}(\nu)}{h_{\rm P}\nu}
 \frac{2h_{\rm P}\nu^{3}}{c^{2}}
 {\rm e}^{-h_{\rm P}\nu/k_{\rm B}T_{\rm M}}d\nu.}
\label{eq:spon}
\end{equation}

To generalize the stimulated recombination rate from the LTE rate to the
non-LTE rate, we rescale the LTE ion density to the
actual ion density, and  replace the LTE
radiation field by the actual radiation field
$J(\nu,t)$ because that is what is `stimulating' the recombination.
The correction for stimulated recombination depends on $T_{\rm M}$,
because the recombination process is collisional; the term always remains
in the LTE form because equation~(\ref{eq:recombLTE}) was derived from
detailed balance.  So we have
\begin{eqnarray}
\lefteqn{n_{\rm e} n_{\rm c} R_{{\rm c}i}^{\rm stim} =
  4\pi \frac{n_{\rm e} n_{\rm c}}
{(n_{\rm e} n_{\rm c})^{\rm LTE}}n_{i}^{\rm LTE} \times} \nonumber \\
 & & {\displaystyle \int_{\nu_{0}}^{\infty}
 \frac{\sigma_{i{\rm c}}(\nu)}{h_{\rm P}\nu}J(\nu,t)
 {\rm e}^{-h_{\rm P}\nu/k_{\rm B}T_{\rm M}}d\nu.}
\end{eqnarray}
Therefore, the total non-LTE recombination rate $(R_{{\rm c}i}^{\rm spon} +
R_{{\rm c}i}^{\rm stim})$ is
\begin{eqnarray}
\lefteqn{n_{\rm e} n_{\rm c} R_{{\rm c}i} = n_{\rm e} n_{\rm c}
 \left(\frac{n{_i}}{n_{\rm e} n_{\rm c}}\right)^{\rm LTE} \times}\nonumber \\
 & {\displaystyle 4\pi\int_{\nu_{i}}^{\infty}
 \frac{\sigma_{i{\rm c}}(\nu)}{h_{\rm P}\nu}
 \left[\frac{2h_{\rm P}\nu^{3}}{c^{2}}
 + J(\nu,t)\right]{\rm e}^{-h_{\rm P}\nu/k_{\rm B}T_{\rm M}}d\nu.}
\end{eqnarray}
The LTE population ratios
$\left( n_i / {n_{\rm e} n_{\rm c}}\right)^{\rm LTE}$
depend only on $T_{\rm M}$ through the Saha relation:
\begin{equation}
\label{eq:saha}
\left( \frac{n_{i}}{n_{\rm e} n_{\rm c}}\right)^{\rm LTE} =
 \left( \frac{h^{2}}{2\pi m_{\rm e}k_{\rm B}T_{\rm M}} \right)^{3/2}
 \frac{g_i}{2g_{\rm c}} {\rm e}^{E_{i}/k_{\rm B}T_{\rm M}}.
\end{equation}
Here $m_{\rm e}$ is the electron mass,
the atomic parameter $g$ is the degeneracy of the
energy level, and $E_{i}$ is the ionization energy of level $i$.
In the recombination calculation presented in this paper we 
use the Planck function $B(\nu,T_{\rm R})$ instead of the radiation field
$J(\nu,T)$ as described earlier.

In the early Universe
Case B recombination is used. This excludes recombinations to the
ground state and considers the Lyman lines to be
optically thick. An implied assumption necessary to compute the
photoionization rate is that the excited states (n~$\ge 2$) are in equilibrium
with the radiation. Our approach is more general than Case B, because
we don't consider the Lyman lines to be optically thick and
don't assume equilibrium among the excited states. For more details on the
validity of Case B recombination see \S\ref{sec-CaseB}.
To get the total recombination coefficient, we sum
over captures to all
excited levels above the ground state.

To summarize, the form of the total photoionization rate is
\begin{equation}
\label{eq:integral1}
\sum_{i>1}^N n_{i}R_{i{\rm c}} = \sum_{i>1}^N n_{i}4\pi\int_{\nu_{0}}^{\infty}
 \frac{\sigma_{i{\rm c}}(\nu)}{h_{\rm P}\nu}B(\nu,T_{\rm R})d\nu,
\end{equation}
and the total recombination rate is
\begin{eqnarray}
\lefteqn{\sum_{i>1}^N n_{\rm e} n_{\rm c} R_{{\rm c}i}
  = n_{\rm e}n_{\rm c}\sum_{i>1}^N
 \left(\frac{n_{i}}{n_{\rm e} n_{\rm c}}\right)^{\rm LTE} \times}\nonumber \\
 & \!\!\!{\displaystyle 4\pi\!\int_{\nu_{i}}^{\infty}
 \frac{\sigma_{i{\rm c}}(\nu)}{h_{\rm P}\nu}
 \left[\frac{2h_{\rm P}\nu^{3}}{c^{2}} + B(\nu,T_{\rm R})\right]
 {\rm e}^{-h_{\rm P}\nu/k_{\rm B}T_{\rm M}}d\nu.}
\label{eq:integral2}
\end{eqnarray}

\subsubsection{Comparison With the `Standard' Recombination
\label{sec-StandardRecomb}
Calculation of Hydrogen}
The `standard' recombination calculation refers to the calculation
widely used today and first derived by Peebles (1968,
1993) and Zel'dovich and collaborators (1968, 1983),
updated with the most recent parameters and recombination
coefficient (HSSW). See also \S\ref{sec-3level}.

For a 300-level H atom in our new recombination calculation, the
expressions (\ref{eq:integral1}) and (\ref{eq:integral2}) include
300 integrals at each redshift step.
The standard recombination calculation does not go through this
time-consuming task but avoids it entirely by using a
pre-calculated recombination coefficient that is a single expression dependent on $T_{\rm M}$ only.
The recombination coefficient to each excited state $i$ is defined by
\begin{equation}
\alpha_{i}(T_{\rm M}) = R_{{\rm c}i}^{\rm spon}.
\end{equation}
Here $\alpha$ is a function of $T_{\rm M}$, because spontaneous recombination
is a collisional process, as described previously.
The total Case B recombination coefficient ($\alpha_{\rm B}$) is obtained from
\begin{equation}
\alpha_{\rm B}(T_{\rm M}) = \sum_{i>1}^{N} \alpha_{i}(T_{\rm M}).
\end{equation}
We will refer to $\alpha_{\rm B}(T_{\rm M})$ as the `pre-calculated
recombination coefficient' because the recombination to each atomic level $i$
and the summation over $i$ are pre-calculated for LTE conditions.
See Hummer (1994) for an example of how these recombination
coefficients are calculated. 
Some more elaborate derivations of $\alpha(T_{\rm M}) = f(T_{\rm M}, n)$
have also been tried (e.g.~Boschan \& Biltzinger 1998).

The standard recombination calculation uses a photoionization
coefficient $\beta_{\rm B}(T_{\rm M})$ which is derived
from detailed balance using the recombination rate: 
\begin{equation}
{(n_i \beta_i)}^{\rm LTE} = ({n_{\rm e} n_{\rm p} \alpha_i})^{\rm LTE}.
\end{equation}
To get the non-LTE rate, one uses the actual populations $n_i$,
\begin{equation}
n_i \beta_i(T_{\rm M}) =
  n_i{\left(\frac{n_{\rm e} n_{\rm p}}{n_i}\right)}^{\rm LTE}
  \alpha_i(T_{\rm M}),
\end{equation}
or with the Saha relation (equation~(\ref{eq:saha})),
\begin{equation}
\label{eq:beta_saha}
n_i \beta_i(T_{\rm M}) = n_i \left( \frac{2\pi m_{\rm e}k_{\rm
B}T_{\rm M}}{h_{\rm P}^2} \right)^{3/2}
\frac{2}{g_{i}} {\rm e}^{-E_{i}/k_{\rm B}T_{\rm M}} \alpha_i(T_{\rm M}).
\end{equation}
Constants and variables are as described before.
To get the total photoionization rate, $\beta_i$ is summed over all
excited levels. Because 
the `standard' calculation avoids use of all levels $i$
explicitly, the $n_i$ are assumed to be in equilibrium with the
radiation, and thus can be related to the first excited state number
density $n_{2s}$ by the Boltzmann relation,
\begin{equation}
\label{eq:boltz}
n_i = n_{2s} \frac{g_i}{g_{2s}} {\rm e}^{-(E_2-E_i)/k_{\rm B}T_{\rm M}}.
\end{equation}
With this relation, the total photoionization rate is
\begin{equation}
\sum_{i>1}^N n_i \beta_i = n_{2s} \alpha_{\rm B}
 {\rm e}^{-E_{2s}/k_{\rm B}T_{\rm M}}
 \left( \frac{2\pi m_{\rm e}k_{\rm B}T_{\rm M}}{h_{\rm P}^2} \right)^{3/2} 
 \equiv n_{2s} \beta_{\rm B}.
\end{equation}
In this expression for the total photoionization rate, the excited
states are populated according to a Boltzmann distribution. $T_{\rm M}$
is used instead of $T_{\rm R}$, because the Saha and Boltzmann
equilibrium used in the derivation are collisional processes. The
expression says nothing about the excited levels being in equilibrium 
with the continuum, because the actual values of $n_{\rm e}$, $n_1$ and
$n_{2s}$ are used, and the $n_i$ are proportional to $n_{2s}$.

To summarize, the standard calculation uses a single expression for 
each of the total recombination rate and the total photoionization rate
that is dependent on
$T_{\rm M}$ only. The
total photoionization rate is
\begin{eqnarray}
\lefteqn{\sum_{i>1}^N n_i R_{i{\rm c}} = n_{2s} \beta_{\rm B}(T_{\rm M})}
 \nonumber \\
 & = n_{2s} \alpha_{\rm B}(T_{\rm M}) {\rm e}^{-E_{2s}/k_{\rm B}T_{\rm M}}
 \left(2\pi m_{\rm e} k_{\rm B} T_{\rm M} \right)^{3/2}/h_{\rm P}^3,
\label{eq:Betatot}
\end{eqnarray}
and the total recombination rate is
\begin{equation}
\sum_{i>1}^N n_{\rm e} n_{\rm p} R_{{\rm c}i} = n_{\rm e} n_{\rm p}
 \alpha_{\rm B}(T_{\rm M}).
\end{equation}
Comparing the right hand side of equation~(\ref{eq:Betatot}) to our
level-by-level total photoionization rate (equation~(\ref{eq:integral1})),
the main improvement in our method over the standard one is clear:
we use the actual excited level populations $n_i$,
{\it assuming no equilibrium distribution among them}. In this way
we can test the validity of the equilibrium assumption.
Far less important is that the standard recombination treatment cannot
distinguish
between $T_{\rm R}$ and $T_{\rm M}$, even though photoionization and
stimulated recombination are functions of 
$T_{\rm R}$ while spontaneous recombination is a function of $T_{\rm
M}$, as shown in
equations~(\ref{eq:integral1}) and~(\ref{eq:integral2}). The
non-equilibrium of excited states is important at the 
$10\%$ level in the residual ionization fraction for $z\lesssim800$, while using $T_{\rm R}$ in
photoionization and photoexcitation is only important at the few percent
level for $z\lesssim300$ (for typical cosmological models). 
Note that although the pre-calculated recombination coefficient
includes spontaneous recombination only,
stimulated recombination (as a function of $T_{\rm M}$)
is still included as negative
photoionization via detailed balance (see
equation~(\ref{eq:db2})). 

\subsubsection{Photoexcitation}
\label{sec-Sobolev}

In the expanding Universe, redshifting of the photons must be taken
into account (see equation~(\ref{eq:radfield})). Line photons emitted
at one position may be redshifted 
out of
interaction frequency (redshifted more than the width of the line) by
the time they reach another position in the 
flow.
We use the Sobolev escape probability to account for this, a method
which was first used for the expanding Universe by Dell'Antonio and
Rybicki (1993). 
The Sobolev escape probability (Sobolev 1946), also sometimes called the
large-velocity gradient approximation, is not an approximation but is
an exact, simple solution to the multi-level radiative transfer in the
case of a large velocity gradient. It is this solution which allows
the explicit inclusion of the line distortions to the radiation field
-- without it our detailed approach to the recombination problem would
be intractable. 
We will call the net bound-bound rate for each line transition $\Delta
R_{ji}$, where $j$ is the upper level and $i$ is the lower level: 
\begin{equation}
\label{eq:delrp}
\Delta R_{ji} = p_{ij} \left\{n_j\left[A_{ji} + B_{ji} B(\nu_{ij}, t)\right]
 - n_i B_{ij} B(\nu_{ij},t)\right\}.
\end{equation}
Here the terms $A_{ji}, B_{ji}, B_{ij}$ are the Einstein coefficients;
the escape probability $p_{ij}$ is the probability that photons associated
with this transition will `escape' without being further scattered or absorbed.
If $p_{ij} = 1$ the photons produced in the line transition escape to
infinity -- they contribute no
distortion to the radiation field. If $p_{ij} = 0$ no photons
escape to infinity; all of them get reabsorbed and the line is
optically thick. This is the case of primary distortions to the radiation
field, and the Planck function cannot be used for the line radiation.
In general $p_{ij} \ll 1$ for the Lyman lines and
$p_{ij} = 1$ for all other line transitions. With this method we have
described the redshifting of photons through the resonance lines and
found a simple solution to the radiative transfer problem for all
bound-bound transitions. The rest of this section is devoted to deriving
$p_{ij}$. 

For the case of no cosmological redshifting, the radiative rates per
cm$^{3}$ for transitions between
excited states of an atom are
\begin{eqnarray}
\label{eq:bbup}
n_{i}R_{ij} &=& n_{i}B_{ij} \overline{J} \\
\label{eq:bbdown}
{\rm and}\quad n_{j}R_{ji} &=& n_{j}A_{ji} + n_{j}B_{ji}\overline{J},
\end{eqnarray}
where
\begin{equation}
\label{eq:Jbar}
\overline{J} = \int_{0}^{\infty}J(\nu,t)\phi(\nu)d\nu,
\end{equation}
and $\phi(\nu)$ is the line profile function with its area normalized by
\begin{equation}
\label{eq:normalize}
 \int_{0}^{\infty}\phi(\nu)d\nu = 1.
\end{equation}
The line profile $\phi(\nu)$ is taken to be a Voigt function that includes
natural and Doppler broadening. In principle, equation~(\ref{eq:Jbar})
is the correct approach. In practise we take $\phi(\nu)$ as a delta
function, and use $J(\nu,t)$ instead of $\overline{J}$. The smooth
radiation field is essentially constant over the width of the line and 
so the line shape is not important; we get the same results using
$\overline{J}$ or $J(\nu,t).$

The Sobolev escape probability considers
the distance over which the expansion of the medium
induces a velocity difference equal to the thermal velocity (for the
case of a Doppler width): $L = v_{\rm th}/|{v^{\prime}}|$,
where $v_{\rm th}$ is the
thermal velocity width and $v^{\prime}$ the velocity gradient. The
theory is valid when this distance $L$ is much smaller than typical
scales of macroscopic variation of other quantities.

We follow Rybicki (1984) in the derivation of the
Sobolev escape probability. The general definition of escape
probability is given by the exponential extinction law,
\begin{equation}
p_{ij} = \exp[-\tau(\nu_{ij})],
\end{equation}
where $\nu_{ij}$ is the frequency for a given line transition, and
$\tau(\nu_{ij})$ is the monochromatic optical depth forward along a
ray from a given point to the boundary of the medium.  Here $\tau(\nu_{ij})$
is defined by
\begin{equation}
d\tau(\nu_{ij}) = -{\tilde k} \phi(\nu_{ij})dl,
\end{equation}
where ${\tilde k}$ is the integrated line absorption coefficient, so that the
monochromatic absorption coefficient or opacity is
$\kappa={\tilde k}\phi(\nu_{ij})$, and $l$ is the distance along the ray
from the emission point ($l=0$).
Rewriting the optical depth for a line profile function
(which has units of inverse frequency) of the dimensionless
frequency variable $x = (\nu - \nu_{ij}) / \Delta $,
with $\Delta$ the width of the line in Doppler units,
and $\nu_{ij}$ the central line frequency,
we have
\begin{equation}
d\tau(\nu_{ij}) = -\frac{\tilde k}{\Delta} \phi(x)dl.
\end{equation}
Here $\kappa$ is the absorption coefficient (defined in
equation~(\ref{eq:radtrans})),
with ${\tilde k}$ just dividing out the line profile
function, and
\begin{equation}
{\tilde k} = \frac{h_{\rm P}\nu}{4\pi}(n_iB_{ij} - n_jB_{ji}).
\end{equation}
Using the Einstein relations
$g_i B_{ij} = g_j B_{ji}$ and 
$A_{ji} = (2 h \nu^3/c^2) B_{ji}$
the absorption coefficient can also be written as
\begin{equation}
{\tilde k} = \frac{A_{ji}\lambda_{ij}^2}{8\pi}
 \left(n_i\frac{g_j}{g_i} - n_j\right).
\end{equation}

Note that distances along a ray $l$ correspond to shifts in
frequency $x$. This is due to the Doppler effect induced by the velocity
gradient and is the essence of the Sobolev escape probability approach.
For example, the Ly$\,\alpha$ photons cannot be reabsorbed in the
Ly$\,\alpha$ line if they redshift out of the frequency interaction range.
This case will happen at some frequency $x$, or at some distance $l$
from the photon emission point, where because of the expansion the
photons have redshifted out of the frequency interaction range.
For an expanding medium with a constant velocity gradient $v^\prime = dv/dl$,
the escape probability along a ray is then
\begin{equation}
p_{ij} = \exp\left[-\frac{\tilde k}{\Delta} \int_{0}^{\infty}\!
 \phi(x - l/L)dl \right] \equiv \exp
 \left[-\tau_{\rm S}\! \int_{-\infty}^{x}\!\phi({\tilde x})d{\tilde x} \right].
\end{equation}
The velocity field has in effect introduced an intrinsic escape mechanism
for photons; beyond the interaction limit with a given atomic transition, the photons can no longer be absorbed
by the material, even if it is of infinite extent, but escape
freely to infinity (Mihalas 1978).
Here the Sobolev optical thickness along the ray is defined by
\begin{equation}
\label{eq:Tau}
\tau_{\rm S} \equiv \frac{\tilde k}{\Delta}L,
\end{equation}
where L is the Sobolev length,
\begin{equation}
L = v_{\rm th}/|{v^{\prime}}| =
  \left.\sqrt{\frac{3k_{\rm B} T_{\rm M}}{m_{\rm atom}}}\right/|{v^{\prime}}|,
\end{equation}
and $\Delta$ is the width of the line, which in the case of Doppler
broadening is
\begin{equation}
\Delta = \frac{\nu_0}{c}\sqrt{\frac{3k_{\rm B} T_{\rm M}}{m_{\rm atom}}}.
\end{equation}
With these definitions, equation~(\ref{eq:Tau}) becomes
\begin{equation}
\tau_{\rm S} = \frac{\lambda_{ij} {\tilde k}}{|{v^{\prime}}|}.
\end{equation}
In the expanding Universe, the velocity gradient $v^{\prime}$ is given by the
Hubble expansion rate $H(z)$, and using the above definition for ${\tilde k}$,
\begin{equation}
\label{eq:sobtau}
\tau_{\rm S} = \frac{A_{ji} {\lambda_{ij}}^3 \left[n_i (g_j/g_i) - n_j\right]}
 {8\pi H(z)}.
\end{equation}
 
To find the Sobolev escape probability for the ray, we average over the
initial frequencies $x$, using the line profile function $\phi(x)$
from equation~(\ref{eq:normalize}),
\begin{displaymath}
p_{ij} =
\int_{-\infty}^{\infty}\!dx\phi(x)\exp\left[-\tau_{\rm S}\!
\int_{-\infty}^{x}\!\phi({\tilde x})d{\tilde x} \right]\! =\!
\int_{0}^{1}\!d\zeta\,\exp(-\tau_{\rm S}\zeta), \\
\end{displaymath}
\begin{equation}
\label{eq:indofphi}
{\rm i.e.}\quad p_{ij} = \frac{1 - \exp(-\tau_{\rm S})}{\tau_{\rm S}}.
\end{equation}
Note that this expression is independent of the line profile shape $\phi(x)$.
The escape probability $p_{ij}$ is defined as a frequency average at a
single point.
Finally, we must average over angle, but in the case of the isotropically 
expanding Universe, the angle-averaged Sobolev escape probability
takes the same form as $p_{ij}$ above.
For further details on the Sobolev escape probability see Rybicki
(1984) or Mihalas (1978), \S14.2. 
 
How does the Sobolev optical depth relate to the usual meaning of
optical depth? The optical depth for a specific
line at a specific redshift point is equivalent to the Sobolev optical
depth,
\begin{equation}
d\tau(\nu_{ij}) = -\tau_{\rm S} \phi(x)\frac{dl}{L}.
\end{equation}
If no other line or continua photons are redshifted into that
frequency range before or after the redshift point, then the optical
depth at a given frequency 
today (i.e.~summed over all redshift points) will be equivalent to the
Sobolev optical depth at that past point. Generally, behaviors in frequency
and space are interchangeable in a medium with a velocity gradient.
 
In order to derive an expression for the bound-bound rate
equations, we must consider the mean radiation field $\overline{J}$ in
the line.  For the case of spectral distortions $\overline{J}$ does not
equal the Planck function at the line frequency. We 
use the core saturation method (Rybicki 1984) to get $\overline{J}$ using
$p_{ij}$; from this we get
the net rate of deexcitations in that transition ($j\rightarrow i$), 
given in equation~(\ref{eq:delrp}).
In general, only the Lyman lines of H and \ion{He}{2}, and the
\ion{He}{1} n$^1$p--1$^1$s lines have $p_{ij} < 1$.
With this solution we have accomplished two things: described the
redshifting of photons through the resonance lines, and found a simple
solution to the radiative transfer problem for all bound-bound lines.

Peculiar velocities during the recombination era may cause line
broadening of the same order of magnitude as thermal broadening over
certain scales (A.~Loeb, private communication).
Because we use $J(\nu) = B(\nu,T)$, the
radiation field is essentially constant over the width of the line and
so the line shape is 
not important. Similarly, the peculiar velocities will not affect the
Sobolev escape probability because it is independent of line shape
(equation~(\ref{eq:indofphi})). If 
peculiar velocities were angle dependent there would be an effect on the
escape probability which is an angle-averaged function.
Only for computing spectral distortions to the CMB, where the line
shape is important, should
line broadening from peculiar velocities be included.

\subsubsection{Chemistry}
\label{sec-Chem}
Hydrogen molecular chemistry has been included because
it may affect the residual electron densities at low redshift ($z<200$).
During the recombination epoch, the H$_2$ formation reactions include
the H$^-$ processes
\begin{equation}
{\rm H}  +  {\rm e}^{-} \longleftrightarrow {\rm H}{^-} + \gamma,
\end{equation}
\begin{equation}
{\rm H}^{-} + {\rm H} \longleftrightarrow {\rm H}_{2} + {\rm e}^{-},
\end{equation}
and the $\mathrm{H_2}^+$ processes
\begin{equation}
{\rm H} + {\rm H}^{+} \longleftrightarrow {{\rm H}_2}^{+} + \gamma,
\end{equation}
\begin{equation}
{{\rm H}_2}^{+} + {\rm H} \longleftrightarrow {\rm H}_2 + {\rm H}^{+},
\end{equation}
together with
\begin{equation}
{\rm H} + {\rm H}_2 \longleftrightarrow {\rm H} + {\rm H} + {\rm H},
\end{equation}
and
\begin{equation}
{\rm H}_2 + {\rm e}^{-} \longleftrightarrow {\rm H} + {\rm H} + {\rm e}^{-}.
\end{equation}
(see~Lepp \& Shull 1984). The direct three-body process for H$_2$ formation
is significant only  at much higher densities.
The most recent rate coefficients and cross sections are listed with
their references in Appendix A (see also Cen~1992; Puy et al.~1993;
Tegmark et al.~1997; Abel et al.~1997; Galli \& Palla~1998). 

We have not included the molecular chemistry of Li, He, or D. 
In general, molecular chemistry only becomes important
at values of $z < 200$. Recent detailed analyses of H, D, and He chemistry
(Stancil et al.~1996a) and of Li and H chemistry (Stancil et al.~1996b;
Stancil \& Dalgarno~1997)
presented improved relative abundances
of all atomic, ionic and molecular species. The values are certainly too small
to have any significant effect on the CMB power spectrum.

There are two reasons for this.  The visibility function is
${\rm e}^{-\tau}\slantfrac{{d\tau}}{{dz}}$, where $\tau$ is the optical depth.
The main component of the optical depth is Thomson scattering by
free electrons. Because the populations of Li, LiH, HeH$^+$, HD, and
the other species are so small relative to the electron density,
they do not affect the
contributions from Thomson scattering. Secondly, these atomic species
and their molecules themselves make no contribution to the visibility
function because they have no strong opacities. HD has
only a weak dipole moment. And while LiH has a very strong dipole moment,
its opacity during this epoch is expected to be
negligible because of its tiny ($< 10^{-18}$) fractional abundance (Stancil,
et al.~1996b).  Similarly, the fractional abundance of
H$_2\mathrm{D}^+$ ($< 10^{-22}$) is too small to have an effect on
the CMB spectrum (Stancil et al.~1996a). For more details,
see Palla et al.~(1995) and Galli \& Palla (1998).
An interesting additional point is that 
because of the smaller energy gap between n=2 and the continuum
in \ion{Li}{1}, it actually recombines at a slightly lower redshift than
hydrogen does (see e.g.~Galli \& Palla 1998).  Of course this has
no significant cosmological effects.

We have also excluded atomic D from the calculation. D, like H,
has an atomic opacity much lower
than Thomson scattering for the recombination era conditions. D parallels H
in its reactions with electrons and protons, and recombines in the same
way and at the same time as H (see Stancil et al.~1996a). Although
the abundance of D is small ([D/H] $\simeq 10^{-5}$), its Lyman
photons are still trapped because
they are shared with hydrogen. This is seen, for example, by the
ratio of the isotopic shift of D Ly$\,\alpha$ to the width of the H
Ly$\,\alpha$ line, on
the order of $10^{-2}$. Therefore by excluding D we expect no change in the
ionization fraction, and hence none in the visibility function.

While the non-hydrogen chemistry is still extremely important for
cooling and triggering the collapse of primordial gas clouds, it is
not relevant for CMB power spectrum observations at the level measurable by
\MAP\ and \Planck.

\subsection{Expansion of the Universe}
The differential equations in time for the number densities and matter
temperature must be
converted to differential equations in redshift by multiplying by a
factor of $dz/dt$ . The redshift $z$ is related to time by the
expression
\begin{equation}
 \frac{dz}{dt} = -(1+z)  H(z),
\end{equation}
with scale factor
\begin{equation}
 a(t) = \frac{1}{1 + z}.
\end{equation}
Here $H(z)={\dot a}/a$ is the Hubble factor
\begin{eqnarray}
\lefteqn{H(z)^2 = H_0^2 \Big[\Omega_0(1+z)^4/(1+z_{\rm eq}) }
 \nonumber \\
 & & \qquad \mbox{} + \Omega_0(1+z)^3 + \Omega_K(1+z)^2 + \Omega_{\Lambda}\Big],
\end{eqnarray}
where $\Omega_0$ is the density contribution, $\Omega_K$ is the
curvature contribution and $\Omega_{\Lambda}$ is
the contribution associated with the cosmological constant,
with $\Omega_0 + \Omega_{\rm R} + \Omega_K + \Omega_{\Lambda} = 1$,
and $\Omega_{\rm R}=\Omega_0/(1+z_{\rm eq})$.
Here $z_{\rm eq}$ is the redshift of matter-radiation equality,
\begin{equation}
 1+z_{\rm eq} = \Omega_0\frac{3(cH_{0}){^2}}{8\pi G(1+f_{\nu})U},
\end{equation}
with $f_{\nu}$ the
neutrino contribution to the energy density in relativistic species
($f_\nu\simeq0.68$ for three massless neutrino types),
$G$ the gravitational constant, $U$ the photon energy density, and
$H_{0}$ the Hubble constant today, which will be written as
$100\,h\,{\rm km}\,{\rm s}^{-1}{\rm Mpc}^{-1}$.
Since we are interested in redshifts $z\sim z_{\rm eq}$,
it is crucial to include the radiation contribution explicitly (see HSSW).
 
\subsection{Matter Temperature}
The important processes that are considered in following the matter
temperature are Compton cooling, adiabatic cooling, and Bremsstrahlung
cooling. Less important but also included are photoionization
heating, photorecombination cooling, radiative and collisional line
cooling, collisional ionization cooling, and collisional recombination
cooling.
Note that throughout the relevant time period, collisions and Coulomb
scattering hold all the matter species at very nearly the same temperature.
`Matter' here means protons (and other nuclei), plus electrons, plus
neutral atoms; dark matter is assumed to be decoupled.

Compton cooling is a major source of energy
transfer between electrons and photons. It is described by
the rate of transfer of energy per unit volume between photons and
free electrons when the electrons are near thermal equilibrium with
the photons:
\begin{equation}
\label{eq:E1}
 \frac{dE_{{\rm e},\gamma}}{dt} = 
 \frac{4{\sigma_{\rm T}}Un_{\rm e}k_{\rm B}}{m_{\rm e}{c}}
 {(T_{\rm R} - T_{\rm M})},
\end{equation}
\begin{equation}
\label{eq:cool1}
{\rm or}\qquad \frac{dT_{\rm M}}{dt} =
 \frac{8}{3}\frac{\sigma_{\rm T} U n_{\rm e}}{m_{\rm e}{c}n_{\rm tot}}
 (T_{\rm R} - T_{\rm M}),
\end{equation}
(Weymann 1965), 
where $E_{{\rm e},\gamma}$ is the electron energy density, $k_{\rm
B}$, $m_{\rm e}$ and $c$ are constants as before,
$\sigma_{\rm T}$ is the Thomson scattering cross section, $T_{\rm R}$ is the
radiation temperature and $T_{\rm M}$ is the electron or matter temperature.
To get from equation~(\ref{eq:E1}) to equation~(\ref{eq:cool1}) we use
the energy of all particles; collisions among all particles keep them
at the same temperature.
Here $n_{\rm tot}$ represents the total
number density of particles, which includes all of the species mentioned in
\S\ref{sec-species}, while $U$ represents the radiation energy density
(integrated over all frequencies) in units of ${\rm ergs}\,{\rm cm}^{-3}$:
\begin{equation}
 U = \int_{0}^{\infty}u(\nu)d\nu,
\end{equation}
where $u(\nu,t)= 4\pi{J(\nu,t)}/{c}$.
In thermal equilibrium the radiation field has a frequency distribution
given by the Planck function, $J(\nu,t)= B(\nu,T_{\rm R})$, and
thus, in thermal equilibrium the energy density is
\begin{equation}
 u(\nu,t) = \frac{4\pi}{c}B_{\nu}(T_{\rm R}),
\end{equation}
and the total energy density $U$ is given by Stefan's law
\begin{equation}
 U = \frac{8\pi{h_{\rm P}}}{c^3}\int_{0}^{\infty}
 ({\rm e}^{h_{\rm P}\nu/k_{\rm B}T_{\rm R}} - 1)^{-1}
 \nu^{3}d\nu = a_{\rm R}T_{\rm R}^{4}.
\end{equation}
The spectrum of the CMB remains close to blackbody because the heat
capacity of the radiation is very much larger than that of the matter
(Peebles 1993), i.e.~there are vastly more photons than baryons.

Adiabatic cooling due to the expansion of the Universe is described by
\begin{equation} 
\label{eq:adiabaticcooling}
\frac{dT_{\rm M}}{dt} = -2H(t)T_{\rm M},
\end{equation}
since $\gamma=\slantfrac{5}{3}$ for an ideal gas implies
$T_{\rm M}\propto(1+z)^2$. \newline
The following cooling and heating
processes are often represented by approximate expressions.
We used the exact forms, with the exception of Bremsstrahlung
cooling, and the negligible collisional cooling. \newline
Bremsstrahlung, or free-free cooling:
\begin{equation}
\Lambda_{\rm brem} = \frac{2^5\pi e^6 Z^2}
 {3^{3/2}h_{\rm P}m_{\rm e} c^3}
 \left(\frac{2\pi k_{\rm B}T}{m_{\rm e}}\right)^{1/2}
 \!\!\!\!g_{\rm ff}n_{\rm e}(n_{\rm p}+n_{\rm He II} + 4 n_{\rm He III}),
\end{equation}
where $g_{\rm ff}$ is the free-free Gaunt factor (Seaton 1960), $n_{\rm
p}$ is the number density of protons, $n_{\rm He II}$ and $n_{\rm He
III}$ the number density of singly and double ionized helium
respectively, and other
symbols are as previously described. \newline
Photoionization heating:
\begin{equation}
\Pi_{\rm p\!-\!i} = \sum_{i=1}^N n_{i}4\pi\int_{\nu_{0}}^{\infty}
 \frac{\alpha_{i{\rm c}}(\nu)}{h_{\rm P}\nu}B(\nu,T_{\rm R})
 h_{\rm P}(\nu - \nu_0)d\nu.
\end{equation}
Photorecombination cooling:
\begin{eqnarray}
\lefteqn{\Lambda_{\rm p\!-\!r} = \sum_{i=1}^N 4\pi n_{\rm e} n_{\rm c}
 \left(\frac{n_{i}}{n_{\rm e} n_{\rm c}}\right)^{\rm LTE} \times}\nonumber \\
 & \!\!\!\!\!\!\!\!\!\!{\displaystyle
 \int_{\nu_{i}}^{\infty}\!\!\frac{\alpha_{i}(\nu)}{h_{\rm P}\nu}\!\!
 \left[\frac{2h_{\rm P}\nu^{3}}{c^{2}}\,{+}\,B(\nu,T_{\rm R})\!\right]\!\!
 {\rm e}^{-h_{\rm P}\nu/k_{\rm B}T_{\rm M}}h_{\rm P}(\nu\,{-}\,\nu_0)d\nu}.
\end{eqnarray}
Line cooling:
\begin{equation}
\Lambda_{\rm line} = h_{\rm P}\nu_0[n_jR_{ji} - n_iR_{ij}].
\end{equation}
Collisional ionization cooling:
\begin{equation}
\Lambda_{\rm c\!-\!i} = h_{\rm P}\nu_0C_{i{\rm c}}.
\end{equation}
Collisional recombination heating:
\begin{equation}
\Lambda_{\rm c\!-\!r} = h_{\rm P}\nu_0C_{{\rm c}i}.
\end{equation}
Here $\nu_0$ is the frequency at the ionization edge.
We used approximations for collisional ionization and recombination
cooling because these collisional processes are essentially
negligible during the recombination
era. $C_{i{\rm c}}$ and $C_{{\rm c}i}$ are the collisional ionization and
recombination rates respectively, computed as in e.g.~Mihalas (1978),~\S5.4.

Thus, with 
\begin{equation}
 \frac{dT_{\rm M}}{dz} = \frac{dt}{dz}\frac{dT_{\rm M}}{dt},
\end{equation}
the total rate of change of matter temperature
with respect to redshift becomes
\begin{eqnarray}
\lefteqn{(1+z)\frac{dT_{\rm M}}{dz} = \frac{8\sigma_{\rm T}U}{3H(z)m_{\rm e}c}\,
  \frac{n_{\rm e}}{n_{\rm e}+n_{\rm H}+n_{\rm He}}\,(T_{\rm M} - T_{\rm R})}
  \nonumber \\
  & \!\!\!\!\!\!\!\!\!\!\mbox{}+ 2T_{\rm M} + {\displaystyle
   \frac{2\!\left(\Lambda_{\rm brem}\,{-}\,\Pi_{\rm p\!-\!i}
  \,{+}\,\Lambda_{\rm p\!-\!r}\,{+}\,\Lambda_{\rm c\!-\!i}
  \,{+}\,\Lambda_{\rm c\!-\!r}
  \,{+}\,\Lambda_{\rm line}\right)}{3 k_{\rm B} n_{\rm tot} H(z)}.}
\label{eq:cooling}
\end{eqnarray}
Here $n_{\rm He}$ is the total number density of helium, and 
the denominator $n_{\rm e}+n_{\rm H}+n_{\rm He}$ ($n_{\rm tot}$ from
equation~(\ref{eq:cool1}))
takes into account the fact that the energy is shared among all the
available matter particles. All the terms except adiabatic cooling
in equation~(\ref{eq:cooling}) involve matter energy conversion into
photons. In particular, Compton and Bremsstrahlung cooling are the
most important, and they
can be thought of as keeping $T_{\rm M}$ very close to $T_{\rm R}$
until their time scales
become long compared with the Hubble time, and
thereafter the matter cools as $T_{\rm M}\propto(1+z)^2$.
Previous recombination calculations only included Compton and adiabatic
cooling, however the additional terms add improvements only at the
$10^{-3}\%$ level in the ionization fraction.
The reason for the negligible improvement is that it 
makes little difference which mechanism keeps $T_{\rm M}$ close to
$T_{\rm R}$ early on,
and adiabatic cooling still becomes important at the same time.

\subsection{Summary of Equations}
The system of equations to be simultaneously integrated in redshift are:
\begin{eqnarray}
(1+z)\frac{dn_{i}(z)}{dz} = 
-&\!\!\! {\displaystyle\frac{1}{H(z)}}
\Big\{\left[n_{\rm e}(z) n_{\rm c}(z) P_{{\rm c}i} - n_i(z) P_{ic}\right]
 \nonumber \\
  & \quad \mbox{} + {\rm \sum_{{\it j}=1}^{N}} \Delta R_{ji}\Big\} + 3 n_i(z),
\label{eq:sum1}
\end{eqnarray}
\begin{displaymath}
{\displaystyle (1+z)\frac{dT_{\rm M}}{dz} =
  \frac{8\sigma_{\rm T}U(J_{\nu,z})}{3H(z)m_{\rm e}c}\,
  \frac{n_{\rm e}}{n_{\rm e} + n_{\rm H} + n_{\rm He}}\,(T_{\rm M} - T_{\rm R})}
\end{displaymath}
\vspace{-3mm}
\begin{equation}
\!\!\mbox{}+ 2T_{\rm M}
 - \frac{2\left(\Lambda_{\rm brem} +
 \Pi_{\rm p\!-\!i} + \Lambda_{\rm p\!-\!r} + \Lambda_{\rm c\!-\!i}
  + \Lambda_{\rm c\!-\!r} + \Lambda_{\rm line}\right)}
  {3 k_{\rm B}n_{\rm tot}H(z)},
\label{eq:sum2}
\end{equation}
and
\begin{equation}
 (1+z)\frac{dJ(\nu,z)}{dz} =
 3J(\nu,z)
 -\frac{c}{H(z)}\Big[j(\nu,z) - \kappa(\nu,z)J(\nu,z)\Big].
\label{eq:sum3}
\end{equation}

For $J(\nu,z) = B(\nu,z) = B(\nu,T_{\rm R})$ (see~\S\ref{sec-radfield}),
equation~(\ref{eq:sum3}) can
be omitted because the expansion of the Universe preserves the thermal
spectrum of non-interacting radiation, and we can use the Sobolev escape
probability method for the primary spectral line distortions.
The system of coupled equations~(\ref{eq:sum1}) that we use contains up to 609
separate equations, 300 for H (one for each of a maximum of 300 levels we
considered), 200 for \ion{He}{1}, 100 for \ion{He}{2}, 1 for \ion{He}{3},
1 for electrons, 1 for protons, and 1 for each of the 5 molecular or ionic H
species.
This system of equations, along with (\ref{eq:sum2}), is extremely stiff,
that is
the dependent variables are changing on very different time scales.
We used the Bader-Deuflhard semi-implicit numerical integration scheme,
which is described in Press et al.~(1992).
To test the numerical integration we checked at each time step that the
total charge and total number of particles are conserved to one part
in $10^{7}$.

\section{Results and Discussion}
By an `effective 3-level' H atom we mean a hydrogen atom that includes
the ground state,
first excited state, and continuum. In an effective 3-level atom, the
energy levels between n=2 and 
the continuum are accounted for by a recombination coefficient which
includes recombinations to those levels.
This should be distinguished from an {\it actual\/} 3-level atom, which would
completely neglect all levels above n=2, and would be a hopeless
approximation.  Good accuracy is obtained by considering an n-level atom,
where n is large enough.  In practice we find that a 300-level atom is
more than adequate.
We do not explicitly include angular momentum states $\ell$, whose
effect we expect to be negligible.
In contrast to the effective 3-level H atom, the 300-level H
atom has no
recombination coefficient with `extra' levels.
The `standard' recombination calculation refers to the calculation
with the effective 3-level atom that is
widely used today and first derived by Peebles (1968) and Zel'dovich
et al.~(1968), updated with the most recent parameters and recombination
coefficient (HSSW). 

The primordial He abundance was taken to be $Y_{\rm P}=0.24$ by mass
(Schramm \& Turner 1998). The present-day CMB temperature $T_{0}$ was
taken to be $2.728\,$K, the central value determined by the FIRAS
experiment (Fixsen et al.~1996).

\subsection{The `Effective 3-level' Hydrogen Atom}
\label{sec-3level}
For comparison with the standard recombination calculation that only
includes hydrogen (see~Peebles 1968, 1993; Scott 1988), we reduce our chemical
reaction network to an effective 3-level atom, i.e.~a two-level
hydrogen atom plus continuum. 
The higher atomic energy levels are included by way of the recombination
coefficient, which can effectively include recombination to hundreds of levels.
The following reactions are included:
\begin{eqnarray}
{\rm H}_{\rm n=2,\ell=2s} + \gamma &\longleftrightarrow&
         {\rm e}{^-} + {\rm H}{^+} \nonumber\\
{\rm H}_{\rm n=1} + \gamma &\longleftrightarrow&
         {\rm H}_{\rm n=2,\ell=2p} \nonumber\\
{\rm H}_{\rm n=1} + 2\gamma &\longleftrightarrow&
         {\rm H}_{\rm n=2,\ell=2s}.\nonumber
\end{eqnarray}

As described in Peebles (1993), we omit the recombinations and
photoionizations to the ground state because any recombination directly
to the ground state will emit a photon with energy $>13.6\,$eV, where
there are few blackbody photons, and this will immediately re-ionize a
neighboring H atom.
We include the two-photon rate from the $2s$ state with the rate
$\Lambda_{2s{-}1s} = 8.22458\,$s$^{-1}$ (Goldman 1989).
The most accurate total Case B recombination coefficient is
by Hummer (1994) and is fitted by the function
\begin{equation}
\alpha_{\rm B} = 10^{-13}\frac{at^{b}}{1 + ct^{d}} \, \mathrm{cm^{3}s^{-1}},
\end{equation}
where $a=4.309$, $b=-0.6166$, $c=0.6703$, $d=0.5300$ and
$t= T_{\rm M}/10^{4}\,$K (P{\'e}quignot et al.~1991; see also Verner
\& Ferland~1996).

Consideration of detailed balance in the effective 3-level atom leads to
a single ordinary differential equation for the ionization fraction:
\begin{eqnarray}
\lefteqn{{dx_{\rm e}\over dz}  =
 \frac{\big[x_{\rm e}^2 n_{\rm H} \alpha_{\rm B}
 - \beta_{\rm B} (1-x_{\rm e})
   {\rm e}^{-h_{\rm P}\nu_{2s}/k_{\rm B}T_{\rm M}}\big]} {H(z)(1+z)}\,\times}
 \nonumber \\
  & & \frac{\big[1 + K \Lambda_{2s{-}1s} n_{\rm H}(1-x_{\rm e})\big]}
 {\big[1+K \Lambda_{2s{-}1s} n_{\rm H} (1-x_{\rm e})
 + K \beta_{\rm B} n_{\rm H}(1-x_{\rm e})\big]}
\label{eq:standard_xe}
\end{eqnarray}
(see e.g.~Peebles~1968; extra terms included in Jones \& Wyse 1985,
for example, are negligible).   Here $x_{\rm e}$ is the residual
ionization fraction, that is the
number of electrons compared to the total number of hydrogen nuclei
($n_{\rm H}$). Here the Case B recombination coefficient
$\alpha_{\rm B}=\alpha_{\rm B}(T_{\rm M})$, the total photoionization rate
$\beta_{\rm B}=\alpha_{\rm B} (2\pi m_{\rm e} k_{\rm B}
T_{\rm M}/h_{\rm P}^2)^{3/2} \exp(-E_{2s}/k_{\rm B}T_{\rm M})$ as described in
\S\ref{sec-StandardRecomb}, $\nu_{2s}$ is the frequency of the $2s$ level from
the ground state,
and the redshifting rate $K\equiv \lambda_\alpha^3/(8\pi H(z))$, where
$\lambda_\alpha$ is the Ly$\,\alpha$ rest wavelength.
Note that $T_{\rm M}$ is used in equation~(\ref{eq:standard_xe}) and
in $\beta_{\rm B}$, because the temperature terms come from detailed balance
derivations that use Boltzmann and Saha equilibrium distributions, which are
collisional descriptions.
In the past, this equation has been solved (for $x_{\rm e}(z)$)
simultaneously with
a form of equation~(\ref{eq:cooling}) containing only adiabatic and
Compton cooling.  We refer to the approach of
equation~(\ref{eq:standard_xe}) as the `standard calculation.'

For the comparison test with the standard recombination calculation,
we also use an effective recombination coefficient, but three equations to
describe the three reactions listed above. That is, we
simplified equation~(\ref{eq:sum1}) to three equations, one for the
ground state population ($n_1$), one for the first excited state
population ($n_2$), and one for the
electrons (for H recombination $n_e = n_p$).
\begin{equation}
(1+z)\frac{dn_{1}(z)}{dz} =
-\frac{1}{H(z)}
\left[\Delta R_{2p{-}1s} + \Delta R_{2s{-}1s}\right] + 3 n_{1}
\end{equation}
\vspace{-3mm}
\begin{eqnarray}
(1+z)\frac{dn_{2}(z)}{dz} =& \!\!\!\!\mbox{}-{\displaystyle\frac{1}{H(z)}}
\Big[(n_{\rm e}(z) n_{\rm p}(z) \alpha_{\rm B}
 - n_{2s}(z) \beta_{\rm B} \nonumber \\
 & \ \mbox{} - \Delta R_{2p{-}1s} - \Delta R_{2s{-}1s}\Big] + 3 n_2
\end{eqnarray}
\vspace{-3mm}
\begin{equation}
(1+z)\frac{dn_{\rm e}(z)}{dz} =
-\frac{1}{H(z)}
\big[(n_{2s}(z) \beta_{\rm B} - n_{\rm e}(z) n_{\rm p}(z) \alpha_{\rm B}
\big] + 3 n_{\rm e}
\end{equation}
The remaining
physical difference between our effective 3-level atom approach and that of
the standard calculation is the treatment of the redshifting of H Ly$\,\alpha$
photons (included in the $\Delta R_{2p{-}1s}$ terms). In our calculation the redshifting is accounted for by 
the Sobolev escape probability (see~\S\ref{sec-Sobolev}).
Following Peebles (1968, 1993), the standard calculation accounts for the
redshifting by
approximating the intensity distribution as a step, and in effect takes the
ratio of the redshifting of the photons through the line to the expansion
scale that produces the same amount of redshifting.
It can be shown that Peebles' step method considered as an escape
probability scales as $1/ \tau_{\rm S}$, where $\tau_{\rm S}$ is the
Sobolev optical depth.
For high Sobolev optical depth, which holds for H Ly$\,\alpha$ during
recombination for any cosmological model (see
Fig.~7),
the Sobolev escape probability also scales as $1/\tau_{\rm S}$:
\begin{equation} 
 \lim_{\tau_{\rm S} \gg 1} p_{ij} = \lim_{\tau_{\rm S} \gg 1}
 \frac{1}{\tau_{\rm S}}(1 - {\rm e}^{-\tau_{\rm S}}) = \frac{1}{\tau_{\rm S}}.
\end{equation}
Therefore the two approximations are equivalent for Ly$\,\alpha$,
although we would expect differences for lines with $\tau_{\rm S}\lesssim1$,
where $p\to1$.
Because we treat recombination in the same way as Peebles, 
no individual
treatment of other lines is permitted, and therefore there are no other
differences between the two calculations for this simple case. 
Note that with Peebles' step method to compute $\Delta R_{2p{-}1s}$,
and the assumption that $n_1 = n_{\rm H} - n_{\rm p}$, the above equations will
reduce to the single ODE equation~(\ref{eq:standard_xe}).

The results from our effective 3-level recombination calculation are
shown in
Fig.~1,
plotted along with values from a
separate code as used in HSSW, which represents the standard
recombination calculation updated with the most recent parameters.
The resulting ionization fractions are equal, which shows that our new approach
gives exactly the standard result when reduced to an effective 3-level atom.
Two other results are plotted for comparison, namely values of $x_{\rm e}$
taken from
Peebles (1968) and Jones \& Wyse (1985). Their differences can be largely
accounted for by
the use of an inaccurate recombination coefficient with
$\alpha_{\rm B}(T_{\rm M}) \propto T_{\rm M}^{-1/2}$.

\vspace{2mm}
\centerline{{\vbox{\epsfxsize=8cm\epsfbox{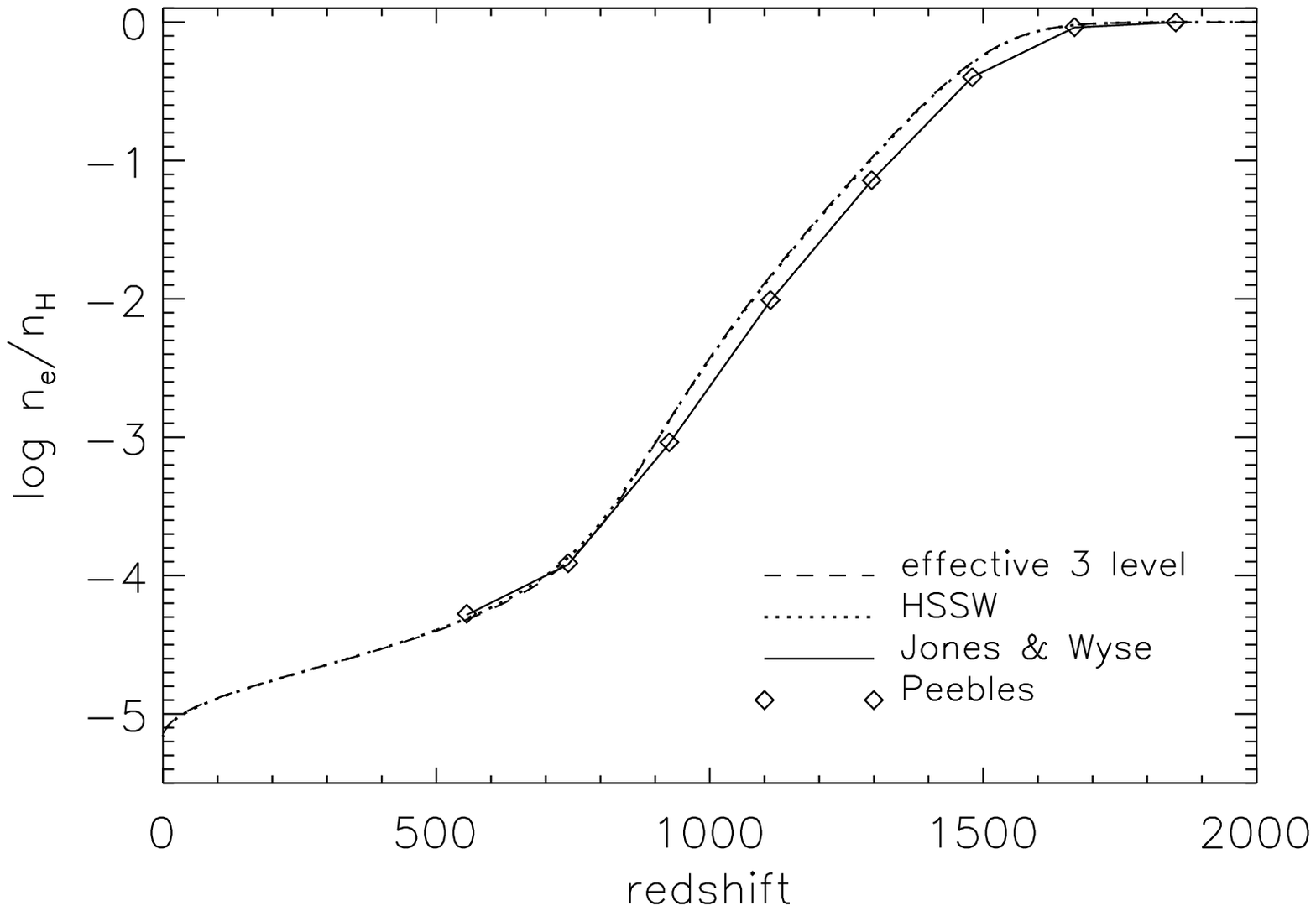}}}}
\baselineskip=8pt
{\footnotesize {\sc Fig.~1.---}
Comparison of effective three-level hydrogen recombination for
the parameters
$\Omega_{\rm tot} = 1.0, \Omega_{\rm B} = 1.0, h=1.0$.
Note that the Jones \& Wyse (1985) and Peebles (1968) curves overlap as does
our curve with the HSSW one.
}
\label{fig:Hrec3level}
\vspace{2mm}

\baselineskip=11pt

As an aside, we note the behavior for $z\lesssim50$ in our curve and the HSSW
one.  This is caused by inaccuracy in the recombination coefficient for very
low temperatures.  The down-turn is entirely artificial and could be removed
by using an expression for $\alpha_{\rm B}(T)$ which is more physical
at small temperatures. The results of our detailed calculations are
not believable
at these redshifts either, since accurate modeling becomes increasingly
difficult due to numerical precision as $T$ approaches zero.
But in any case the optical
depth back to such redshifts is negligible, and the real Universe is
reionized at a similar epoch (between $z=5$ and 50 certainly).

\subsection{Multi-level Hydrogen Atom}
\label{sec-Multi}
The purpose of a multi-level hydrogen atom is to improve the
recombination calculation, by following the population of each atomic
energy level with redshift and by including all bound-bound and bound-free
transitions. This includes recombination to, and photoionization from,
all levels {\it directly\/} as a function of time,
in place of a parameterized recombination and photoionization coefficient. The individual
treatment of all levels in a coupled manner allows for the
development of departures from equilibrium among the states with time,
and feedback on the rate of recombination. Since the accuracy of the
recombination coefficient is probably the single most important effect
in obtaining accurate power spectra (HSSW), it makes sense to
follow the level populations as accurately as possible.

In the multi-level H atom recombination calculation, we do not consider
individual $\ell$
states (with the exception of $2s$ and $2p$), but assume the $\ell$ sublevels have populations proportional to
(2$\ell$ + 1). The $\ell$ sublevels only deviate from this distribution
in extreme non-equilibrium conditions (such as planetary nebulae).
In their H recombination calculation, Dell'Antonio \& Rybicki (1993)
looked for such $\ell$ level deviations for n$\,{\leq}\,10$ and found none.
For n$\,{>}\,10$, the $\ell$ states are even less likely to differ from an
equilibrium distribution, because the energy gaps
between the $\ell$ sublevels are increasingly smaller as n increases.

\vspace{2mm}
\centerline{{\vbox{\epsfxsize=8cm\epsfbox{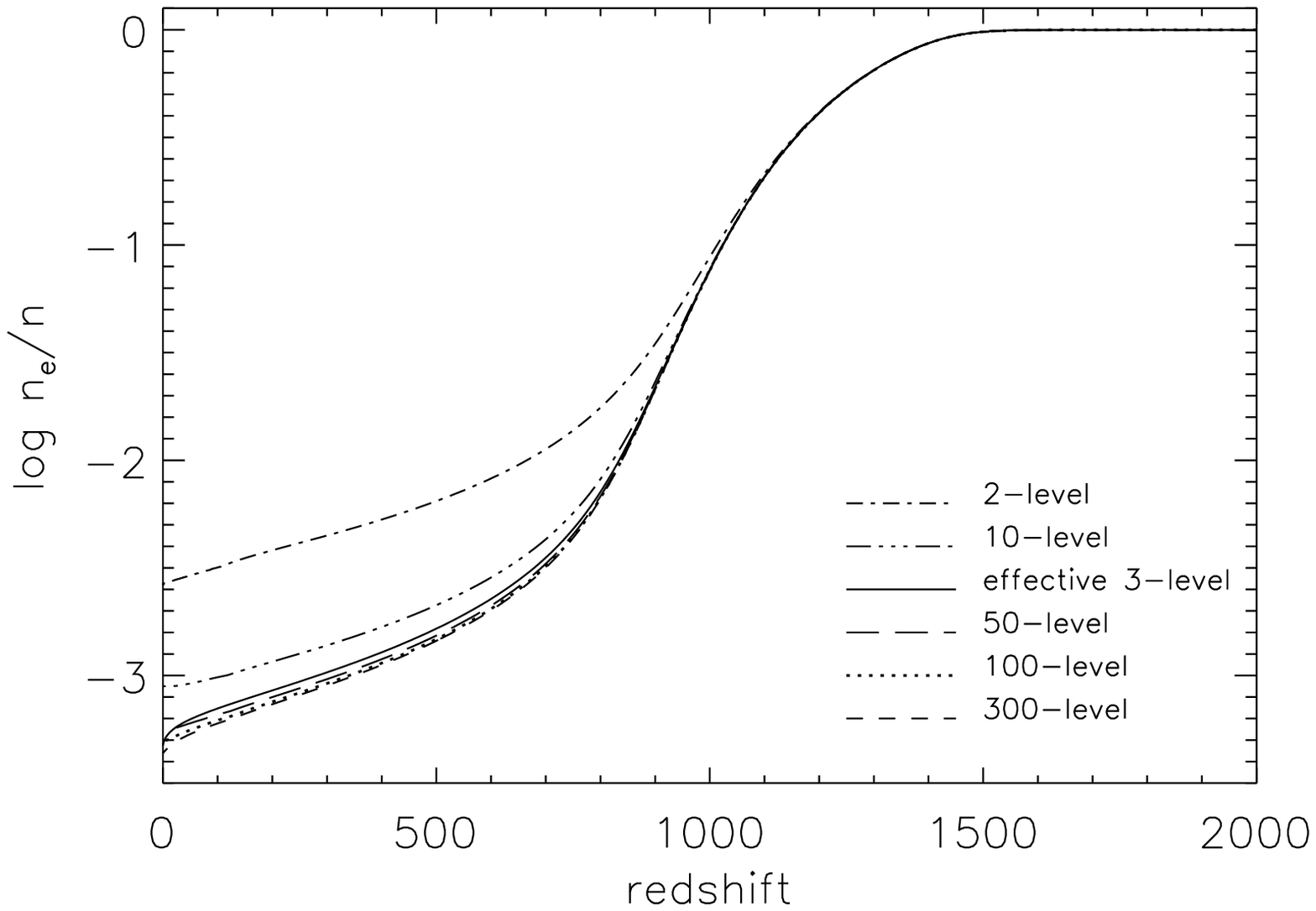}}}}
\baselineskip=8pt
{\footnotesize {\sc Fig.~2.---}
Multi-level hydrogen recombination for the standard CDM
parameters $\Omega_{\rm tot} = 1.0, \Omega_{\rm B} = 0.05, h = 0.5$
(top), and for $\Omega_{\rm tot} = 1.0, \Omega_{\rm B} = 1.0, h = 1.0$
(bottom), both with 
$Y_{\rm P} = 0.24, T_0 = 2.728\,$K. The `effective 3-level' calculation
is essentially the same as in HSSW, and uses a recombination coefficient
which attempts to account for the net effect of all relevant levels.  We find
that we require a model which considers close to 300 levels for full
accuracy.  Note also that although we plot all the way to $z=0$, we know
that the Universe becomes reionized at $z>5$, and that our calculations
(due to numerical precision at low $T$) become unreliable
for $z\lesssim50$ in the upper two curves, and for $z\lesssim20$ in
the other curves.
\label{fig:HrecMulti}
}
\vspace{2mm}

\baselineskip=11pt

\subsubsection{Results From a Multi-level H Atom}
\label{sec-resultsmulti}
Fig.~2
shows the ionization fraction $x_{\rm e}$ from
recombination of a 2-, 10-, 50-, 100- and 300-level H atom, compared with
the standard effective 3-level results. The $x_{\rm e}$ converges
for the highest n-level atom calculations. The effective
3-level atom actually includes about 800 energy levels via the recombination
coefficient (e.g.~Hummer 1994).

\centerline{{\vbox{\epsfxsize=8cm\epsfbox{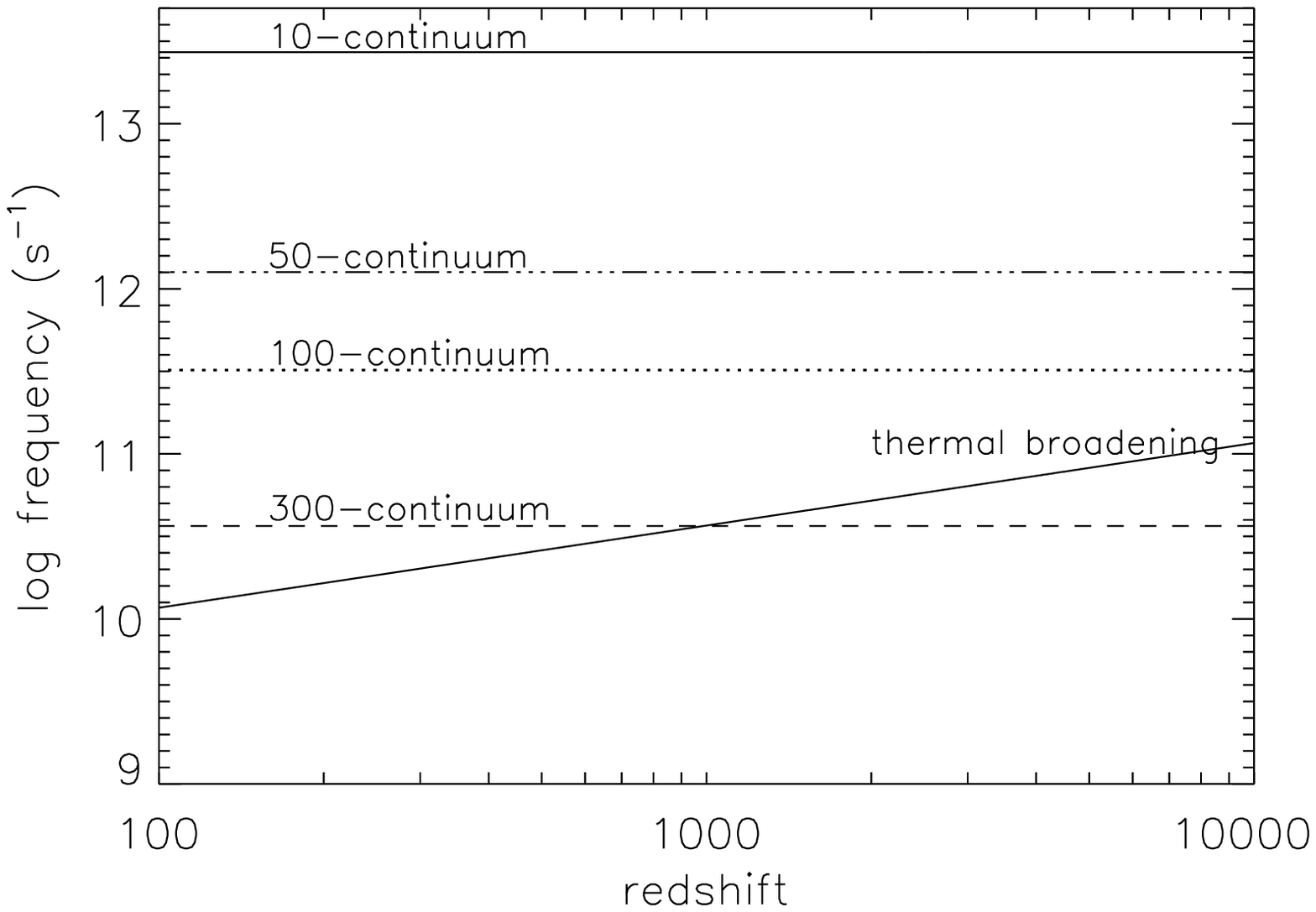}}}}
\baselineskip=8pt
{\footnotesize {\sc Fig.~3.---}
Energy separations between various hydrogen atomic levels and the
continuum.  The solid curve shows the energy width of the same energy
levels due to thermal
broadening. Thermal broadening was calculated using
$\nu (2 k_{\rm B} T_{\rm M} / m_{\rm H} c^2)^{1/2}$. The
frequency of the highest atomic energy level (n=300, with an energy
from the ground state of 109676.547 cm$^{-1}$; the continuum energy
level is 109677.766 cm$^{-1}$) was used for $\nu$, but a
thermal broadening value for any atomic energy level would overlap on
this graph.
\label{fig:thermbroad}
}
\vspace{2mm}

\baselineskip=11pt

Fig.~2
shows that the more levels that are included in the
hydrogen atom, the lower
the residual $x_{\rm e}$. The simple explanation is that the
probability for electron capture increases with more energy levels per
atom. Once captured, the electron can cascade
downwards before being reionized. Together this means 
adding more higher energy levels per atom
increases the rate of recombination.
Eventually $x_{\rm e}$
converges as the atom becomes complete in terms of electron energy
levels, i.e.~when there is no gap between the highest energy level and
the continuum (see
Fig.~3).
Ultimately the uppermost levels
will have gaps to the continuum which are smaller than the thermal
broadening of those levels, and so
energy levels higher than about n=300 do not need to be considered,
except perhaps at the very lowest redshifts.
 
For other reasons entirely, our complete (300-level) H atom recombination
calculation gives an $x_{\rm e}$ lower than that of the
effective 3-level atom calculation. The faster production of
hydrogen atoms is due to non-equilibrium processes in the excited states
of H, made obvious by our new, level-by-level treatment of recombination. The
details are described in \S\ref{sec-noneq} below.

\subsubsection{Faster H Recombination in our Level-by-level
Recombination Calculation}
\label{sec-noneq}

The lower $x_{\rm e}$ in our calculation compared to the standard calculation
is caused by the strong but cool radiation field. Specifically, 
both a faster downward cascade rate and a lower total photoionization rate
contribute to a faster net recombination rate.

\centerline{{\vbox{\epsfxsize=8cm\epsfbox{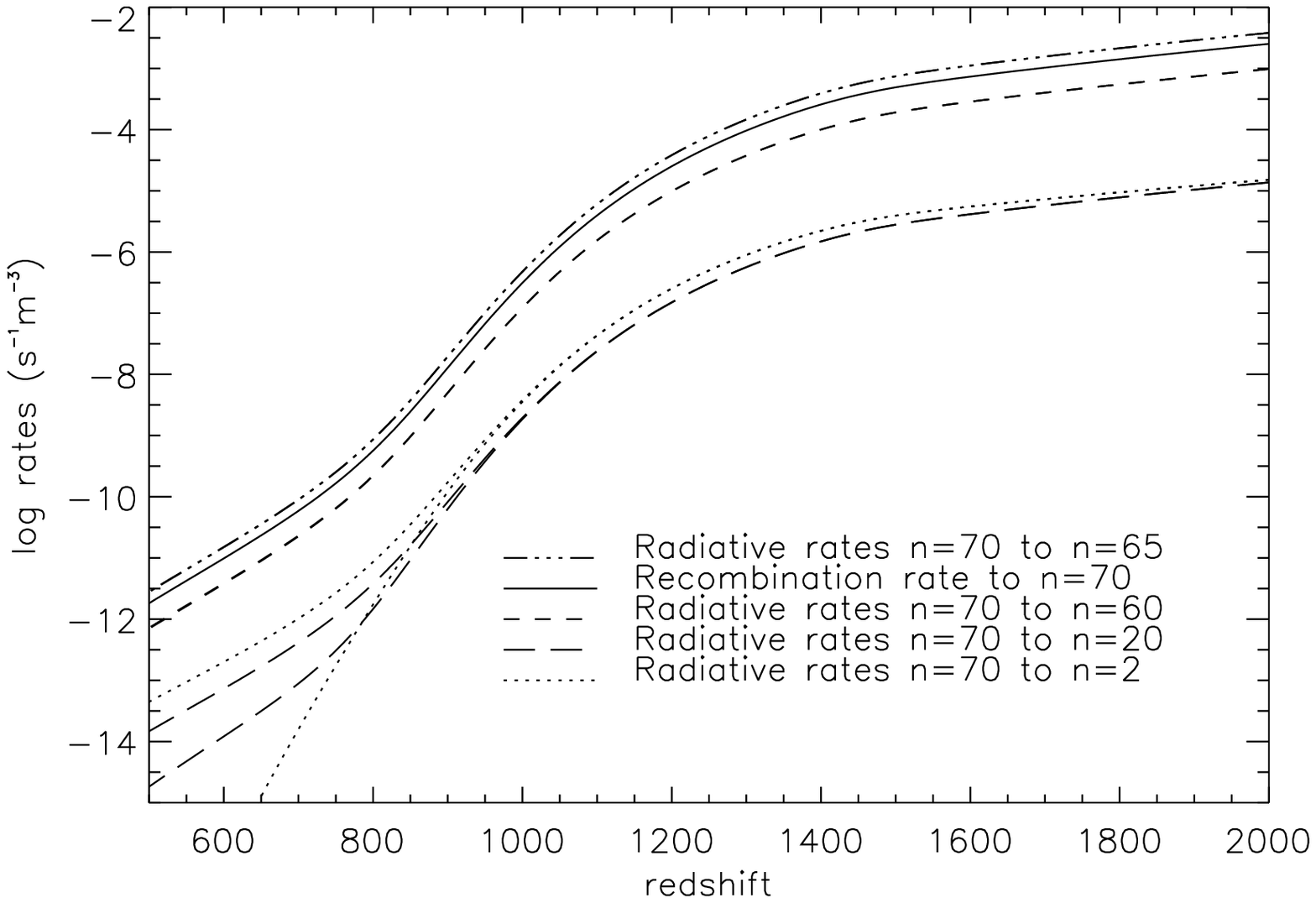}}}}
\baselineskip=8pt
{\footnotesize {\sc Fig.~4.---}
How the bound-bound rates of large energy separation
(e.g. n=70 to n=20) go out of
equilibrium at low $T$, illustrated using an upper level of
n=70 for definiteness. The case of equilibrium corresponds to net
bound-bound rates of zero. Each upward and downward bound-bound rate for
a given transition is represented by the same curve; non-equilibrium occurs
where a single curve separates into two as redshift decreases.
The rates shown are for the sCDM model.
\label{fig:bbrates}
}
\vspace{2mm}

\baselineskip=11pt

By following the population of each
atomic energy level with redshift, we relax the assumption used in the standard
calculation that the excited states are in equilibrium. In addition, we
calculate all bound-bound rates which control equilibrium among
the bound states.
In the standard calculation, equilibrium among the excited states
n~$\geq 2$ is
assumed, meaning that the net bound-bound rates are
zero.
Fig.~4
shows that the net bound-bound
rates are actually different from zero at $z\lesssim1000$.
The reason for this is that at low temperatures, the strong but cool
radiation field means that high energy transitions are rare due to few high
energy photons. More specifically,
photoexcitation and stimulated photodeexcitation for high energy transitions
become rare (e.g.~70--10, 50--4 etc.). 
In this case
spontaneous deexcitation dominates, causing a
faster downward cascade to the n=2 state. 
In addition, the faster downward cascade rate is faster than the
photoionization rate from the upper state, and one might view this as
radiative decay stealing some of the depopulation `flux' from
photoionization. Both the faster downward cascade and the lower
photoionization rate contribute to the faster net recombination rate.

The cool radiation field is strong, so photoexcitations and 
photodeexcitations are rapid among nearby energy
levels (e.g.~70--65, etc.;
see Fig.~4). What we see in Fig.~4 is that with time, after $z\lesssim1000$,
the n=70 energy level becomes progressively decoupled from the distant
lower energy levels (n=2,3,...20,...), but remains tightly coupled to its
nearby `neighbors' (n=60,65, etc.). This explains the departures from an
equilibrium Boltzmann distribution (in the excited states) as seen in the shape of the curves in
Fig.~5.

\centerline{{\vbox{\epsfxsize=8cm\epsfbox{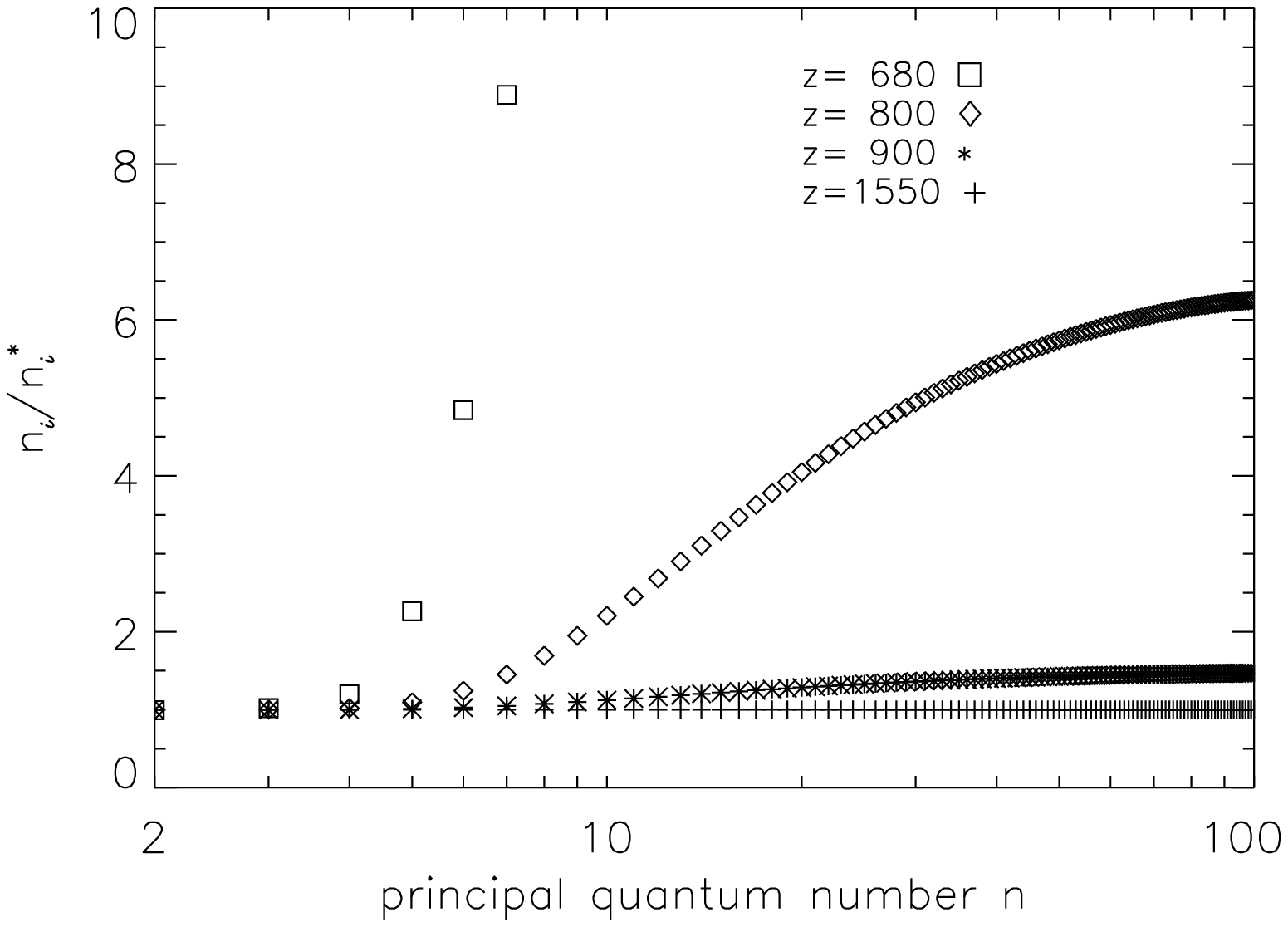}}}}
\centerline{{\vbox{\epsfxsize=8cm\epsfbox{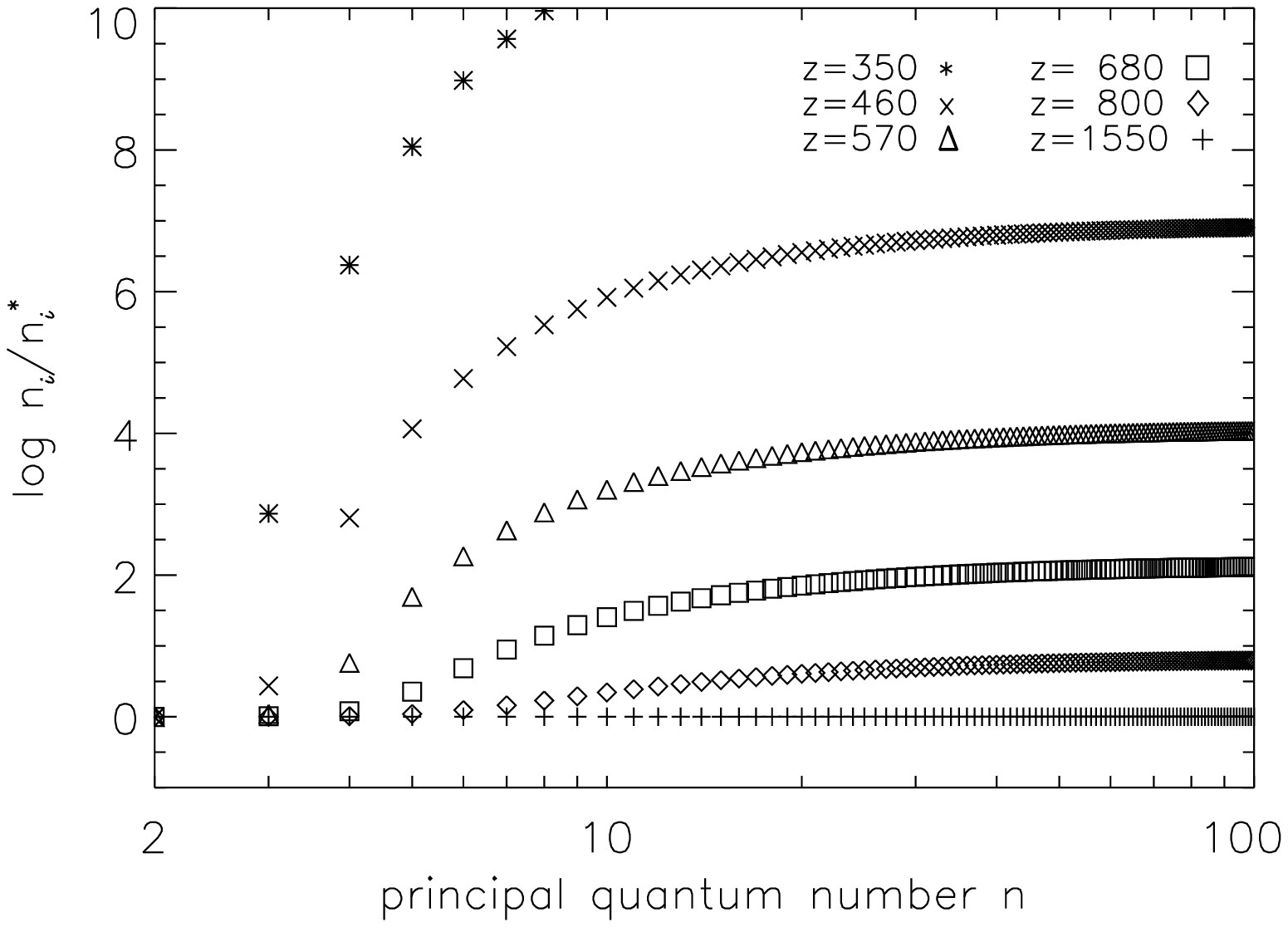}}}}
\baselineskip=8pt
{\footnotesize {\sc Fig.~5.---}
Ratio of the actual number densities of the excited
states ($n_i$) and
number densities for a Boltzmann distribution of excited states
($n_i^*$) at
different redshifts for recombination within the standard CDM model.
See section~\ref{sec-noneq} for details.  We give two different plots with
linear and logarithmic y-axes and different redshift values, to show
a wider range of out-of-equilibrium conditions.  The ratio approaches
a constant for high n at a given $z$, because the high energy (Rydberg)
states are all very close in energy, and thus have similar behavior
(i.e. remain coupled to the radiation field and each other).
\label{fig:Boltz}
}
\vspace{2mm}

\baselineskip=11pt

Fig.~5
illustrates the non-equilibrium of the excited
states by showing the ratio of populations of the excited states
compared to a Boltzmann equilibrium distribution with respect to n=2. 
We find that the
upper levels of the hydrogen atom are not in thermal equilibrium with
the radiation, i.e.~the excited levels are not populated according to
a Boltzmann distribution.  The excited states are in fact overpopulated
relative to a
Boltzmann distribution. This is not a surprise for the population of the
n=2 state, which is strongly overpopulated compared to the n=1 ground
state, and so should be all n~$\geq 2$ states because all Lyman lines remain
optically thick during recombination. What {\em is\/} surprising is that
all excited states develop a further overpopulation {\em with respect to\/}
n=2 and each other. Note that this is not a population inversion.
The recombination rate to a given high level is faster than the downward
cascade rate, and this causes a `bottleneck' creating the
overpopulation.
Fig.~5
shows that all states
are in equilibrium at high redshifts, with the highest states going
out of equilibrium first, followed by lower and lower states as the
redshift decreases.  The factor by which the excited states are
over-populated approaches a constant at high n for a given redshift,
with this factor increasing as $z$ decreases. The ratio is constant
because the high energy level Rydberg states have very similar energy
levels to each other, with a relatively large energy separation from the n=2
state (i.e.~the exponential term in equation~(\ref{eq:boltz}) dominates
over the $g_i$ ratios, and the exponential term is similar for all of
the Rydberg states.) 

Fig.~5
also shows an enormous ratio at low
redshift ($z < 500$) for number densities
of the actual excited states to the number densities of a Boltzmann
distribution of excited states, on the order of $10^6$. At such a low
redshift, there are almost no electrons in
the excited states ($\sim 10^{-20}$ cm$^{-3}$), and so
unlike at higher redshifts, the ratio is only an
illustration of the strong departure from an equilibrium distribution;
the actual populations are very low in any case.

In comparison with the standard
equilibrium capture-cascade calculation for $\alpha_{\rm B}$, the unusual
situation described above (caused by the strong but cool radiation field)
leads to higher effective recombination
rates for the majority of excited states without increasing photoionization
proportionally. This results in a higher net rate of production of neutral
hydrogen atoms, i.e.~a lower $x_{\rm e}$.

\subsubsection{Accurate Recombination vs Recombination Coefficient}
To demonstrate why the non-equilibrium in the excited states of H affects
the recombination rate, we must consider the
difference in our new treatment of recombination compared to the
standard treatment.
An important new benefit of our level-by-level calculation
lies in replacing the recombination coefficient with
a direct calculation of recombination to and photoionization from each level at each redshift step.
In other words, we calculate the recombination rate and the
photoionization rate using
individual level populations and parameters of the excited states $i$,
\begin{eqnarray}
\lefteqn{\sum_{i=1}^N n_{\rm e} n_{\rm p} R_{{\rm c}i}
  = n_{\rm e} n_{\rm p} \sum_{i=1}^N
 {\left(\frac{n_{i}}{n_{\rm e} n_{\rm p}}\right)}^{\rm LTE}\times}\nonumber \\
& \!\!{\displaystyle 4\pi\!\int_{\nu_{i}}^{\infty}
 \frac{\alpha_{i}(\nu)}{h_{\rm P}\nu}
 \left[\frac{2h_{\rm P}\nu^{3}}{c^{2}} + B(\nu,T_{\rm R})\right]
 {\rm e}^{-h_{\rm P}\nu/k_{\rm B}T_{\rm M}}d\nu},
\label{eq:feedback2}
\end{eqnarray}
and similarly for the photoionization rate,
\begin{equation}
\label{eq:feedback}
\sum_{i=1}^N n_{i}R_{i{\rm c}} = \sum_{i=1}^N
 n_{i}4\pi\int_{\nu_{0}}^{\infty}
 \frac{\alpha_{i{\rm c}}(\nu)}{h_{\rm P}\nu}B(\nu,T_{\rm R})d\nu.
\end{equation}
For the standard recombination calculation, the recombination rate is
\begin{equation}
\label{eq:standardrate}
\sum_{i=1}^N n_{\rm e} n_{\rm p} R_{{\rm c}i} = n_{\rm e} n_{\rm p}
 \alpha_{\rm B}(T_{\rm M}),
\end{equation}
and the photoionization rate is
\begin{eqnarray}
\lefteqn{\Sigma_i n_i R_{i{\rm c}} \equiv n_{2s} \beta_{\rm B}(T_{\rm M})}
 \nonumber \\
& = n_{2s} \alpha_{\rm B}(T_{\rm M}) {\rm e}^{-E_{2s}/k_{\rm B}T_{\rm M}}
 \left(2\pi m_{\rm e} k_{\rm B} T_{\rm M} \right)^{3/2}/h_{\rm P}^3.
\label{eq:nofeedback}
\end{eqnarray}
In this last equation, the excited state populations are hidden by the
Boltzmann relation with $n_{2s}$ (see~\S\ref{sec-StandardRecomb}).
The important point here is that our method allows redistribution of the
H level populations over all 300 levels at each redshift step, which feeds
back on the recombination process via
equation~(\ref{eq:feedback}), and leads to the lower $x_{\rm e}$ shown in
Fig.~2.
This
redistribution of the level populations is not possible in the standard 
calculation's equation~(\ref{eq:nofeedback}) which only considers the
populations $n_{\rm e}$, $n_{1}$, and $n_{2s}$, and considers the
excited level populations $n\,{>}\,2s$
to be proportional to $n_{2s}$ in an equilibrium distribution.

A small improvement in our new recombination treatment over
the standard treatment is in our distinguishing the various temperature
dependencies of recombination.  Photoionization and
stimulated recombination are radiative, so they should depend on
$T_{\rm R}$. Spontaneous recombination is collisional and depends on
$T_{\rm M}$ (see~\S\ref{sec-Recomb}). In the standard calculation, the radiative nature of 
recombination and photoionization is
overlooked because both the recombination
coefficient and photoionization coefficient are a function $T_{\rm M}$ only
(equations~(\ref{eq:standardrate}) and (\ref{eq:nofeedback})).
Although adiabatic cooling (equation~(\ref{eq:adiabaticcooling}))
does not dominate until quite low redshifts ($z\lesssim100$), it still
contributes partially to matter cooling
throughout recombination.  The resulting difference between
$T_{\rm M}$ and
$T_{\rm R}$ in the net recombination rate affects $x_{\rm e}$ at the
few percent level at $z\lesssim300$ for the popular cosmologies,
and has an even larger effect for high $\Omega_{\rm B}$ models.

\subsubsection{Collisions}
The standard recombination calculation omits collisional excitation
and ionization because at the relevant temperatures and densities they
are negligible for a two-level hydrogen
atom (Matsuda et al.~1971).
We have found that the collisional processes are also not important for 
the higher levels, even though those electrons are bound with little
energy. In high $\Omega_{\rm B}$ models collisional ionization and
collisional recombination rates for 
the highest energy levels are of the same order of
magnitude as the photoionization and recombination rates, though not
greater than them (see
Fig.~6).

\vspace{2mm}
\centerline{{\vbox{\epsfxsize=8cm\epsfbox{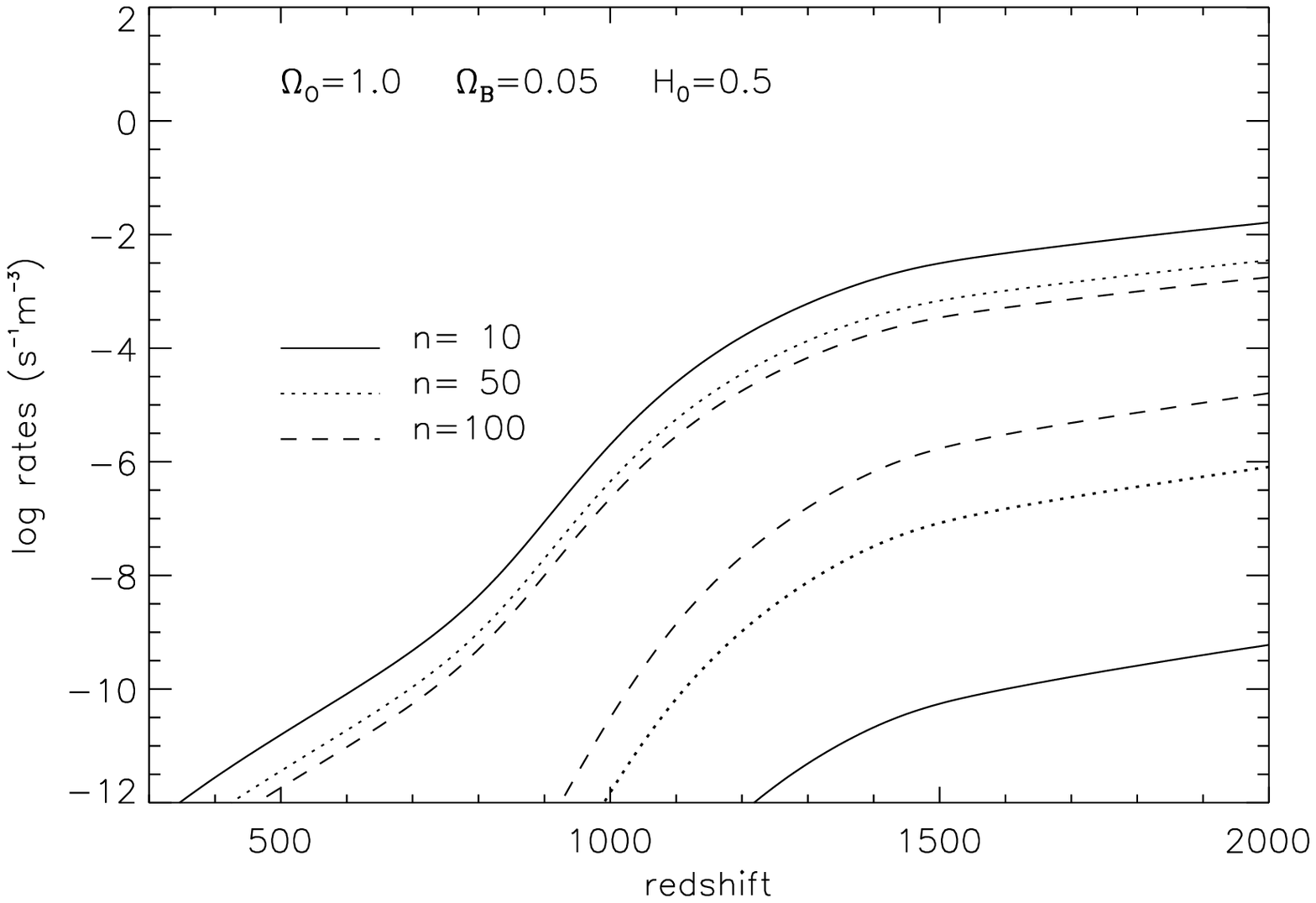}}}}
\centerline{{\vbox{\epsfxsize=8cm\epsfbox{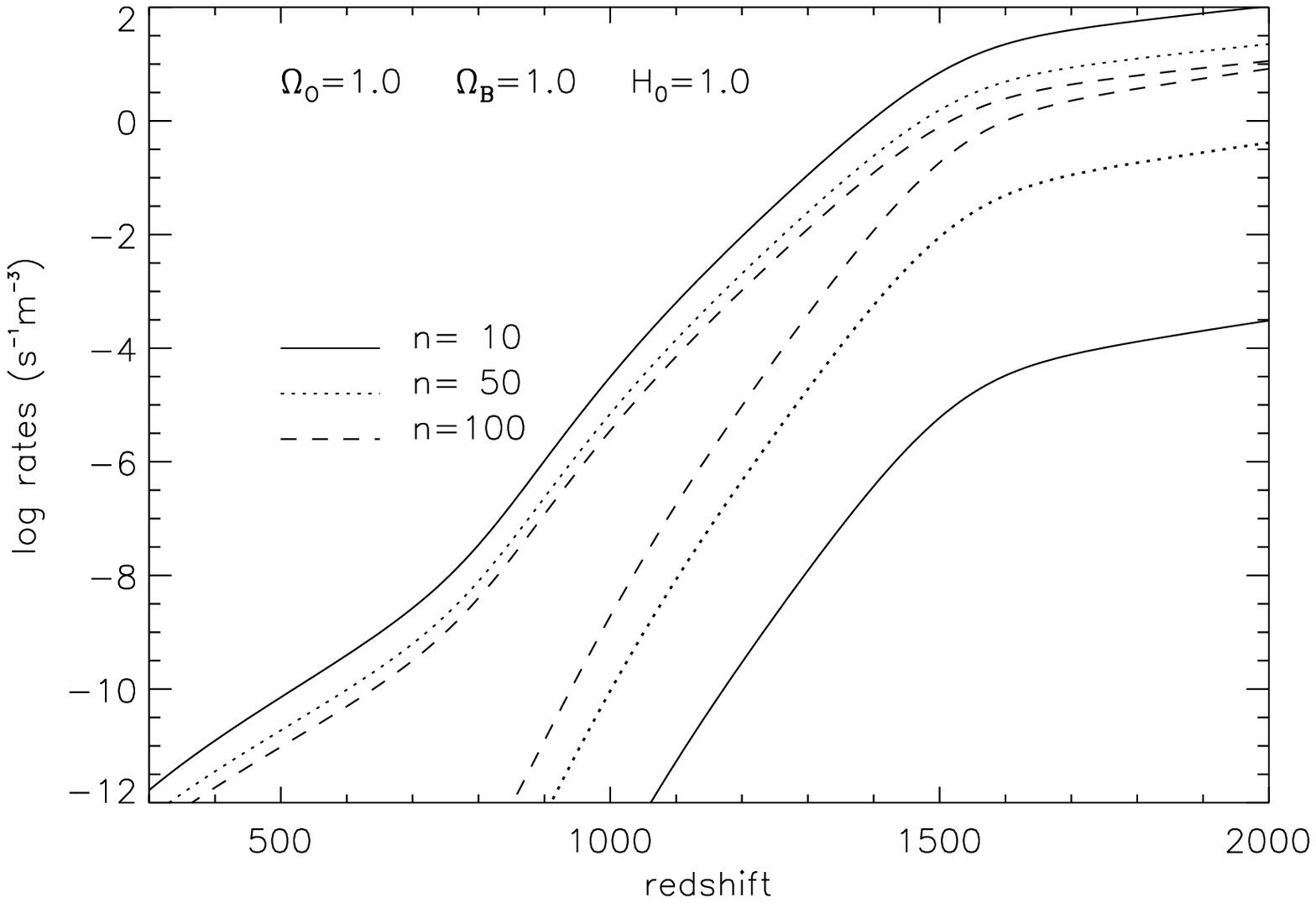}}}}
\baselineskip=8pt
{\footnotesize {\sc Fig.~6.---}
Comparison of ionization rates. The upper curves are
photoionization rates, the lower curves are collisional ionization rates.
The left panel shows a model with the standard CDM parameters, while
the right panel shows an extreme baryon cosmology.
\label{fig:Collrates}
}
\vspace{2mm}

\baselineskip=11pt

\subsubsection{Departures From Case B}
\label{sec-CaseB}
Case B recombination excludes recombination to the ground state and considers
the Lyman lines to be optically thick (i.e.~photons associated with all
permitted radiative transitions to
n=1 are assumed to be instantly reabsorbed).
An implied assumption necessary to compute the
photoionization rate is that the excited states (n~$\ge 2$) are in
equilibrium with the radiation.
Unlike the standard recombination calculation, our method allows departures
from Case B because the Lyman lines are
treated by the Sobolev escape probability method which is valid
for any optical thickness, and our
method allows departures from equilibrium of the excited states
(\S\ref{sec-resultsmulti}).
We find that the excited states depart from equilibrium at redshifts
$\lesssim800$, so Case B does not hold then.
However, our calculations show that for hydrogen all Lyman lines
are indeed
optically thick during all of hydrogen recombination, so
Case B holds for H recombination above redshifts $\simeq 800$.

\vspace{2mm}
\centerline{{\vbox{\epsfxsize=8cm\epsfbox{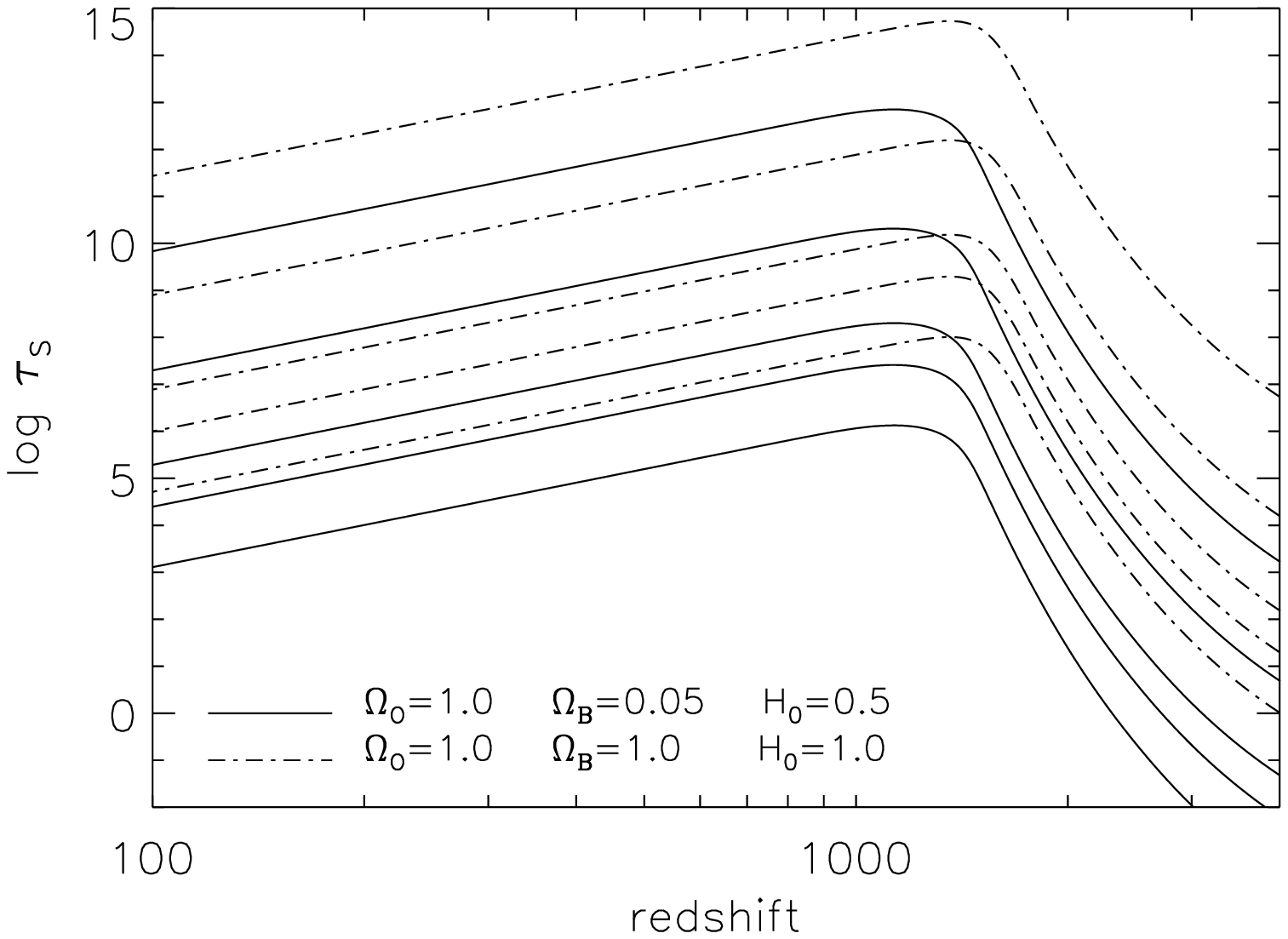}}}}
\baselineskip=8pt
{\footnotesize {\sc Fig.~7.---}
Sobolev optical depth ($\tau_{\rm S}$) for the H Lyman lines for
two different
cosmological models. From upper to lower the curves represent the
optical depth in the Lyman transitions from n=2 (i.e.~Ly$\,\alpha$), n=10,
n=50, n=100, and n=270. The curves show that all the Lyman transitions are
optically thick during H recombination ($z\lesssim2000$)
but some are optically thin ($\tau_{\rm S} < 1$)
earlier, during He recombination ($z \gtrsim$ 2000).
\label{fig:SobolevTau}
}
\vspace{2mm}

\baselineskip=11pt

Fig.~7
shows that the Lyman lines are not
optically thick
at earlier times, e.g.~during helium recombination, where we
find some optically thin H
Lyman lines.  The Sobolev escape probability treats this consistently,
which is necessary because we evolve H, \ion{He}{1}, and \ion{He}{2}
simultaneously.
which is
necessary because we evolve H, \ion{He}{1}, and \ion{He}{2} simultaneously.

\subsubsection{Other Recent Studies}

The previous study that was closest in approach to our own was that of
Dell'Antonio \& Rybicki (1993), who calculated recombination
for a ten-level hydrogen atom in order to estimate the spectral distortions
to the CMB blackbody radiation spectrum. Ten levels are insufficient to
calculate recombination accurately, because the higher energy levels of the
atom are completely ignored (see
Fig.~3).
However the accuracy of the ionization
fraction ($x_{\rm e}$) was sufficient to determine the magnitude
of the spectral distortions. Their recombination model treated individual
levels, but used a recombination coefficient to each level of the form
${T_{\rm M}}^{-1/2}$. Because the form of the recombination
coefficient dominates the H recombination process, our models are not
equivalent, and so there is little use in comparing the results.

More recently, Boschan \& Biltzinger (1998) derived a new
parameterized recombination coefficient to solve the recombination
equation of the standard calculation, and to
generate spectral distortions in the CMB. 
Their calculation differs from ours in that their
recombination coefficient is pre-calculated.  Hence it is not an interactive
part of the calculation, and does not allow the advantages that our
calculation does, mainly the feedback of the non-equilibrium in the excited
states on the net recombination rate. While they include pressure
broadening for a cutoff in the partition function, they neglect
thermal broadening. A more serious problem is their method
of inclusion of stimulated recombination, as originally suggested by
Sasaki \& Takahara (1993), who included stimulated recombination as
positive recombination instead of negative ionization.
The physics (as described in \S\ref{sec-Recomb}) and our
computational results are the same regardless of whether stimulated
recombination is treated as positive recombination or negative ionization.
However, this may not be the case computationally for the standard
calculation, if it is
not treated with care.  We defer a full discussion of these matters to a
separate paper (Seager \& Sasselov, in preparation).

We have also investigated how we can approximate our calculations, so
that other researchers can obtain approximately accurate results without
the need to follow 300 levels in a hydrogen atom.  Because the net effect 
of our new H calculation is a faster recombination (a lower freeze-out
ionization fraction), our results can be reproduced by artificially
speeding up recombination in the standard calculation. Further
details are described in Seager, Sasselov, \& Scott (1999).

\subsection{Helium}
\label{sec-Helium}
We compute helium and hydrogen recombination simultaneously.
The recombination of \ion{He}{3} into \ion{He}{2} and \ion{He}{2} 
into \ion{He}{1}
is calculated in much the same way as hydrogen, with recombination,
photoionization,
redshifting of the n$^1p$--$1^1s$ lines (in H these are the Lyman
lines), inclusion of the $2^1s$--$1^1s$ two-photon rates, collisional
excitation, collisional deexcitation, collisional ionization, and
collisional recombination, as described in 
\S\ref{sec-Recomb}--\ref{sec-Sobolev}.  The multi-level \ion{He}{1} atom
includes the first 4 angular momentum states up to the level n=20, above which
only the principal quantum number energy levels and transitions are used.
Fig.~8
(which shows the levels up to n=4 only)
indicates how much more complicated the
\ion{He}{1} atom is compared with H or the hydrogenic \ion{He}{2}.
Our multi-level \ion{He}{2} atom includes the first 4 angular momentum states
up to the level n=4, above which only the principal quantum number energy
levels and transitions are used.

\centerline{{\vbox{\epsfxsize=8cm\epsfbox{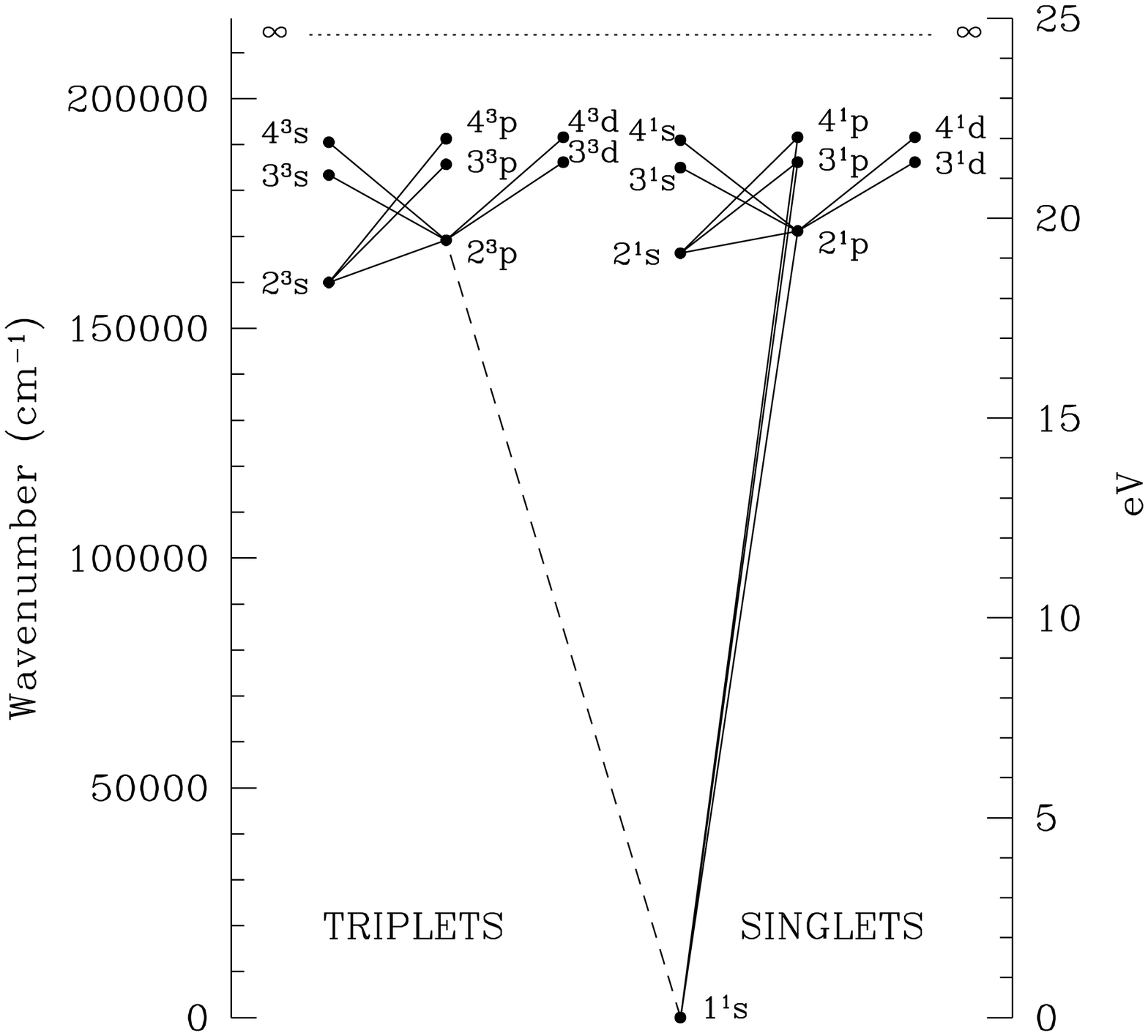}}}}
\baselineskip=8pt
{\footnotesize {\sc Fig.~8.---}
Grotrian diagram for \ion{He}{1}, showing the states with
n~$\le4$ and the continuum.  In practice our model atom explicitly
contains the first 4 angular momentum states up to n=20, and
120 principal quantum number energy levels beyond.
\label{fig:Hegrotrian}
}
\vspace{2mm}

\baselineskip=11pt

Photoionizations from
any \ion{He}{1} excited state are allowed only into the ground state
of \ion{He}{2}, because there are few photons energetic enough ($>40\,$eV)
to do more than that.  Two electron transitions in \ion{He}{1} are
negligible at recombination era temperatures.

Cosmological helium recombination was discussed explicitly in
Matsuda et al.~(1969, 1971, hereafter MST), and Sato, Matsuda \& Takeda (1971),
and to a lesser extent in Lyubarsky \& Sunyaev (1983),
while several
other papers give results, but no details (e.g.~Lepp \& Shull~1984;
Fahr \& Loch~1991; Galli \& Palla~1998).
The main improvement in our calculation over previous treatments of helium is
that we use a multi-level \ion{He}{2} atom, a multi-level \ion{He}{1}
atom with triplets and singlets treated correctly, and evolve the
population of each energy level with redshift by including all
bound-bound and bound-free transitions. This is not possible for the
standard recombination calculation method
(equation~(\ref{eq:standard_xe})) extended to \ion{He}{1}, using an
effective three-level \ion{He}{1} atom with only a singlet ground
state, singlet first excited state and continuum.

\subsubsection{Results From \ion{He}{1} Recombination}

Fig.~9,
shows the ionization fraction $x_{\rm e}$
through \ion{He}{2}, \ion{He}{1} and H recombination, plotted against the
standard H calculation that includes \ion{He}{2} and \ion{He}{1}
recombination via the Saha equation.  For completeness we give the
helium Saha equations here: \\
for \ion{He}{1}$\,\leftrightarrow\,$\ion{He}{2}
\begin{equation}
\label{eq:heonesaha}
{(x_{\rm e}-1)x_{\rm e}\over 1+f_{\rm He}-x_{\rm e}} = 
 4 {(2\pi m_{\rm e} k_{\rm B} T)^{3/2}\over h_{\rm P}^3 n_{\rm H}}
 {\rm e}^{-\chi_{\rm He I}/k_{\rm B}T},
\end{equation}
and for \ion{He}{2}$\,\leftrightarrow\,$\ion{He}{3}
\begin{equation}
\label{eq:hetwosaha}
{(x_{\rm e}-1-f_{\rm He})x_{\rm e}\over 1+2f_{\rm He}-x_{\rm e}} = 
 {(2\pi m_{\rm e} k_{\rm B} T)^{3/2}\over h_{\rm P}^3 n_{\rm H}}
 {\rm e}^{-\chi_{\rm He II}/k_{\rm B}T}.
\end{equation}
Here the $\chi$s are ionization potentials, $n_{\rm H}$ is the total
number density of hydrogen, $f_{\rm He}$ is the total number fraction of
helium to hydrogen $f_{\rm He}=n_{\rm He}/n_{\rm H}=Y_{\rm
P}/4(1-Y_{\rm P})$, and our definition of $x_{\rm e}\equiv n_{\rm e}/n_{\rm H}$
results in the complicated-looking left hand sides.  The extra factor of
4 on the right hand side for \ion{He}{1}$\,\leftrightarrow\,$\ion{He}{2}
arises from the statistical weights factor. 
 
\vspace{2mm}
\centerline{{\vbox{\epsfxsize=8cm\epsfbox{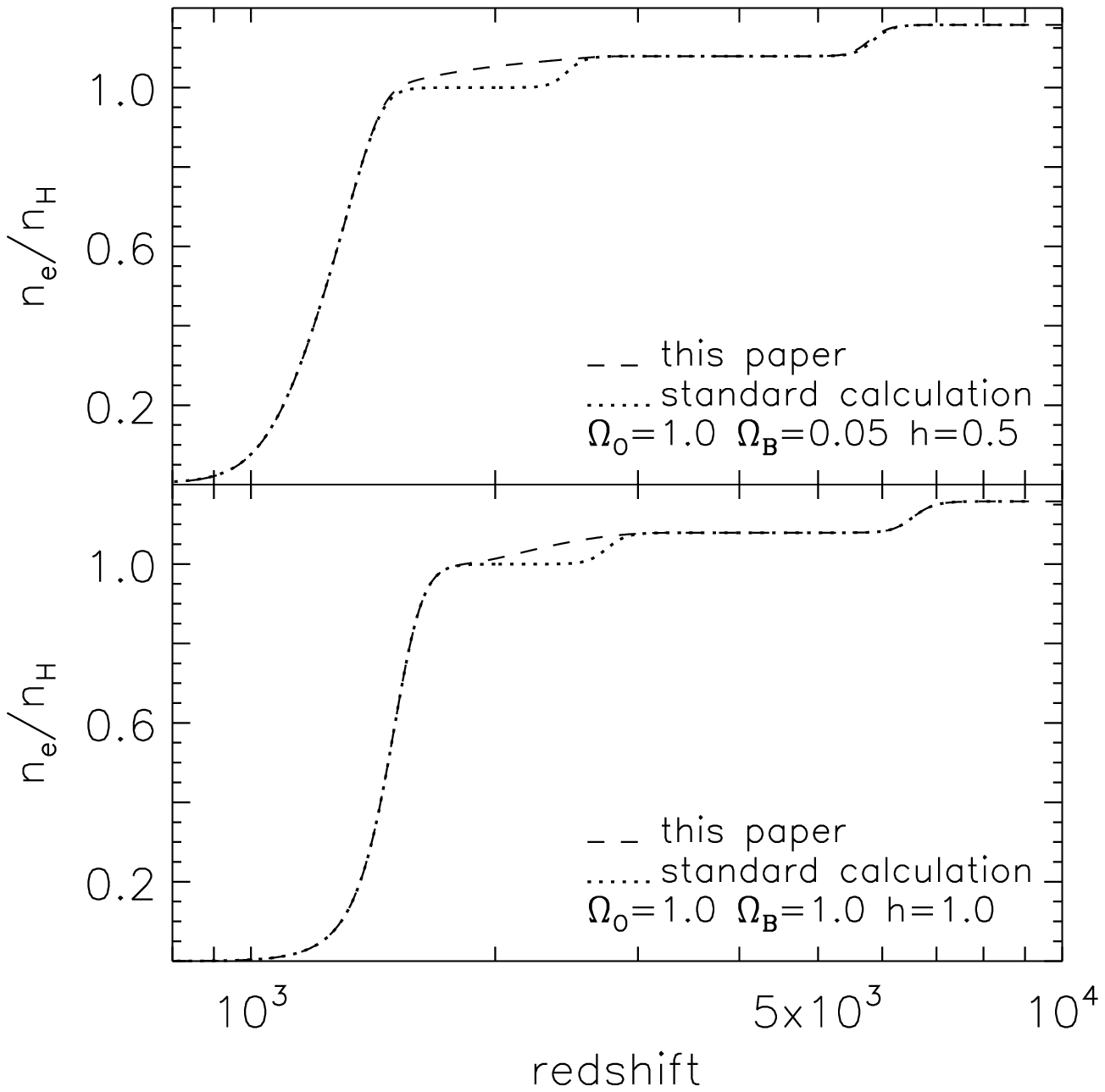}}}}
\baselineskip=8pt
\vspace{6mm}
{\footnotesize {\sc Fig.~9.---}
Helium and hydrogen recombination for two cosmological models with
$Y_{\rm P} = 0.24$ and $T_0 = 2.728\,$K.  The first step from
right to left is recombination of \ion{He}{3} to \ion{He}{2}, the second
step is \ion{He}{2} to \ion{He}{1}, and the third step is H recombination.
\label{fig:Herec}
}
\vspace{2mm}

\baselineskip=11pt

\centerline{{\vbox{\epsfxsize=8cm\epsfbox{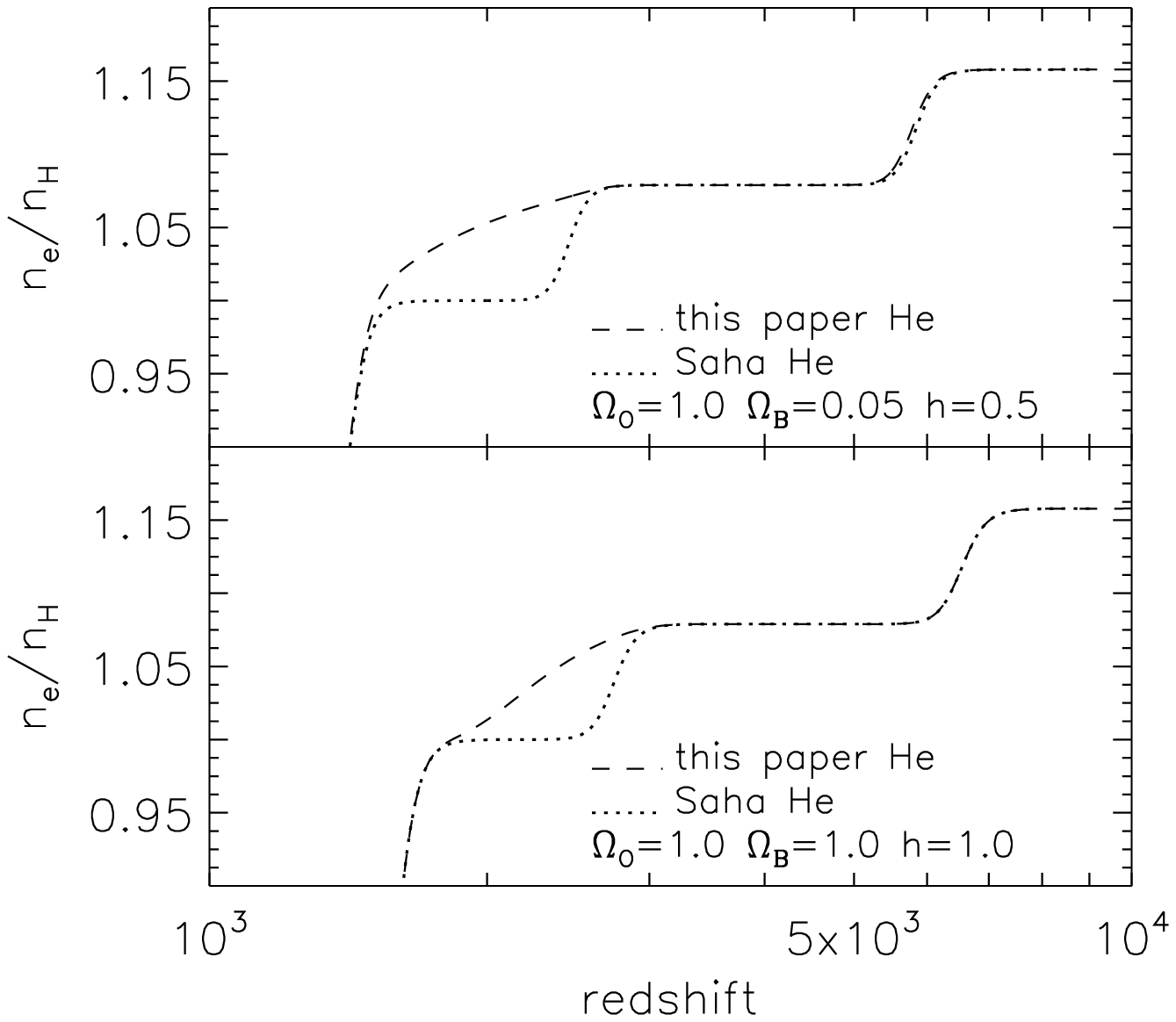}}}}
\baselineskip=8pt
\vspace{3mm}
{\footnotesize {\sc Fig.~10.---}
Details of helium recombination for the standard CDM
cosmology (top figure) and the high $\Omega_{\rm B}$ cosmology (bottom
figure). The dashed lines show our new results, and the dotted lines 
show the results assuming the `standard calculation' (equivalent to
Saha equilibrium).
\label{fig:Herecdetails}
}
\vspace{2mm}

\baselineskip=11pt

While our improved $x_{\rm e}$ agrees fairly closely with Saha recombination
for \ion{He}{2} (see~\S\ref{sec-HeII}), the 
difference in $x_{\rm e}$ from Saha recombination during \ion{He}{1}
recombination is dramatic.
Our new detailed treatment of \ion{He}{1} shows \ion{He}{1}
recombination finishing just after the start of H recombination (see
Fig.~9),
i.e.~significantly delayed compared with the
Saha equilibrium case. This
is different from the earlier calculations (e.g.~MST), in which
\ion{He}{1} recombination is
finished well before H recombination begins. In this previous case,
\ion{He}{1} recombination still affected the CMB anisotropy power
spectrum on small angular scales because the diffusion damping length
grows continuously and is
sensitive to the full thermal history (HSSW). In our new case,
particularly for our low
$\Omega_{\rm B}$ models \ion{He}{1} recombination is still finishing at the
very beginning of H recombination, which further affects the power spectrum
at large angular scales (see~\S\ref{sec-Powerspectrum}).
We show a `blow-up' of the two helium recombination epochs in
Fig.~10.

\subsubsection{Physics of \ion{He}{1} Recombination}
The physics of \ion{He}{1} recombination can be summarized as follows. There
are three major aspects to it:  (1) the \ion{He}{1} has excited states which
are able to retain charge; but (2) being very close to the continuum, the
highly excited states are easily photoionized by the radiation field at
$z \simeq 3000$; then (3) we have a standard hydrogenic-like Case B
recombination, which is unaffected by neutral H removing \ion{He}{1}
$2^1p$--$1^1s$ (resonance line) photons.

\vspace{2mm}
\centerline{{\vbox{\epsfxsize=8cm\epsfbox{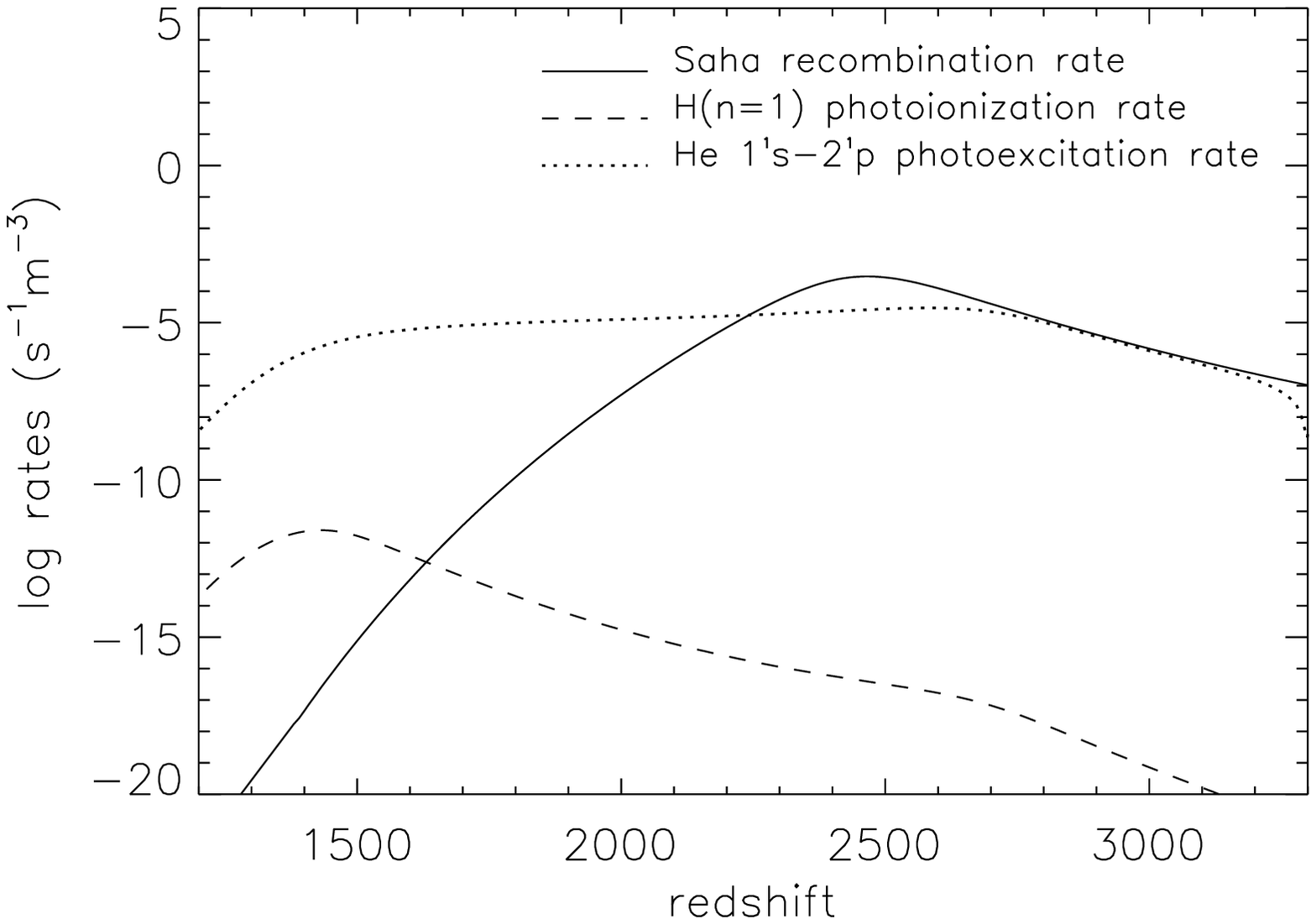}}}}
\baselineskip=8pt
{\footnotesize {\sc Fig.~11.---}:
This figure shows why the Saha equilibrium recombination rate
($R_{\rm Saha}$) for
\ion{He}{1} is not valid. Comparing the dotted and dashed lines,
the photoexcitation rate (i.e.~photoabsorption rate) for He 2$^1p\rm{-}1^1s$
($R_{\rm He}$) is orders
of magnitude greater than the photoionization rate for H ($R_{\rm H}$)
from the same \ion{He}{1} 2$^1p \rm{-} 1^1s$
photon pool; there is no possibility for H to `steal' the
photons to speed up \ion{He}{1} recombination. For
Saha recombination to be valid, $R_{\rm H} > R_{\rm He}$, as well as
$R_{\rm H} \geq R_{\rm Saha}$.  For the sCDM  model shown here, \ion{He}{1}
recombination begins around $z=3000$.
\label{fig:fracnu}
}
\vspace{2mm}

\baselineskip=11pt

The \ion{He}{1} atom has a metastable, i.e.~very slow, set of
states -- the triplets (e.g.~n$^3p$--n$^1s$). Therefore, overall the
excited states of \ion{He}{1} can naturally retain more charge than a simple
hydrogenic system under Boltzmann equilibrium. The situation would resemble
what we found for H recombination with the enhanced populations of the
higher states, and would lead to faster reduction of $x_{\rm e}$.  However, the
high excited states of \ion{He}{1} are much more strongly `packed' towards the
continuum compared to those of H;
the energy difference between the $3p$ levels and the continuum is
$1.6\,$eV for \ion{He}{1} versus $1.5\,$eV for H, 
compared to $24.6\,$eV versus $13.6\,$eV for the ground state--continuum energy difference.
This is
enough to depopulate the triplets (whose `ground state' is n=$2^3s$),
given the much higher radiation temperature during \ion{He}{1} recombination.
Left on its own under these circumstances, \ion{He}{1} would recombine much
like the standard Case~B effective 3-level H atom, i.e.~slower than Saha
recombination. There is one possible obstacle -- it is the existence of some
neutral H, which could `steal'  \ion{He}{1} resonance line photons,
invalidate the effective Case~B and make it a Saha recombination instead.
However, our detailed calculation 
shows that neutral H during \ion{He}{1} recombination is not able to
accomplish that, and the process is {\it not\/} described by Saha equilibrium.

Fig.~11
shows that Saha equilibrium recombination is
invalid for \ion{He}{1}, by comparing the three possible destruction
processes of the \ion{He}{1} $2^1p$--$1^1s$ (in H this is Ly$\,\alpha$)
photons: (1) cosmological redshift; (2) the $2^1s$--$1^1s$ two-photon rate;
and (3) the photoionization rate of the ground state of H by the same
\ion{He}{1} $2^1p$--$1^1s$ photons.
Fig.~11
clearly shows that process (3) is negligible
(in contradiction to the discussion in HSSW).
To be doubly sure that
absorption of these photons by hydrogen is negligible we explicitly
included the relevant rate in our models and found no discernible effects.
In order for \ion{He}{1} recombination to be approximated by Saha equilibrium,
one of the 3 processes described above would have to be 
faster than or equal to the Saha equilibrium rate, which we do not find to be
the case.

The recombination of \ion{He}{1} is slow for the same reasons that H
recombination is,
namely because of the optically thick n$^1p$--$1^1s$ transitions which
slow cascades to the ground state, and the 
exclusion of recombinations to the ground state.
In other words \ion{He}{1} follows a Case B recombination. Because the
`bottleneck' at n=2 controls recombination, it is not surprising
that \ion{He}{1} and H recombination occur at a similar redshift; the
ionization energy of n=2 is similar in both. \ion{He}{1}
recombination is slower than H recombination because of its different
atomic structure. The excited states of \ion{He}{1} are more tightly packed,
and the $2^1p$--$1^1s$ energy difference greater than that of H.
The strong radiation field keeps the ratio of photoionization rate/downward
cascade rate higher than in the H case, resulting in a slower recombination.

We find the strong radiation field also causes the triplet states to
be virtually unpopulated. 
The lack of electrons in triplet states is easily understood by
considering the blackbody radiation spectrum.
At \ion{He}{1} recombination (z $\simeq$ 3000), the blackbody
radiation
peak is around $2\,$eV, so there are around 11 orders of magnitude
more photons that can
ionize the lowest triplet state 2$^3$s ($4.8\,$eV),
than the singlet ground state ($24.6\,$eV), since both are on the
steeply decreasing Wien tail.
It is interesting to note that in planetary nebulae where the young,
hot ionizing star produces most of its energy in the UV, the opposite
occurs: the \ion{He}{1} atoms have few electrons in the singlet
states; instead most of them are in the triplet states.
 
There is one more possible method to speed up \ion{He}{1}
recombination, and that is collisional rates between the triplets and
singlet states. If fast enough, the collisional rates would provide
another channel to keep hold of captured electrons -- by pumping them into the
triplet states faster than they can be reionized. The triplets are 3
times as populated as the singlets due to the statistical weight factors.
By forcing the collisional
rates to be greater than the recombination rates and the bound-bound
radiative rates, we find an extremely fast He I recombination --
approximated by the Saha equilibrium. Essentially we force electrons
from the singlets into the triplets faster than they can cascade
downwards, and faster than they can be photoionized out of the
triplets. In reality, the collisions are
negligible, a few orders of magnitude less than the radiative rates.
It is important to note that
apart from collisions, the singlet and triplet states are {\it only\/}
connected via the n$^3p$--n$^1s$ transitions, which are orders of
magnitude slower than the $2^3s$--$1^1s$ rate.
We note here that MST stated that the collisional rates were high
enough to cause equilibrium between the triplet and singlet states.
One must be careful
to compare all relevant rates, and we keep all of them in our code.
We find the allowed radiative rates
(e.g.~photoexcitation and photodeexcitation) are greater than the
collisional rates. Therefore the allowed radiative rates control the
excited states' population distribution, {\it not\/} the
collisional rates.  In other words, electrons in the singlet states
are jumping between bound singlet states faster than
the collisional rates can send them into the triplet states.

\subsubsection{Effective 3-level calculation for \ion{He}{1}}

We note here that MST used an effective 3-level \ion{He}{1} singlet atom
and calculated
\ion{He}{1} recombination in the same way as the standard H calculation
(equation~(\ref{eq:standard_xe})) 
with the appropriate \ion{He}{1} parameters.  When we follow their
treatment, we get essentially the same result as our multi-level \ion{He}{1} calculation.
We are not sure why MST obtained such a fast \ion{He}{1} recombination. 

As with hydrogen, we have also investigated what is required to
achieve an accurate solution for helium, without modeling the full
suite of atomic processes.  We have found that the use of the
`effective 3-level' equations for helium (as described in MST),
together with an appropriate recombination coefficient for singlets only
(equation~(\ref{eq:hecoefficient})), results in a very accurate
treatment of $x_{\rm e}(z)$ during the time of helium recombination.
In detail it is necessary to follow hydrogen and
helium recombination simultaneously, increasing the number of
differential equations to solve.  However, little accuracy is in fact lost
by treating them independently -- since recombination is governed by dramatic
changes in time scales through Boltzmann factors and the like, and is
affected little by small changes in the number of free electrons at a given
time.  Further details are discussed in Seager, Sasselov, \& Scott (1999).

Although our model does not explicitly use a recombination coefficient,
it does allow us to calculate one easily. 
To aid other researchers it
is worth presenting a fit for the singlet-only Case~B recombination
coefficient for \ion{He}{1} (including recombinations to all states except the ground state) from the data in Hummer \& Storey (1998).  
Hummer \& Storey (1998) compute photoionization cross sections that are more
accurate than the ones we use (Hofsaess 1979), but are not publicly available. Following the functional forms used in the
fits of Verner \& Ferland (1996) we find
\begin{equation}
\label{eq:hecoefficient}
\alpha_{\rm He}=a\!\left[\sqrt{T_{\rm M}\over T_2}
 \left(1+\sqrt{T_{\rm M}\over T_2}\right)^{1-b}\!\!
 \left(1+\sqrt{T_{\rm M}\over T_1}\right)^{1+b}\right]^{-1}\!\!\! m^3 s^{-1},
\end{equation}
with
$a=10^{-16.744}$m$^3$s$^{-1}$, $b=0.711$, $T_1=10^{5.114}\,$K, and $T_2$ fixed
arbitrarily at $3\,$K. This fit is good to $<0.1\%$
over the relevant temperature range (4{,}000--10{,}000\,K), and still
fairly accurate over a much wider range of temperatures.

\subsubsection{\ion{He}{2} Recombination}
\label{sec-HeII}
\ion{He}{2} recombination occurs too early to affect the power spectrum
of CMB anisotropies. For completeness, we mention it briefly
here. \ion{He}{2} recombination is fast because of the very fast two
photon rate.
Fig.~14
shows that for most cosmologies the two-photon
rate is faster than the net recombination rate, meaning that as fast
as electrons are captured from the continuum they can cascade down to
the ground state. Because of this, there is essentially no `bottleneck' at the
n=2 level.  In high baryon models, \ion{He}{2} recombination can be
approximated using the Saha recombination. As shown in
Fig.~9,
\ion{He}{2} recombination is slightly slower than the Saha recombination for
low baryon models. 

\vspace{0.2in}
\centerline{{\vbox{\epsfxsize=8.25cm\epsfbox{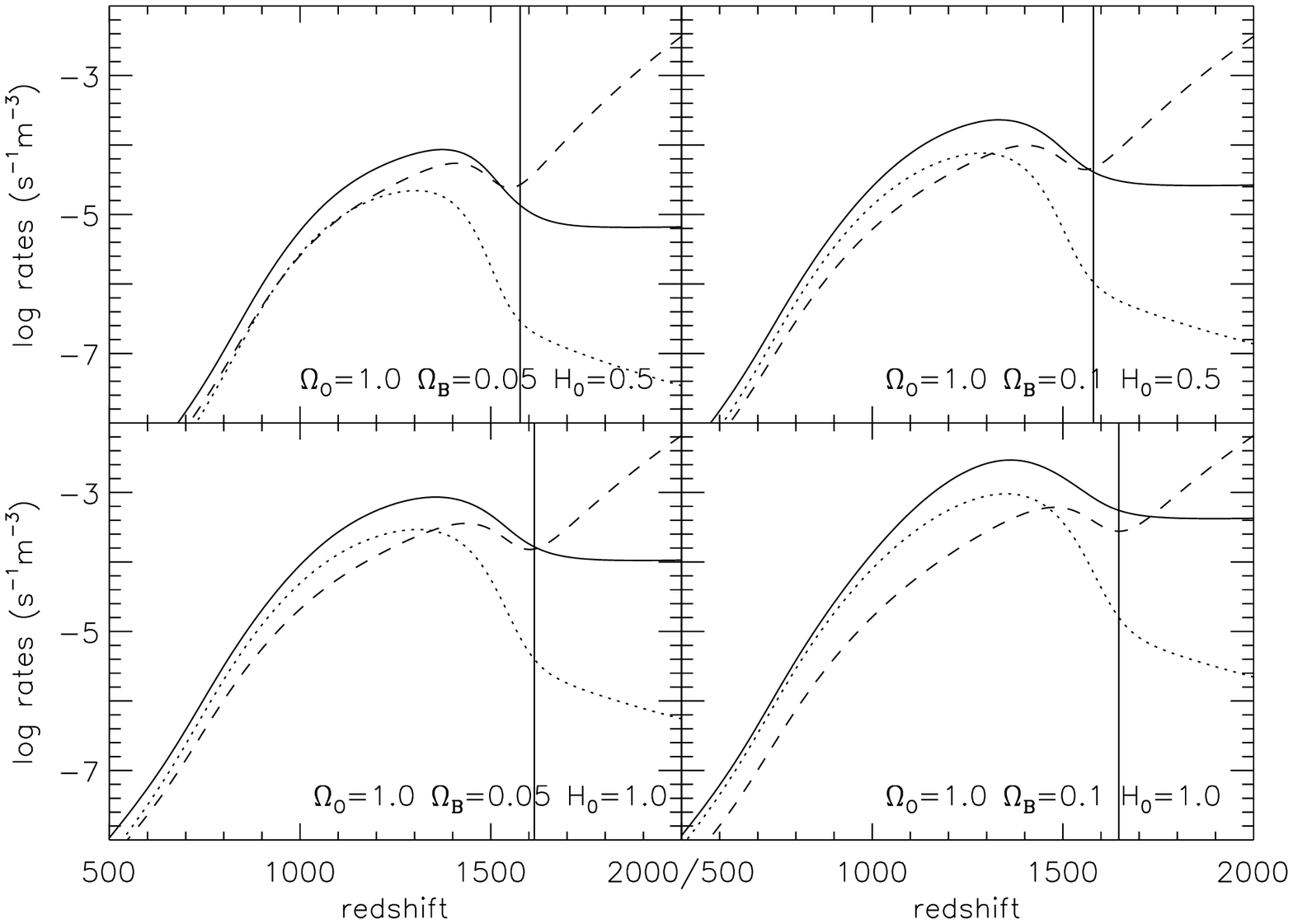}}}}
\baselineskip=8pt
\vspace{0.3in}
{\footnotesize {\sc Fig.~12.---}
What controls H recombination? The net $2p$--$1s$ rate
(dashed) compared to the $2s$--$1s$ two-photon rate (dotted) and the
net recombination rate (solid) for 4 different cosmologies. Except for
low $\Omega_{\rm B}$ and low $h$ models (e.g.~the sCDM model),
the $2s$--$1s$ rate dominates.  The solid vertical line represents where
5\% of the atoms have recombined.
\label{fig:Hrates}
}
\vspace{2mm}

\baselineskip=11pt

\subsubsection{What Controls Recombination?}
H recombination is largely controlled by the $2s$--$1s$
two-photon rate, which except for low-baryon cases, is much faster than
the H Ly$\,\alpha$ rate. The net recombination rate, net $2s$--$1s$
rate, and net Ly$\,\alpha$ rate are compared for different cosmologies
in
Fig.~12.
Figs.~13
and~14
show the same rate
comparison for \ion{He}{1} and \ion{He}{2}.
The three figures all have the same scale on the $x$ and $y$ axes, for
easy comparison. 
\ion{He}{1} recombination is controlled by the $2^1p$--$1^1s$ rate
rather than the $2^1s$--$1^1s$ rate as previously stated (e.g.~MST).
Fig.~13
(for \ion{He}{1}) also illustrates the slow net
recombination rate, which is the primary factor in the slow Case B
\ion{He}{1} recombination.
Fig.~14
also illustrates that \ion{He}{2} in the
high $\Omega_{\rm B}$ and $h$ models has a $2s$--$1s$ rate faster
than the net recombination rate, meaning that there is no slowdown
of recombination due to n$=2$, and the Saha equilibrium approximation
is valid. 

The rates change with cosmological model. Physically this is because
all of the rates are very sensitive to the baryon density. The
$2^1p$--$1^1s$ rates are further affected by the Hubble factor
because the Sobolev approximation
(equations~(\ref{eq:indofphi}) and~(\ref{eq:sobtau})), depends on the
velocity gradient.  Whether most of the atoms in the Universe recombined
via a $2p$--$1s$ or a $2s$--$1s$ two-photon transition depends on the
precise values of the cosmological parameters.  A confident answer to that
question is still not known, given today's parameter uncertainties.

\vspace{0.2in}
\centerline{{\vbox{\epsfxsize=8.25cm\epsfbox{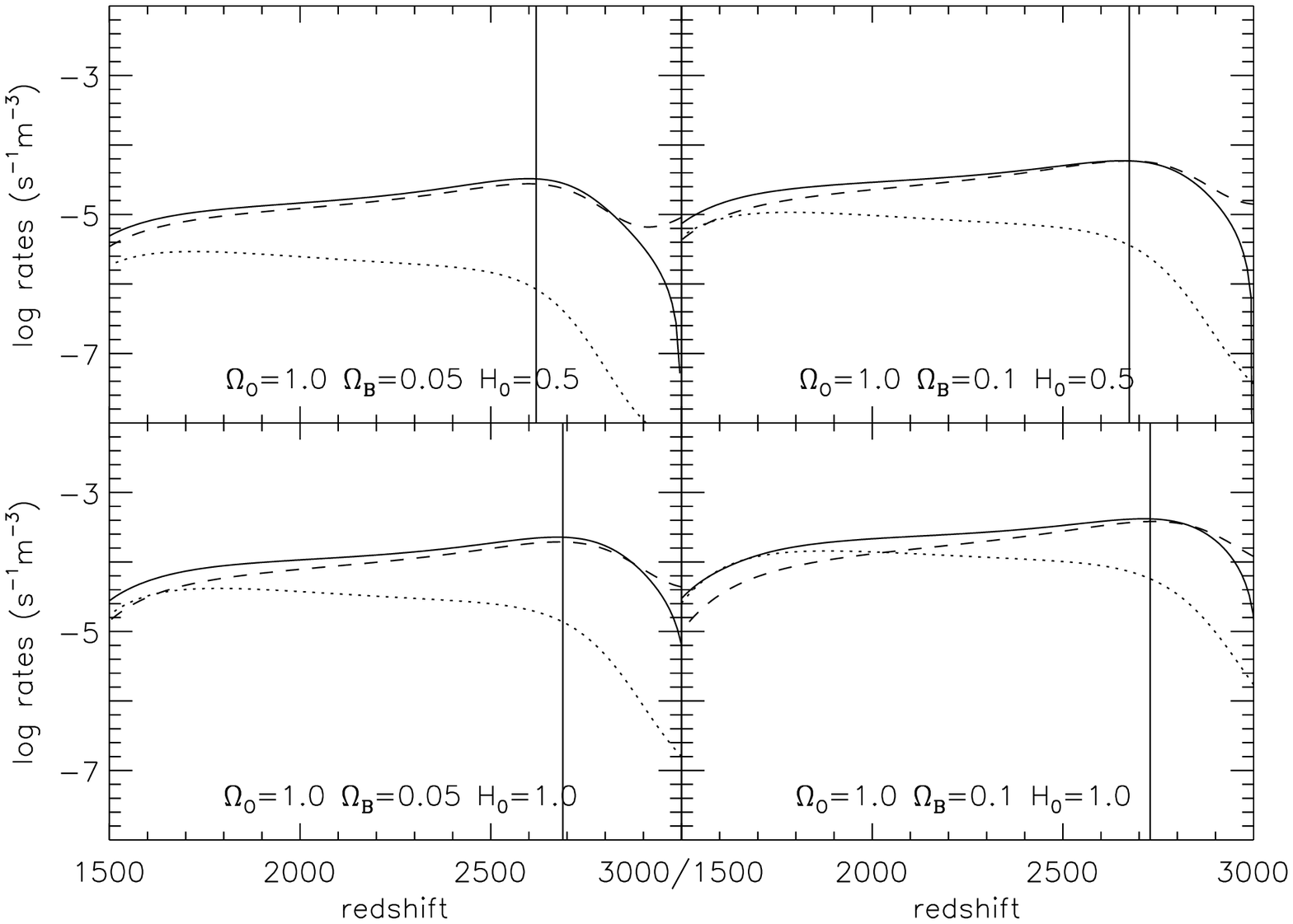}}}}
\baselineskip=8pt
\vspace{0.3in}
{\footnotesize {\sc Fig.~13.---}
What controls \ion{He}{1} recombination? The net
$2^1p$--$1^1s$ rate (dashed) compared to $2^1s$--$1^1s$
two-photon rate (dotted)
and the net recombination rate (solid) for 4
different cosmologies. Except for high $\Omega_{\rm B}$ and high $h$
models, the $2^1p$--$1^1s$ rate dominates, in contrast to H. The solid
vertical line represents where 5\% of the atoms have recombined.
\label{fig:HeIrates}
}
\vspace{2mm}

\baselineskip=11pt

\vspace{0.2in}
\centerline{{\vbox{\epsfxsize=8.5cm\epsfbox{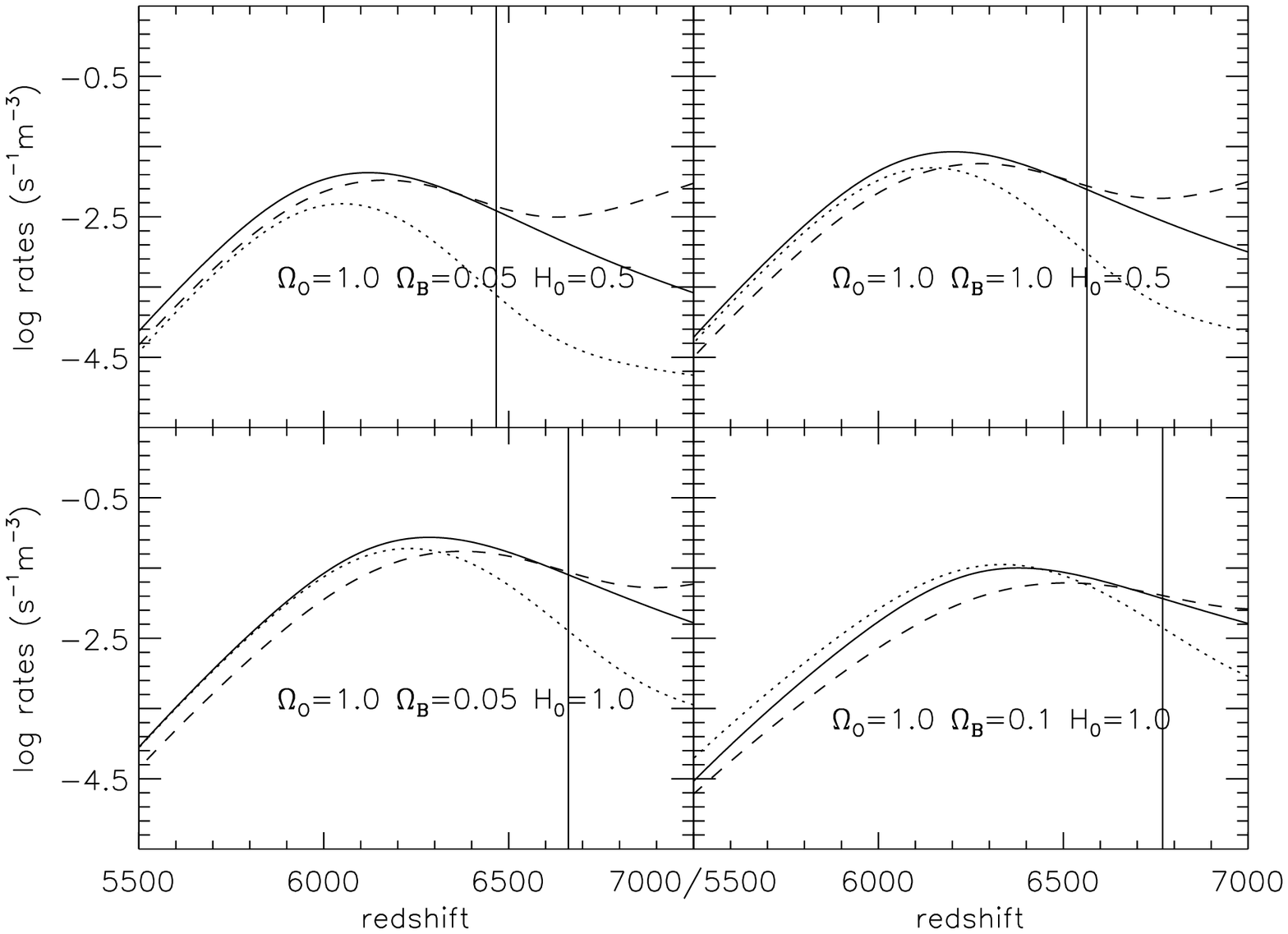}}}}
\baselineskip=8pt
\vspace{0.25in}
{\footnotesize {\sc Fig.~14.---}
What controls \ion{He}{2} recombination? The net $2p$--$1s$
rate (dashed) compared to the $2s$--$1s$ two-photon rate (dotted) and
the net recombination rate (solid) for 4 different cosmologies. Except
for low $\Omega_{\rm B}$ and low $h$ 
models, the $2s$--$1s$ rate dominates during recombination, and the
$2p$--$1s$ at the start of recombination. The solid vertical line
represents where 5\% of the atoms have recombined.
\label{fig:HeIIrates}
}
\vspace{2mm}

\baselineskip=11pt

\subsection{Atomic Data and Estimate of Uncertainties}
\label{sec-AtomicData} 
Our approach in this work has been to include all relevant degrees of freedom
of the recombining matter in a consistent and coupled manner. This requires
special attention to the quality of the atomic data used. The challenge
lies in building a consistent model for $all$ energy levels and transitions,
not just for the low-lying ones, which are often better known experimentally
and theoretically.

\subsubsection{H and \ion{He}{2}}
Hydrogen (and the hydrogenic ion of helium) have exactly known rate
coefficients for radiative processes from precise quantum-mechanical
calculations (uncertainties below 1\%). We use exact values for the
bound-bound radiative transitions and for radiative recombination,
as in e.g.~Hummer (1994). For more details see Hummer \& Storey (1987),
but also Brocklehurst (1970) and Johnson (1972). In particular, the
rate of radiative recombination to level n of a hydrogenic ion can 
be evaluated from the photoionization (bound-free) cross section for
level n, ${\sigma}_{{\rm nc}}(\nu)$, with the standard assumption of detailed
balance (see~\S\ref{sec-Recomb}).
For hydrogenic bound-free cross sections, we follow in essence
Seaton's work (Seaton 1959) with its 
asymptotic expansion for the Gaunt factor (see Brocklehurst 1970).
Note that the weak dependence of the Gaunt factor on wavelength has
a noticeable effect in our final recombination rate calculation.
Given our application, we do not require the resolution of resonances,
as achieved for a few transitions by the Opacity Project (TOPbase,
Canto et al.~1993).
Like Hummer (1994), we work with the n-levels assuming that the
$\ell$-sublevels have populations proportional to (2$\ell$+1). The
resulting uncertainties for hydrogenic radiative rates at low
temperatures (T~$\leq~$10$^5$K) certainly do not exceed the 1\% level.

Collisional rate coefficients cannot be calculated exactly.
So, compared to the hydrogenic radiative rate coefficients, the
situation for the bound-bound collisional rates
and collisional ionization is poor, with errors typically about 6\%,
and as high as 20\% in some cases (Percival \& Richards 1978).
A number of methods are used to evaluate
electron-impact excitation cross sections of hydrogen-like ions (Fisher
et al.~1997). These most recent values compare well to the older
sources (Johnson 1972; Percival \& Richards 1978). The helium ion, \ion{He}{2},
is hydrogenic and was treated accordingly. We basically
followed Hummer \& Storey (1987) and Hummer (1994) in building the model
atom. For our application,
collisional processes are negligible, so that the large
uncertainties that still persist for the collisional rates have no
impact on our results.

\subsubsection{\ion{He}{1}}
Helium, in its neutral state, poses a challenge for building a multi-level
atomic model of high precision. Unlike atomic hydrogen, no exact solutions
to the Schr\"odinger equation are available for helium.  However, very high
precision approximations are now available (Drake 1992, 1994)
which we have used.  These approximations are essentially exact for all
practical purposes. The largest 
relativistic correction comes from singlet-triplet mixing between states
with the same n, $L$, and $J$, but is still small.
Transition rates were calculated following
the recent comprehensive \ion{He}{1} model built by Smits (1996) and some
values in Theodosiou (1987).  The source of our photoionization cross
sections was TOPbase (Cunto et al.~1993) and Hofsaess (1979) for small n;
above n=10 we used scaled hydrogenic values.
New detailed calculations (Hummer \& Storey 1998)
show that the \ion{He}{1} 
photoionization cross sections become strictly hydrogenic at about n~$>20$.
The uncertainties in the \ion{He}{1} radiative rates are at the 5\%
level and below.

The situation with the collisional rates for \ion{He}{1} is predictably much
worse than for \ion{He}{1} radiative rates, with good R-matrix
calculations existing only for n~$\leq 5$ (Sawey \&
Berrington 1993). The collisional rates at large n are a crucial ingredient
in determining the amount of singlet-triplet mixing, but fortunately
collisions are not very important for the low density conditions in the early
Universe, so the large uncertainty in these rates does not effect
our calculation. The Born approximation,
which assumes proportionality to the radiative transition rates, is used
(see Smits 1996) to calculate the collisional cross sections for large n.

For the $2^1s$--$1^1s$ two-photon rate for \ion{He}{1} we used the value
$\Lambda_{\rm He I}=51.3\,{\rm s}^{-1}$ (Drake et al. 1969)
which differs from a previously used value (Dalgarno~1966) by
$\sim10\%$. An uncertainty even of this magnitude
would still make little difference in the final results.
For the \ion{He}{2} $2s$--$1s$ two-photon rate we used the value
$\Lambda_{\rm He I}=526.5\,{\rm s}^{-1}$ (for hydrogenic ions this is
essentially $Z^6$ times the value for H) from Lipeles et al.~(1965).
Dielectronic recombination for \ion{He}{1} is not at all
important during \ion{He}{1} recombination. While dielectronic
recombination dominates at temperatures above
$6\times10^4$K, for the range of temperatures relevant here it
is at least 10 orders of magnitude below the radiative recombination
rate (using the fit referred to in Abel et al.~1997).

\subsubsection{Combined Error From Atomic Data}
We have gathered together the uncertainties in the atomic data in
order to estimate the resulting uncertainty in our derivation of $x_{\rm e}$.
The atomic data with the dominant effect on our calculation are the set
of bound-free cross sections for all H and \ion{He}{1} levels -- not so much 
any individual values, but the overall consistency of the sets (which
are taken from different sources). The differences between our
model atom and Hummer's (1994) reflect the uncertainty in the atomic
data. To test the effect on our hydrogenic results, we compared the $x_{\rm e}$
results of an effective 3-level atom using Hummer's (1994)
recombination coefficient with the
results using a recombination coefficient calculated with our own model H
atom. We find maximum
differences of 1\% at $z=300$, which corresponds to measurable effects on
CMB anisotropies of much less than 1\%.

The error in $x_{\rm e}$ arising from \ion{He}{1} is more difficult to
calculate.  We estimate it to be considerably less than 1\%, because the low
level (n~$\leq$~4)
bound-bound and bound-free radiative rates dominate \ion{He}{1}
recombination, and as described above, those data are accurate.

\subsection{Secondary Distortions in the Radiation Field}
\label{sec-Distortions}
In our recombination calculation we follow `secondary' distortions in the radiation
field that could affect the recombination process at a later time. The
secondary distortions are caused by the primary distortions that are
frozen into the radiation field. At a later time they are redshifted
into interaction frequency with other atomic transitions.
Explicitly, we follow: \newline
(1) H Ly$\,\alpha$ photons; \newline
(2) H $2s$--$1s$ photons. \newline
By the time of H recombination these photons have been redshifted
into an energy range where they
could photoionize H(n=2). In addition we follow: \newline
(3) \ion{He}{1} $2^1p$--$1^1s$; \newline 
(4) \ion{He}{1} $2^1s$--$1^1s$. \newline
By the time of H recombination these photons have been
redshifted into an energy range which could photoionize H(n=1).
And finally we also follow: \newline
(5) \ion{He}{2} Ly$\,\alpha$ photons; \newline
(6) \ion{He}{2} $2s$--$1s$ photons. \newline
By the time of H recombination these photons have been redshifted
into an energy range
which could photoionize H(n=1). These \ion{He}{2} photons bypass
\ion{He}{1} because the photons have not been
redshifted into a suitable energy range for interaction.

Here we only attempt to investigate the maximum effects of secondary spectral
distortions. To that end we do not include additional distortions
which are smaller.  For example,
Lyman lines other than (1), (3), and (5), whose distortions are smaller 
than Ly$\,\alpha$,
will produce a comparably smaller feedback on photoionization.
The \ion{He}{1} singlet recombination photons could theoretically
photoionize \ion{He}{1} triplet states, but as previously discussed
there are virtually no electrons in the triplet states, so this
process is also negligible. Another possible effect is due to the
similar energy levels of H and \ion{He}{2}: $\Delta E_{\rm He II} =
4\Delta E_{\rm H}$.  For example, the transition from
\ion{He}{2} (n=4) to (n=2) produces the same frequency photons as the
transition from H~(n=2) to (n=1).  These transitions are
theoretically competing for photons, and this effect can be important for other
astrophysical situations (e.g.~planetary nebulae) where H and \ion{He}{2}
simultaneously exist.  However, any such effect is negligible for primeval
recombination because during \ion{He}{2} recombination the amount of
neutral H is very small ([H/\ion{He}{2}] $< 10^{-8}$), and during
H recombination, there is almost no \ion{He}{2}
([\ion{He}{2}/H]$< 10^{-10}$).

Because we are only investigating maximum effects we assume the photons
were emitted at line center and are redshifted undisturbed until their
interaction with H(n=1) or H(n=2) as described above. We also 
assume two photons at half the energy for the $2s$--$1s$ transitions,
compared to the $2p$--$1s$ transitions.
The distorting photons emitted at a time $z_{\rm em}$ are
absorbed at a later time $z$, where
\begin{equation}
z = z_{\rm em} \nu_{\rm edge}/\nu_{\rm em}.
\end{equation}
Here $\nu_{\rm edge}$ is the photoionization edge frequency where the
photons are being absorbed, and $\nu_{\rm em}$ is the photon's
frequency at emission.
The distortions are calculated as
\begin{eqnarray}
\lefteqn{J(\nu,z) = {h \nu(z) c}\,
p_{ij}(z_{\rm em})\,\times}\nonumber \\
& & \qquad \bigg\{ n_j(z_{\rm em})\Big[A_{ji}
+ B_{ji}B\big(\nu_{\rm em},T_{\rm R}(z_{\rm em})\big)\Big]\nonumber \\
& & \qquad\qquad\quad \mbox{}- n_i(z_{\rm em})B_{ij}
 B\big(\nu_{\rm em},T_{\rm R}(z_{\rm em})\big)\bigg\} ,
\end{eqnarray}
where: $B(\nu_{\rm em},T_{\rm R}(z_{\rm em}))$ is the Planck function at the
time of emission; $A_{ji}, B_{ji}$ and $B_{ij}$ are the Einstein
coefficients; $p_{ij}$ is the Sobolev escape probability for the
line; and the other variables are as described previously.

The distortions (1) and (2) were previously discussed by Rybicki
\& Dell'Antonio (1992). They pointed out that the effect from (1) should
be small, because the Ly$\,\alpha$ distortion must be redshifted by at
least a factor of 3 to have any effect. This means that the Ly$\,\alpha$
photons produced at $z\lesssim2500$
will only affect the Balmer continuum at $z\lesssim800$ when the
recombination process (and any possibility of photoionization) is
almost entirely over.

We find that including the distortions (1) through (6) improves
$x_{\rm e}$ during H
recombination at
less than the $0.01\%$ level.
This difference is far too small to make a
significant change in the power spectrum, and it is negligible
compared to the major improvements in this paper, which are 
the level-by-level treatment of H, \ion{He}{1}, and \ion{He}{2},
allowing departures of the excited state populations from
an equilibrium distribution, calculating recombination directly to
each excited state, and the correct treatment of
\ion{He}{1} triplet and singlet states.
However, the removal of these distorting photons by photoionization
must be taken into account when calculating spectral distortions to
the CMB blackbody, which we plan to study in a later paper.

\subsection{Chemistry}
Including the detailed hydrogen chemistry (see~\S\ref{sec-Chem})
marginally affects the fractional abundances of protons and electrons at
low $z$.  However, the correction is of the order
$10^{-2}x_{\rm e}$ at $z<150$.
This change in the electron density
would change the Thomson scattering optical depth by $\sim 10^{-5}$,
too little to make a difference in the CMB power spectrum.

\vspace{3mm}
\centerline{{\vbox{\epsfxsize=8cm\epsfbox{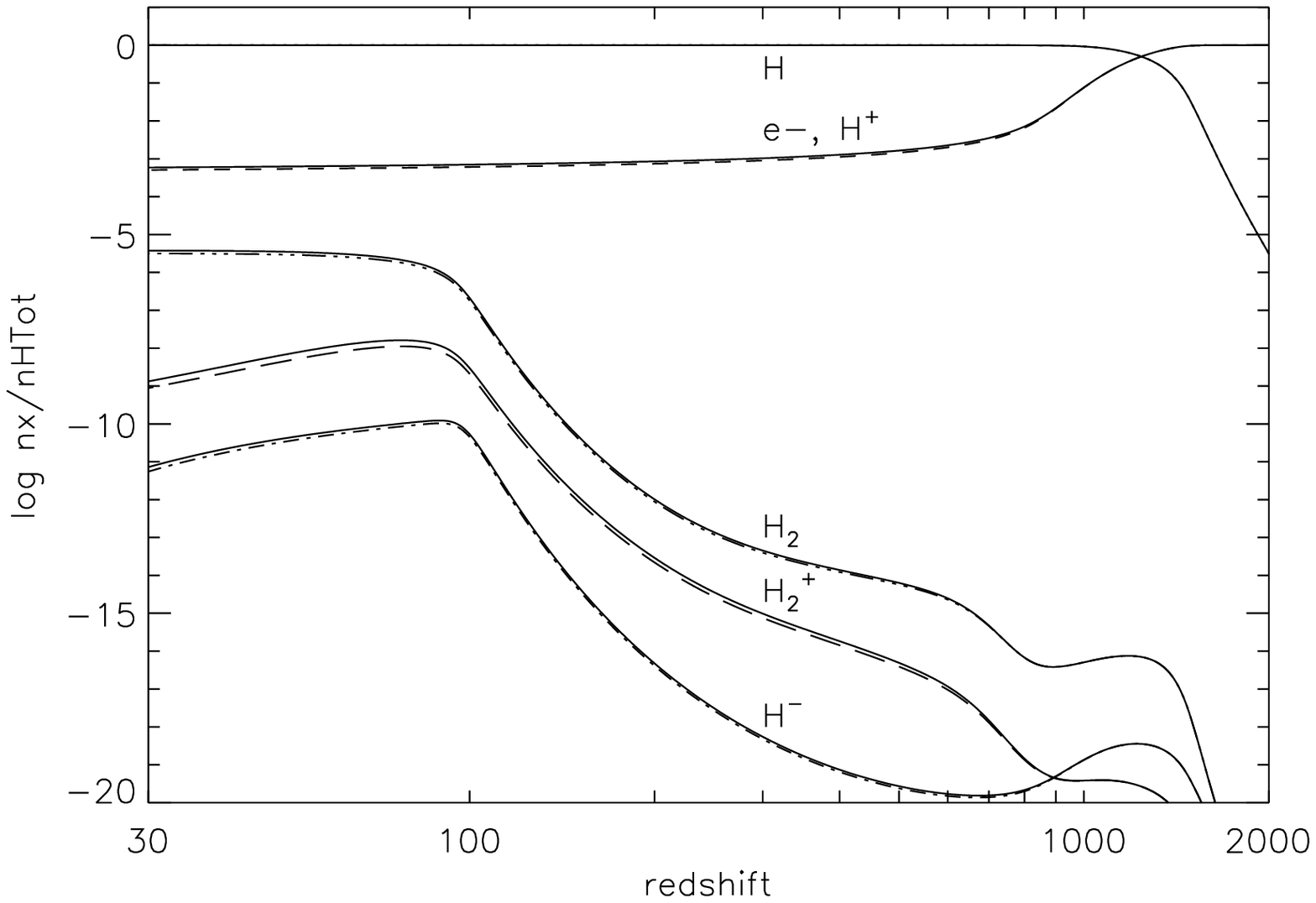}}}}
\centerline{{\vbox{\epsfxsize=8cm\epsfbox{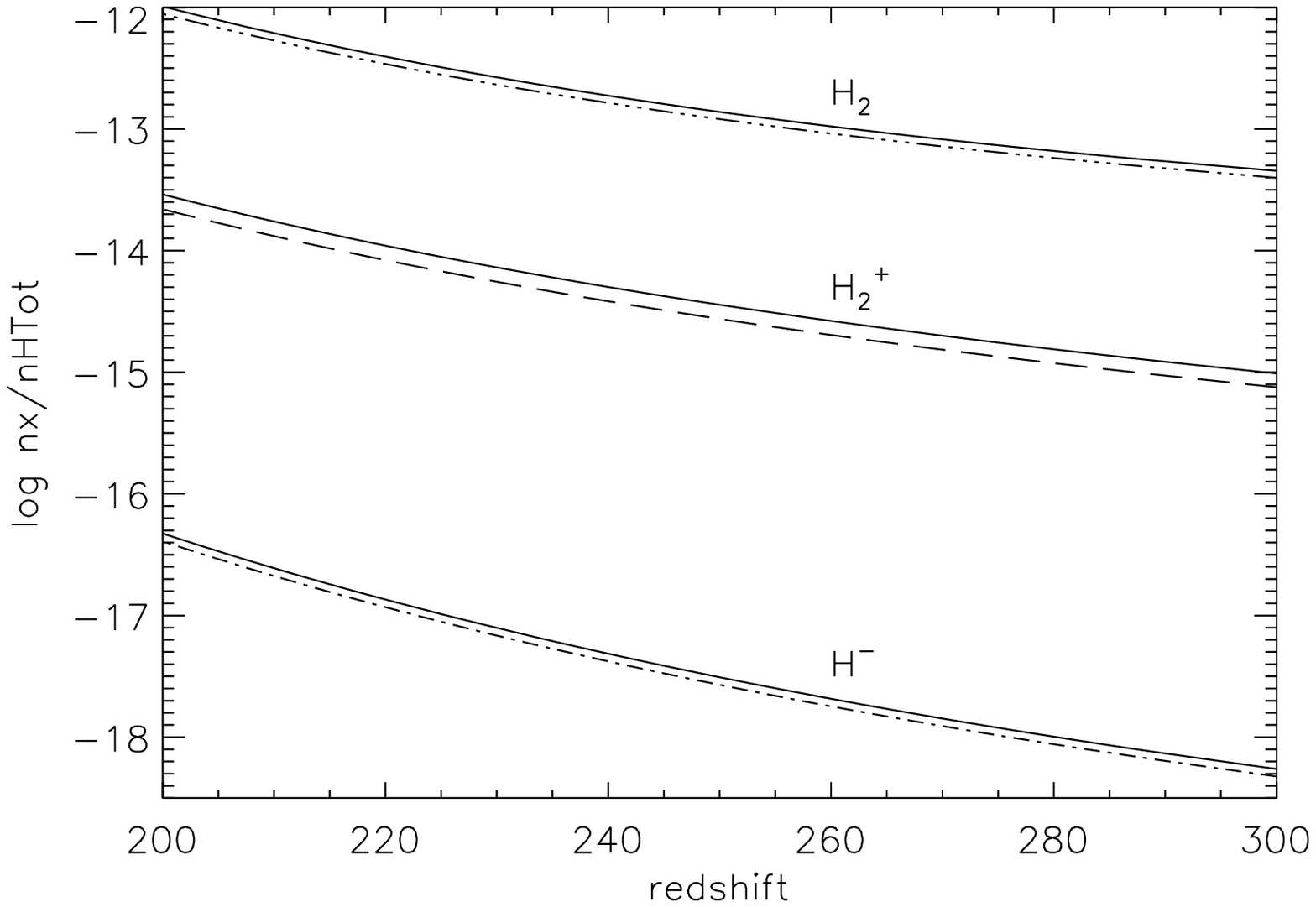}}}}
\baselineskip=8pt
{\footnotesize {\sc Fig.~15.---}
The effect of the improved treatment of recombination on H
chemistry. Shown is the standard CDM model. Solid lines are values
from the standard calculation, dashed and dotted lines from our
improved results.
\label{fig:chem}
}
\vspace{2mm}

\baselineskip=11pt

On the other hand, as shown in
Fig.~15,
the different
$x_{\rm e}(z)$ that we find will lead to 
fractional changes of similar size in molecular abundances at low $z$, since
H$_2$ for example is formed via H$^{-}$ which is affected by the residual
free electron density.  The delay in
\ion{He}{1} recombination compared to previous studies causes a
similar delay in formation of He molecules (P.~Stancil, private
communication).  However, with the exception of He$_2^+$, no
changes are greater than those caused by the residual
free electron density at freeze-out. Since molecules can be important
for the cooling of primordial gas clouds and the formation of the first
objects in the Universe, the precise determination of molecular abundances is
an important issue (e.g.~Lepp \& Shull 1984, Tegmark et al.~1997, Abel
et al.~1997, Galli \& Palla~1998).  However, the roughly 10-20\% change
in the abundance of some chemical species is probably less than other
uncertainties in the reaction rates (A.~Dalgarno, private communication).
With this in mind, we suspect no drastic implications for theories of
structure formation. 

\centerline{{\vbox{\epsfxsize=8cm\epsfbox{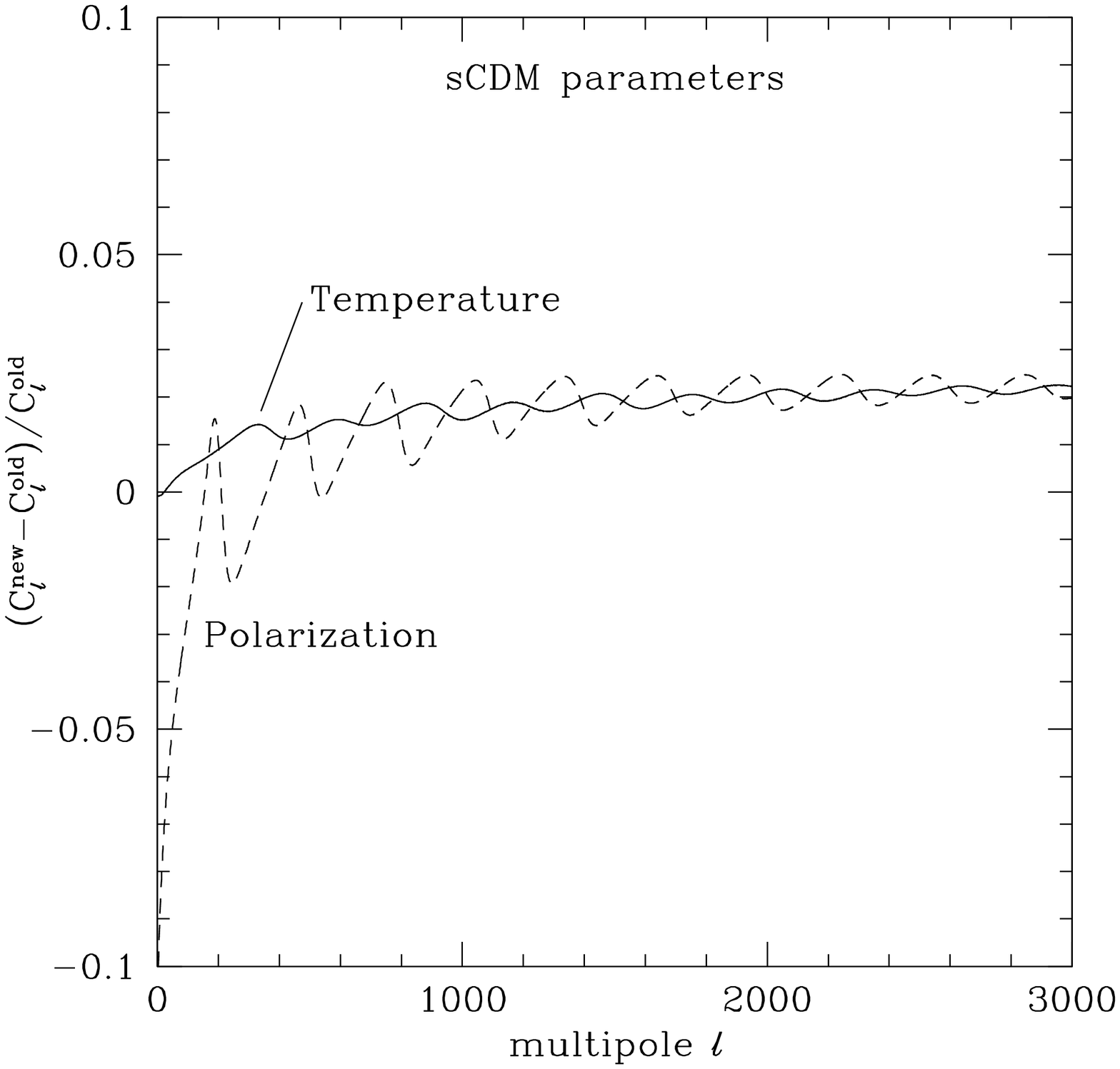}}}}
\centerline{{\vbox{\epsfxsize=8cm\epsfbox{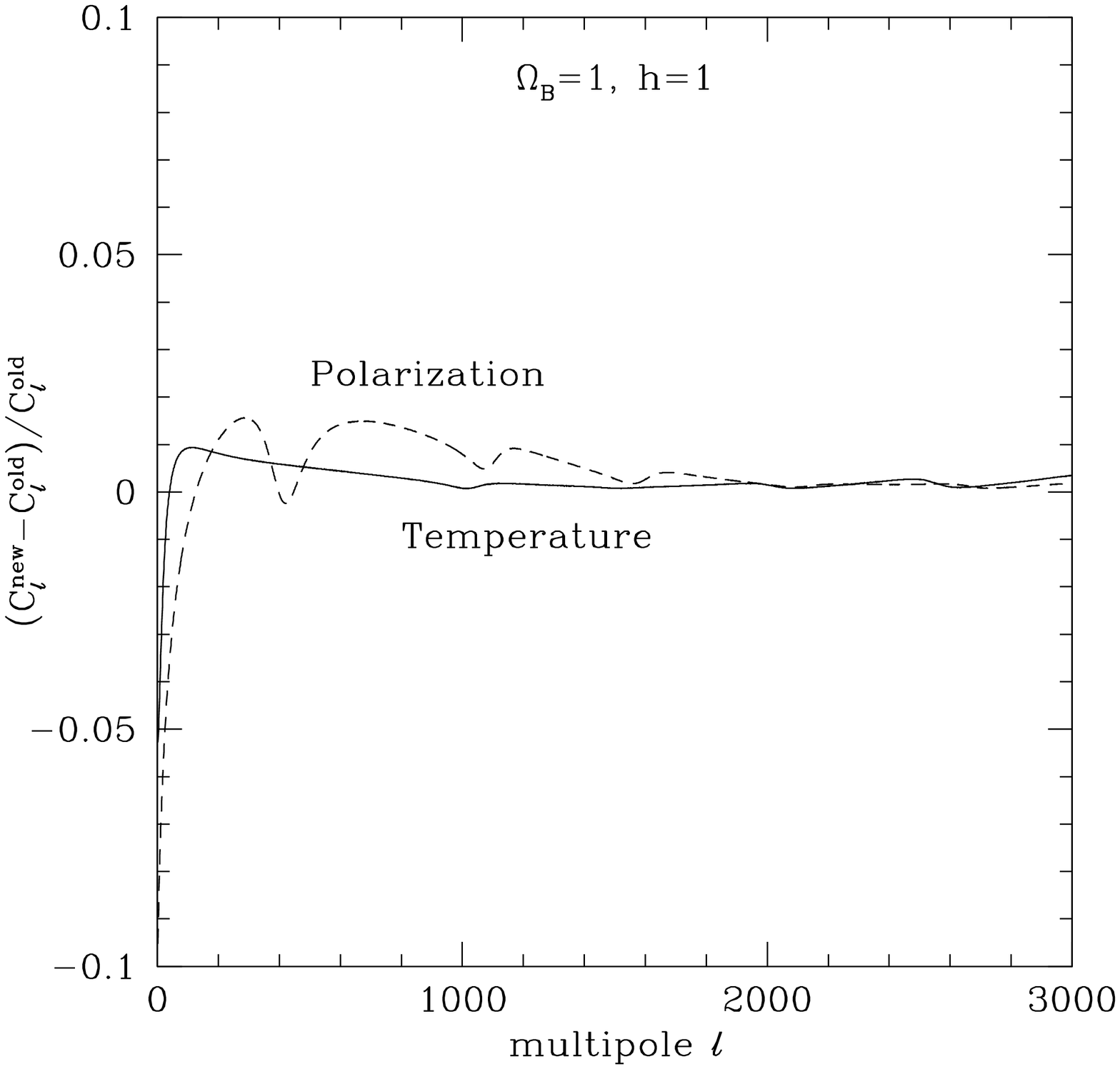}}}}
\baselineskip=8pt
{\footnotesize {\sc Fig.~16.---}
Differences in CMB power spectra arising from the improved treatment
of hydrogen for: (a) the standard CDM parameters
$\Omega_{\rm tot} = 1.0$, $\Omega_{\rm B} = 0.05$, $h =0.5$,
$Y_{\rm P} = 0.24$, $T_0 = 2.728\,$K; and (b) an extreme baryon model
with $\Omega_{\rm B}=1$, $h=1.0$ (for which there is relatively little
effect).  The fractional
difference plotted is between our new hydrogen recombination calculation and
the standard hydrogen recombination calculation (e.g.~in HSSW), with the
sense of $C_\ell^{\rm new}-C_\ell^{\rm old}$, and with the two calculations
normalized to have the same amplitude for the initial conditions.
The solid lines are for
temperature, while the dashed lines are for the (`E'-mode)
polarization power spectrum.
\label{fig:h_powerspectrum}
}
\vspace{2mm}

\baselineskip=11pt

\subsection{Power Spectrum}
\label{sec-Powerspectrum}
Even relatively small differences in the recombination history of the
Universe can have potentially measurable effects on the CMB
anisotropies. And so we might expect our two main changes (one in H
and one in He) to be noticeable in the power spectrum.
As a first example
Fig.~16
compares the difference in the anisotropy power spectrum derived
from our new $x_{\rm e}(z)$ to that derived from the standard
recombination $x_{\rm e}$ (essentially identical to that described in
HSSW), for hydrogen recombination only. Here the $C_{\ell}$s are
squares of the amplitudes in a spherical 
harmonic decomposition of anisotropies on the sky (the azimuthal index
$m$ depends on the choice of axis, and so is irrelevant for an
isotropic Universe).  They represent the power and angular scale of
the CMB anisotropies by describing the rms temperatures at fixed
angular separations averaged over the whole sky (see e.g.~White, Scott,
\& Silk~1994). These $C_{\ell}$s
depend on the ionization fraction $x_{\rm e}$
through the precise shape of the thickness of the photon last
scattering surface (i.e.~the visibility function).
Since the detailed shape of the power spectrum may allow
determination of fundamental cosmological parameters,
the significance of the change in $x_{\rm e}$ is evident.
To determine the effect of the change in $x_{\rm e}$ we have
used the code {\tt cmbfast} written and made available by Seljak
\& Zaldarriaga (1996), with a slight modification to allow for the input
of an arbitrary recombination history.

The dominant physical affect arising from the new H calculation comes
from the change in $x_{\rm e}$ at low $z$.  A process seldom mentioned in
discussions of CMB anisotropy physics (which are otherwise quite
comprehensive, e.g.~Hu, Silk, \& Sugiyama~1997) is that the low-$z$
tail of the visibility function results in {\it partial erasure\/} of the
anisotropies produced at $z\sim1000$.  The optical depth in Thomson
scattering back to, for example, $z=800$
($\tau=c\sigma_{\rm T}\int n_{\rm e}(dt/dz)\,dz$)
can be several percent.  This partial rescattering of the photons
leads to partial erasure of the $C_\ell$s by an amount ${\rm e}^{-2\tau}$.  Let
us look at the standard CDM calculation first
(Fig.~16(a)).
Our change in the optical depth back to $z\simeq800$ (see
Fig.~2)
is around 1\% less than that obtained
using the standard
calculation, and so we find that the anisotropies suffer less partial erasure
by about 2\%.
There is no effect on angular scales larger than the horizon at the
scattering epoch (here redshifts of several hundred), so that all multipoles
are effected except for the lowest hundred or so $\ell$s.  Hence this
effect is largely a change in the overall normalization of the power spectrum,
with some additional differences at low $\ell$ which will be masked by the
`cosmic variance'.  In addition there are smaller effects due to changes in
the {\it generation\/} of anisotropies in the low-$z$ tail, giving small
changes in the acoustic peaks, which can be seen as wiggles in the figure.
Since the partial erasing effect is essentially unchanged in the case of the
$\Omega_{\rm B}=h=1.0$ model, these otherwise sub-dominant effects are more
obvious in
Fig.~16(b).

Differences in the power spectra are rather small in absolute terms,
so
Fig.~16
plots the relative difference.  We
have shown this for our two chosen models, one being standard Cold Dark Matter
(a), which we will refer to as sCDM,
and the other being an extreme baryon-only model (b).  These models are
meant to be representative only, and changes in cosmological parameters will
result in curves which differ in detail.  We describe how to calculate an
approximately correct recombination history for arbitrary models in a
separate paper (Seager, Sasselov, \& Scott~1999). Since the main effect is
similar to an overall amplitude change, we normalized our CMB power spectra
to have
the same large-scale matter power spectrum, which is equivalent to normalizing
to the same amplitude for the initial conditions.  The amplitude of the
effect of our new H calculation clearly depends on the cosmology.  For the high
$\Omega_{\rm B}$ and $h$ case, the freeze-out value of $x_{\rm e}$ is
much smaller (around $10^{-5}$),
and since the fractional change in $x_{\rm e}$ is similarly $\sim10$\%,
the absolute change in ionization fraction is much lower than for the sCDM
model.  The integrated optical depth is directly proportional to
$\Omega_{\rm B}h\Delta x_{\rm e}$, which is small, despite the increase in
$\Omega_{\rm B}$ and $h$.  Hence we see a much smaller increase from our
hydrogen improvement in
Fig.~16(b).
The normalization
change is rather difficult to see, since it is masked by relatively small
changes around the power spectrum peaks, giving wiggles in the difference
spectrum.

The dashed lines in
Fig.~16
show the effect
on the power spectrum for CMB polarization.  In standard models polarization
is typically at the level of a few percent of the anisotropy signal, and
so will be difficult to measure in detail (see Hu \& White~1997b for a
discussion of CMB polarization).
We show the results here to indicate that there
are further observational consequences of our improved recombination
calculation (explicitly we have plotted the `E' mode of polarization, see
e.g.~Seljak 1997).  The effect of our improvements on the polarization can be
understood similarly through the visibility function.  Since the polarization
power spectrum tends to have sharper acoustic peaks, the wiggles in the
difference spectrum are more pronounced than for the temperature anisotropies.
Note that the large relative differences at low $\ell$
are actually very small in absolute terms, since the polarization
signal is so small there.  The polarization-temperature correlation power
spectrum and the `B' mode of polarization (for models with gravity
waves) could also be plotted, but little extra insight is gained, and
so we avoid this for the sake of clarity.

The other major difference we find compared with previous treatments
is in the delayed recombination of \ion{He}{1}.
In
Fig.~17
we show the effect of our new \ion{He}{1} calculation, again as a fractional
change in the CMB anisotropy power spectrum versus multipole
$\ell$. The change in the recombination of \ion{He}{1} affects the density
of free electrons just before
hydrogen recombination, which in turn affects the diffusion of the
photons and baryons, and hence the damping scale for the acoustic
oscillations which give rise to the peaks in the power spectrum.
The phases of the acoustic oscillations will also be affected
somewhat, which shows up in the wiggles in the difference spectrum.
For CDM-like models the
main effect is the change in the damping scale, since we now think there are
more free electrons at $z\sim1500$--2000.  The resulting change in the
$C_\ell$s is essentially the same as assuming the wrong angular scale for the
damping of the anisotropies (see Hu \& White~1997a), which is
the same physical effect that HSSW found in arguing
for the need to include \ion{He}{1} recombination {\it at all\/} for obtaining
percent accuracy in the $C_\ell$s. The effect of this improved \ion{He}{1} on
the power spectrum will depend on the background cosmology through the
baryon density ($\propto\Omega_{\rm B}h^2$) and the horizon size at
last scattering through $\Omega_0h^2$ -- hence there is no simple
fitting formula, and it is necessary to calculate the effect on the
anisotropy damping tail for each cosmological model considered.

\centerline{{\vbox{\epsfxsize=8cm\epsfbox{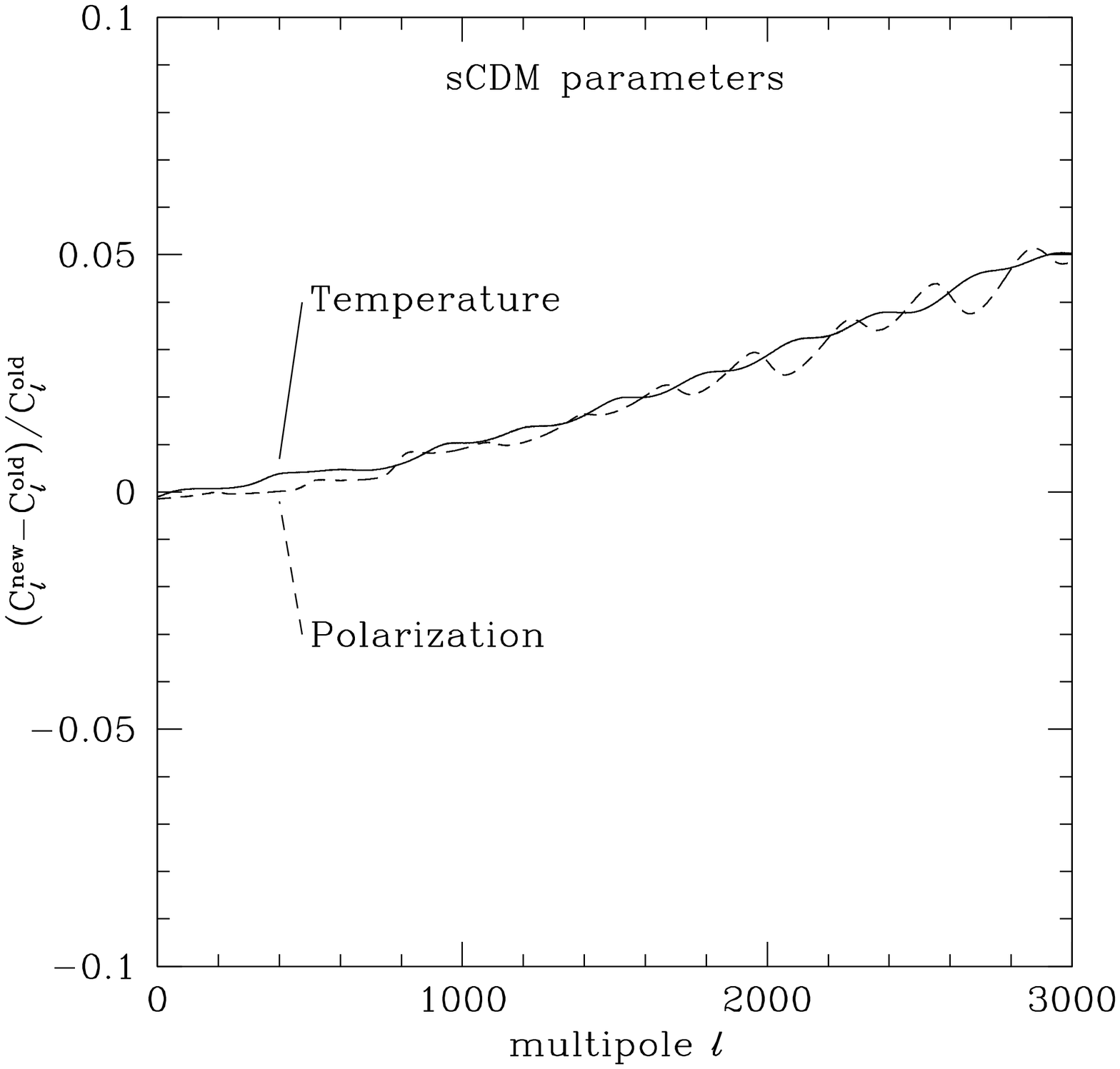}}}}
\centerline{{\vbox{\epsfxsize=8cm\epsfbox{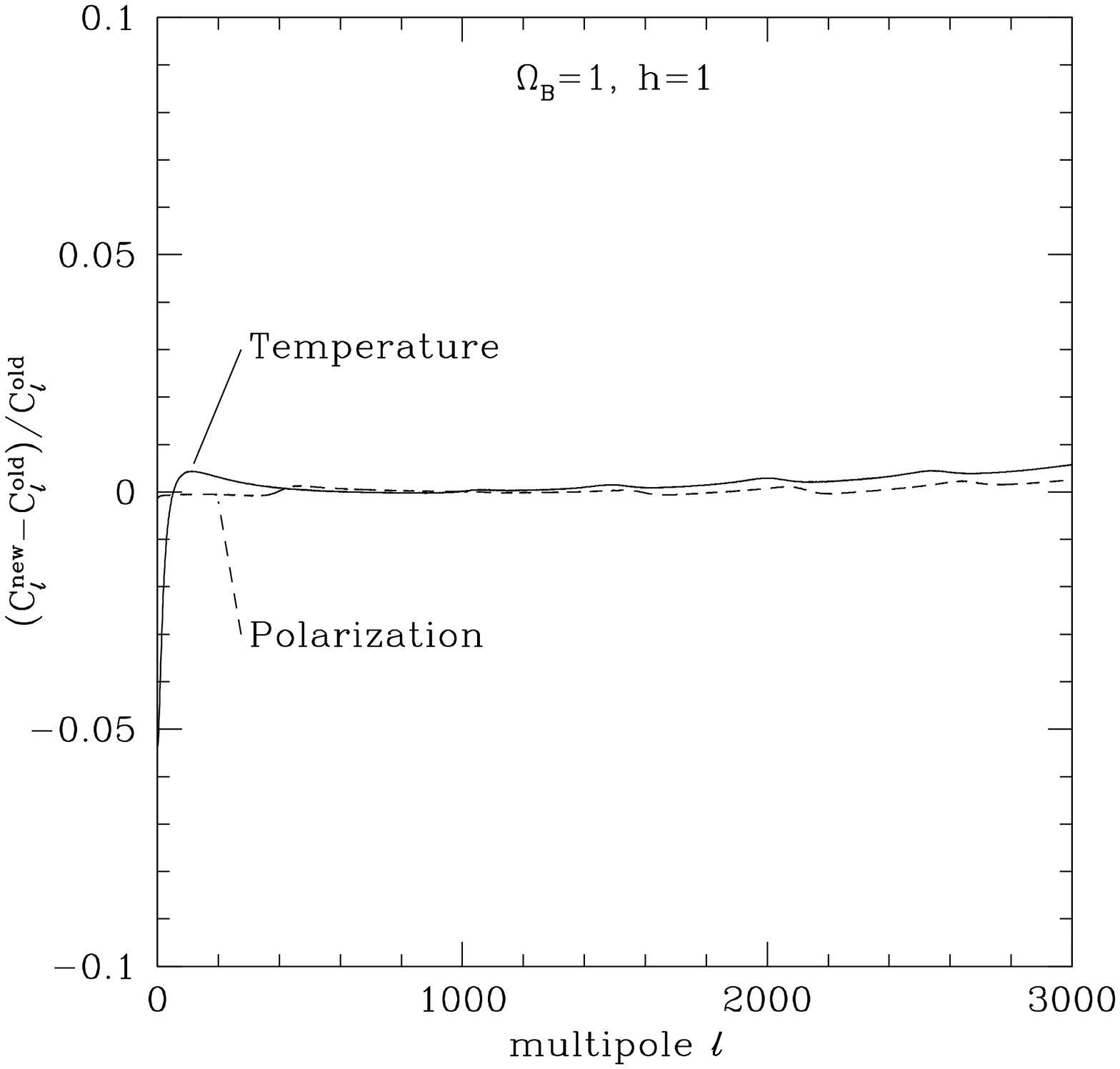}}}}
\baselineskip=8pt
{\footnotesize {\sc Fig.~17.---}
The effect of the improved treatment of helium on the CMB power
spectra for: (a) the standard CDM model; and (b) the extreme baryonic model
(for which there is essentially no change).  The fractional
difference plotted is between our new helium recombination calculation and
the assumption that helium follows Saha equilibrium (as in HSSW), with
the same `effective 3-level' hydrogen recombination used in both cases.
Again the sense is $C_\ell^{\rm new}-C_\ell^{\rm old}$, solid lines are
temperature, and dashed lines are polarization.
\label{fig:he_powerspectrum}
}
\vspace{2mm}

\baselineskip=11pt

There are really two parts to the \ion{He}{1} effect.  Firstly the extra
$x_{\rm e}$ makes the tight coupling regime tighter, so that the
photon mean free path is shorter, and the length scale for diffusion
is smaller.  Secondly, the effective damping scale comes from an
average over the visibility function, so an increase in the high-$z$
tail also leads to a smaller damping scale. The CMB anisotropies can
be thought of as a series of acoustic peaks multiplied by a roughly
exponential damping envelope, with the characteristic multipole of the
cut-off being determined by the damping length scale. As a result of
this smaller damping scale, the high $\ell$ part of the power
spectrum is {\it less\/} suppressed, and so we see an increase in
Fig.~17(a)
towards high $\ell$.  For the
$\Omega_{\rm B}=h=1$ model
(Fig.~17(b))
we
see only a very small effect at the highest $\ell$s.
This is easily understood by examining
Fig.~10,
where we see that \ion{He}{1} recombination is
pushed back to higher redshifts than for the sCDM case, and also that H
recombination happens earlier, which, together with the higher
$\Omega_{\rm B}$ and $h$,
shifts the peak of the visibility function to lower redshifts relative
to the recombination curve.  Hence the the high-$z$ tail of the
visibility function is much less affected in this case, and our
improved \ion{He}{1} calculation has essentially negligible effect.  However,
for less extreme models we find that the \ion{He}{1} effect is always at least
marginally significant.

\centerline{{\vbox{\epsfxsize=8cm\epsfbox{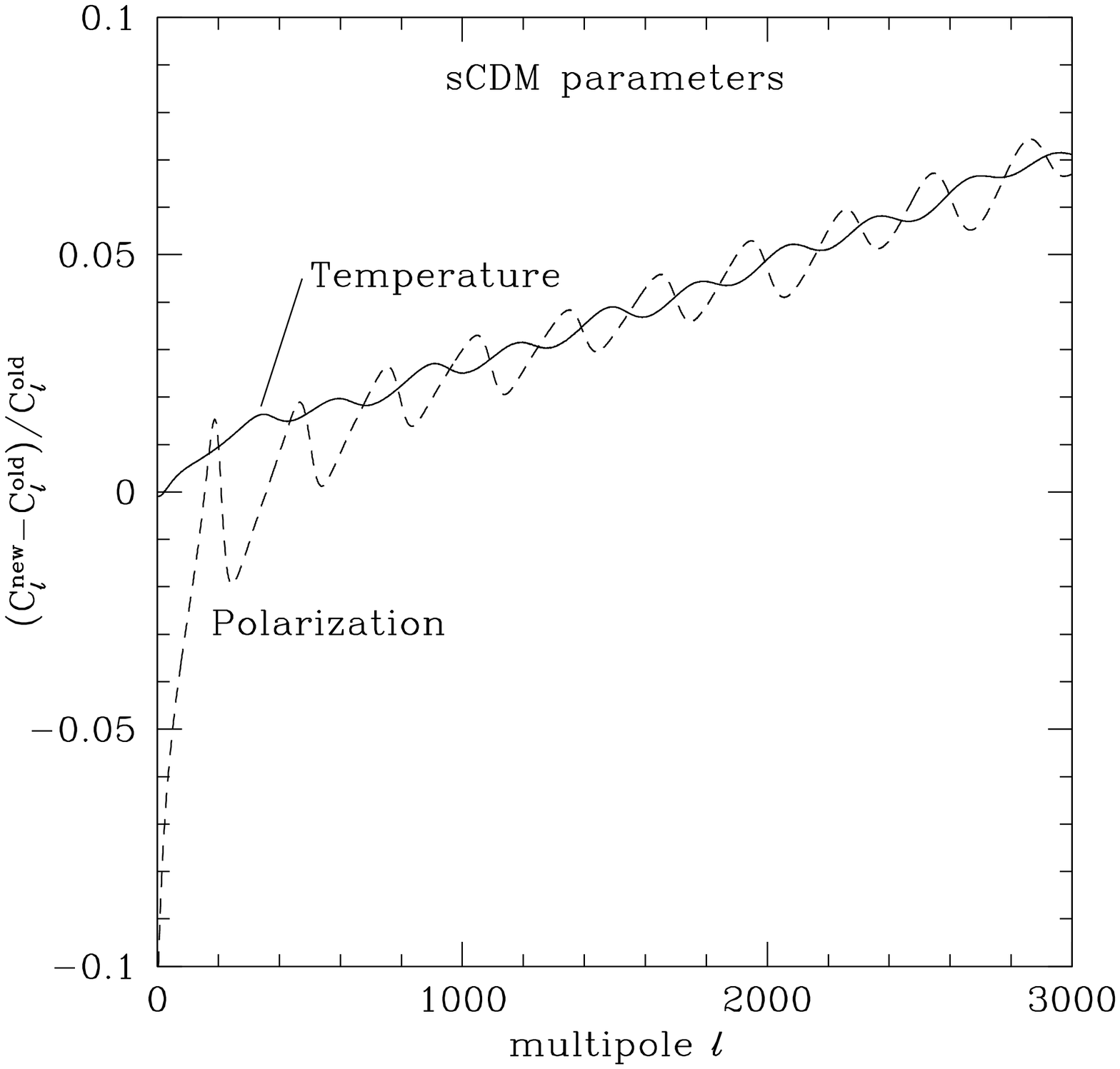}}}}
\centerline{{\vbox{\epsfxsize=8cm\epsfbox{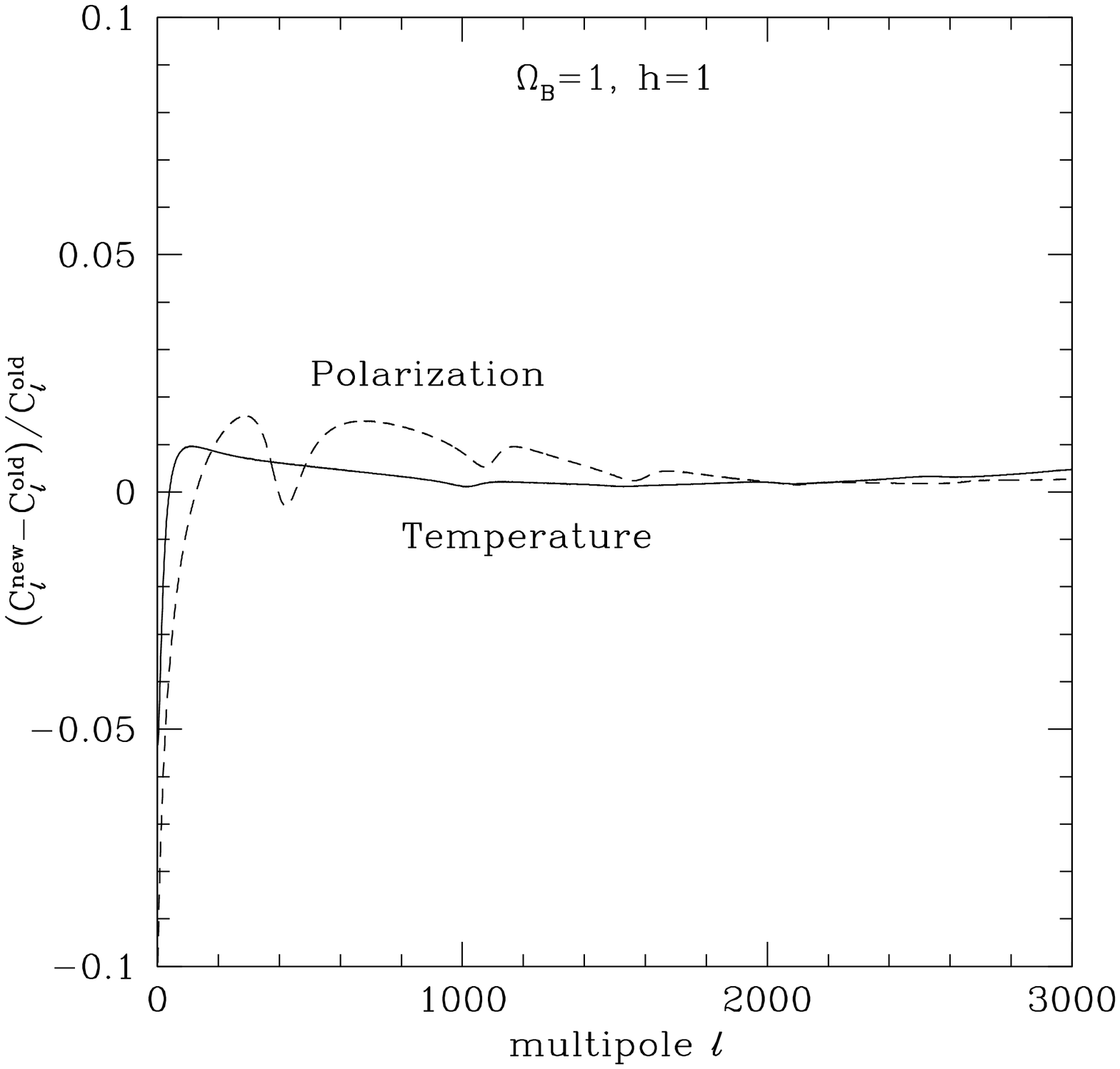}}}}
\baselineskip=8pt
{\footnotesize {\sc Fig.~18.---}
The total effect of our improvements on the CMB power
spectra for: (a) the standard CDM model; and (b) an extreme baryonic model.
These plots are essentially the sum of the separate effects of hydrogen
and helium.
\label{fig:tot_powerspectrum}
}
\vspace{2mm}

\baselineskip=11pt

Taking the two main effects together, for the sCDM model,
we find that they have essentially the
same sign, so that the total effect of our new calculation is more
dramatic (see
Fig.~18(a)).
Both effects lead to a slight increase in the anisotropies,
particularly at small angular scales (high $\ell$).  Although the
exact details will depend on the underlying cosmological model, we see
the same general trend for most parameters we have looked at.  The
change can be
higher than 5\% for the smallest scales, although it should be
remembered that the absolute amplitude of the anisotropies at such scales is
actually quite low.  Nevertheless, changes of this size are well above the
level which is relevant for future determinations of the power
spectrum.  As a rough measure, the cosmic variance at $\ell\sim1000$
is about 3\%. Hence
a 1\% change over a range of say 1000 multipoles is something like a
$10\sigma$ effect for a cosmic variance limited experiment.
What we can see is that although the effects are far from astonishing,
they are at a level which is potentially measurable.  Hence our improvements
are significant in terms of using future CMB data-sets to infer the values
of cosmological parameters; if not properly taken into account
these subtle effects in the
atomic physics of hydrogen and helium might introduce biases in the
determination of fundamental parameters.

\subsection{Spectral Distortions to the CMB}
\label{sec-Future}
With the model described in this paper we plan to calculate spectral
distortions to the blackbody radiation of the CMB today (see also
Dubrovich~1975; Lyubarsky \& Sunyaev 1983;
Fahr \& Loch 1991; Dell'Antonio \& Rybicki~1993; Burdyuzha \& Chekmezov 1994;
Dubrovich \& Stolyarov 1995, 1997; Boschan \& Biltzinger~1998).
The main emission from Ly$\,\alpha$ and the two-photon process will
be in the far-infrared part of the spectrum.
Transitions among the very high energy levels, which have very
small energy separations, may produce spectral distortions in the
radio. Although far weaker than distortions by the lower Lyman lines,
they will be in a spectral region less contaminated by background
sources.

Detecting such distortions will not be easy, since they are generally swamped
by Galactic infra-red or radio emission and other foregrounds.
Our new calculation
does not yield any vast improvement in the prospects for detection.  However,
confirmation of the presence of these recombination lines would be a
definitive piece of supporting evidence for the whole Big Bang paradigm.
Moreover, detailed measurement of the lines, if ever possible to carry out,
would be a direct diagnostic of the recombination process.  For these
reasons we will present spectral results elsewhere.

\section{Conclusions}

One point we would like to stress is that
our detailed calculation agrees very well
with the results of the effective 3-level atom.  This underscores the
tremendous achievement of Peebles, Zel'dovich and colleagues in so fully
understanding cosmic recombination 30 years ago.  However, the great goal
of modern cosmology is to determine the cosmological parameters to an
unprecedented level of precision, and in order to do so it is now
necessary to understand very basic things, like recombination, much more
accurately.  

We have shown that improvements upon previous recombination calculations
result in a roughly 10\% change in $x_{\rm e}$ at low redshift for most
cosmological models, plus a substantial delay in \ion{He}{1}
recombination, resulting in a few percent change in the CMB
power spectrum at small angular scales.
Specifically, the low redshift difference in $x_{\rm e}$ is due to the
H excited states' departure from an equilibrium distribution.  This in turn
comes from the level-by-level treatment of a 300-level H atom, which
includes all bound-bound radiative rates, and which allows
feedback of the disequilibrium of the excited states on
the recombination process.  The large improvement in
$x_{\rm e}$ during \ion{He}{1} recombination comes from the
correct treatment of the atomic levels, including
triplet and singlet states. While
it was already understood that \ion{He}{1} recombination would
affect the power spectrum at high multipoles (HSSW), our improved
\ion{He}{1} recombination affects even the start of H recombination for
traditional low $\Omega_{\rm B}$ models.  There is thus a substantially
bigger change in the $C_\ell$s, reaching to larger angular scales.

Careful use of $T_{\rm M}$ rather than $T_{\rm R}$ can also have
noticeable consequences, as to a lesser extent can the treatment of
Ly$\,\alpha$ redshifting using the Sobolev escape probability.
Our other new contributions to the recombination calculation produce negligible 
differences in $x_{\rm e}$. Collisional excitation and ionization for
H, \ion{He}{1} and for \ion{He}{2} are of little importance. Inclusion
of additional cooling and heating terms in the evolution of $T_{\rm M}$ also
produce little change in $x_{\rm e}$. The largest spectral distortions do not
feed back on the recombination process to a level greater than
$0.01\%$ in $x_{\rm e}$.  Finally, the H chemistry occurs too low in
redshift to make any noticeable difference in the CMB power spectrum. 

Although we have tried to be careful to consider every process we can
think of, it is certainly possible that other subtle effects remain to
be uncovered. We hope that we do not have to wait another 30
years for the next piece of substantial progress in understanding how
the Universe became neutral.

\acknowledgments
The program {\tt recfast}, which performs approximate calculations of the
recombination history is available at
{\tt http://www.astro.ubc.ca/people/scott/recfast.html} (FORTRAN version) and
at {\tt http://cfa-www.harvard.edu/ ${\sim}$sasselov/rec/} (C version).
We would like to thank George Rybicki, Ian Dell'Antonio, Avi Loeb, and
Han Uitenbroek for many useful conversations.  Also David Hummer and
Alex Dalgarno for discussions on the atomic physics, Alex Dalgarno and
Phil Stancil for discussions on the chemistry, Martin White, Wayne Hu
and Uro{\v s} Seljak for discussion on the cosmology, and Jim Peebles
for discussions on several aspects of this work.  We thank the referee
for a careful reading of the manuscript. Our study of effects
on CMB anisotropies was made much easier through the availability of
Matias Zaldarriaga and Uro{\v s} Seljak's code {\tt cmbfast}. DS is
supported by the Canadian Natural Sciences and Engineering Research
Council.




\appendix
\section{REACTION RATES AND CROSS SECTIONS}

\noindent Where cross sections are listed instead of reaction rates,
we calculate the reaction rate through the integrals for
photoionization (or photodissociation) and recombination as described
in \S\ref{sec-Recomb}.

\noindent
[1] De Jong 1972; [2] Hirasawa 1969; [3] Karpas et al.~1979;
[4] Shapiro \& Kang 1987; [5] Dove \& Mandy 1986;
[6] Gredel \& Dalgarno 1995; [7] Wishart 1979 and Broad \&
Reinhardt 1976.

\begin{tabular}{l l r}
\\ \hline
Reaction & Rate Coefficient (cm$^3$s$^{-1}$) & Reference \\ \hline

H$^{-}$ + H $\longrightarrow$ H$_{2}$ + e$^{-}$ & 1.30 x 10$^{-9}$ &
[1] \\

H$_{2}$ + e$^{-} \longrightarrow$ H$^{-}$ + H  &
2.70 x $10^{-8}T_{\rm M}^{-3/2}\exp(-43000/T_{\rm M})$ & [2] \\

H${_2}^{+}$ + H $\longrightarrow$ H$_2$ + H$^{+}$
& 6.40 x 10$^{-10}$ & [3] \\

H$_2$ + H$^{+} \longrightarrow$ H${_2}^{+}$ + H
& 2.4 x $10^{-9}\exp(-21200/T_{\rm M})$ & [4] \\

H + H${_2} \longleftrightarrow$ H + H + H
& 1.0 x 10$^{-10}\exp(-52000/T_{\rm M})$ & [5] \\

H${_2}$ + e$^- \longleftrightarrow$ H + H + e$^-$
& 2.0 x 10$^{-9} \left(T_{\rm M}/300\right)^{0.5}\exp(-116300/T_{\rm
M})$ & [6] \\

\\ \hline
\\ \hline
Reaction & Cross Section (cm$^2$) & Reference \\ \hline

H${_2}^{+}  + \gamma \longleftrightarrow$ H + H$^{+}$ &
 See expression in reference. & [4]\\

H${^-}  + \gamma \longleftrightarrow$ H  +  e$^{-}$ &
 See table in reference. & [7]\\

H  + $\gamma \longleftrightarrow$ H$^{+}$  +  e$^{-}$ &
 See section~\ref{sec-AtomicData}\\

He  + $\gamma \longleftrightarrow$ He$^{+}$  +  e$^{-}$ &
 See section~\ref{sec-AtomicData}\\

He$^{+}  + \gamma \longleftrightarrow$ He$^{++}$  +  e$^{-}$ &
 See section~\ref{sec-AtomicData}\\
\\ \hline
\\
\end{tabular}

\end{document}